\documentclass{tCPH2e}

\pdfoutput=1

\usepackage{epsfig}
\usepackage{diagrams} 
\usepackage{color}   
\usepackage{graphicx}
\usepackage{wrapfig}
\usepackage{amssymb}     
\usepackage{amsfonts}      

\DeclareMathSymbol{\C}{\mathbin}{AMSb}{"43}
\DeclareMathSymbol{\N}{\mathbin}{AMSb}{"4E}
\DeclareMathSymbol{\Z}{\mathbin}{AMSb}{"5A}
\DeclareMathSymbol{\R}{\mathbin}{AMSb}{"52}
\DeclareMathSymbol{\Q}{\mathbin}{AMSb}{"51}

\newcommand{\bit}{\begin{itemize}}
\newcommand{\eit}{\end{itemize}\par\noindent}
\newcommand{\ben}{\begin{enumerate}}
\newcommand{\een}{\end{enumerate}\par\noindent}
\newcommand{\beq}{\begin{equation}}
\newcommand{\eeq}{\end{equation}\par\noindent}
\newcommand{\beqa}{\begin{eqnarray*}}
\newcommand{\eeqa}{\end{eqnarray*}\par\noindent}
\newcommand{\beqn}{\begin{eqnarray}}
\newcommand{\eeqn}{\end{eqnarray}\par\noindent}

\def\bR{\begin{color}{red}}
\def\bB{\begin{color}{blue}}
\def\bM{\begin{color}{magenta}} 
\def\bC{\begin{color}{cyan}}
\def\bW{\begin{color}{white}}
\def\bBl{\begin{color}{black}}
\def\bG{\begin{color}{green}}
\def\bY{\begin{color}{yellow}}
\def\e{\end{color}}

\newcommand{\A}{$\bB A\e$}
\newcommand{\B}{$\bB B\e$}
\newcommand{\Aa}{\bB A\e}
\newcommand{\Bb}{\bB B\e}
\newcommand{\Cc}{\bB C\e}
\newcommand{\Dd}{\bB D\e}
\newcommand{\Ee}{\bB E\e}
\newcommand{\Ff}{\bB F\e}
\newcommand{\Mm}{\bB M\e}
\newcommand{\Xx}{\bB X\e}
\newcommand{\Yy}{\bB Y\e}
\newcommand{\Zz}{\bB Z\e}
\newcommand{\Uu}{\bB U\e}

\newcommand{\II}{{\bB{\rm I}\e}}

\newcommand{\Gcirc}{\bG\circ\e}

\doi{10.1080/0010751YYxxxxxxxx}
 \issn{1366-5812}
\issnp{0010-7514}

\jvol{00} \jnum{00} \jyear{2009} \jmonth{February}

\title{Quantum picturalism}

\author{Bob Coecke$^{\ast}$\thanks{$^\ast$Email: coecke@comlab.ox.ac.uk\vspace{6pt}} 
\\\vspace{6pt}  
{\em{Oxford University Computing Laboratory, Wolfson Building, Parks Road, OX1 3QD Oxford, UK}}
\\\vspace{6pt}
\received{Received 01 02 2009; final version received XX YY ZZZZ}}

\begin{document}

\maketitle

\begin{abstract}

Why did it take us 50 years since the birth of the quantum mechanical formalism to discover that unknown quantum states cannot be cloned? Yet, the proof of the `no-cloning theorem' is easy, and its consequences and potential for applications are immense.  Similarly, why did it take us 60 years to discover the conceptually intriguing and  easily derivable physical phenomenon of `quantum teleportation'?   We claim that the quantum mechanical formalism doesn't support our intuition, nor does it elucidate the key concepts that govern the behaviour of the entities that are subject to the laws of quantum physics. The arrays of complex numbers are kin to the  arrays of 0s and 1s of the early days of computer programming practice.  Using a technical term from computer science, the quantum mechanical formalism is  `low-level'.   In this review we present steps towards a diagrammatic  `high-level' alternative for the Hilbert space formalism, one which appeals to our intuition. 

The diagrammatic language  as it currently stands allows for intuitive reasoning about interacting quantum systems, and trivialises many otherwise involved and tedious computations.  It clearly exposes limitations such as the no-cloning theorem, and phenomena such as quantum teleportation.  As a logic, it supports `automation': it enables a (classical) computer to reason about interacting quantum systems, prove theorems, and design protocols.  It allows for a wider variety of underlying theories, and can be easily modified, having the potential to  provide the required step-stone towards a deeper conceptual understanding of quantum theory, as well as its unification with other physical theories.  Specific applications discussed here are purely diagrammatic proofs of several quantum computational schemes, as well as an analysis of the structural origin of quantum non-locality.

The underlying mathematical foundation of this high-level diagrammatic formalism relies on so-called \em monoidal categories\em, a product of a fairly recent development in mathematics. Its logical underpinning is \em linear logic\em, an even more recent product of research in logic and computer science.   These monoidal categories do not only provide a natural foundation for physical theories, but also for proof theory, logic, programming languages, biology, cooking, ...   So the challenge is to discover the necessary  additional pieces of structure that allow us to predict genuine quantum phenomena.  These additional pieces of structure represent the capabilities nature has provided us with to manipulate entities subject to the laws of quantum theory.

\end{abstract}

\section{Historical context}\label{sec:introduction}

With John von Neumann's ``Mathematische Grundlagen der Quantenmechanik'', published in 1932 \cite{vN}, quantum theory reached maturity, now having been provided with a rigourous mathematical underpinning.   Three year later something remarkable happened.  John von Neumann wrote in a letter to the renowned  American mathematician Garrett Birkhoff the following:  
\begin{quote}
{\normalsize{\it I would like to make a confession which may seem immoral: I do not believe
absolutely in Hilbert space no more} -- sic~\cite{Birk,Redei}}
\end{quote}
In other words, merely three years after completing something that is in many ways the most successful formalism physics has ever known, both in terms of experimental predictions, technological applications, and conceptual challenges, its creator denounced his own brainchild.  However, today,  more than 70 years later, we still teach John von Neumann's Hilbert space formalism to our students.   People did try to come up with alternative formalisms, by relying on physically motivated mathematical structures other than Hilbert spaces. For example,  in 1936 Birkhoff and von Neumann proposed so-called `quantum logic' \cite{BvN}. But quantum logic's disciples failed to convince the wider physics community of this approach's virtues.  There are similar alternative approaches due to Ludwig, Mackie, Jauch-Piron, and Foulis-Randall \cite{CMW}, but neither of these have made it into mainstream physics, nor is there any compelling evidence of their virtue.  

Today, more than 70 years later, we meanwhile did learn  many new things.  For example, we discovered new things about the quantum world and its potential for applications:
\bit
\item 
During the previous century, a vast amount of the ongoing discourse on quantum foundations challenged in some way or another the validity of quantum theory.  The source of this was the community's inability to craft a satisfactory worldview in the light of the following:
\bit
\item[-]
\em Quantum non-locality\em, or, the \em EPR paradox\em, that is: Compound quantum systems which may be far apart exhibit certain correlations that cannot be explained as having been established in the past when the two systems were in close proximity.  Rather, the correlations can only be explained as being instantaneously created over a large distance, hence `non-locality'. But remarkably, these correlations are so delicate that this process does not involve instantaneous transmission of information, and hence does not violate Einstein's theory of relativity.
\item[-]
The  \em quantum measurement problem\em, that is: There is no good explanation of what causes the wavefunction to collapse, and, there is no good explanation of the non-determinism in quantum measurements.  The latter turns out to be closely related to quantum non-locality.
\eit
We refer the reader to  \cite{Bub, Peres} for more details on these issues.  Many took these `paradoxes' or `quantum weirdness' to be tokens of the fact that there is something fundamentally wrong with quantum theory.  But this position that quantum theory is in some way or another `wrong' seems to be increasingly hard to maintain.  Not only have there been impressive experiments which assert quantum theory in all of its aspects, but also, several new quantum phenomena have been observed, which radically alter the way in which we need think about nature, and which raise new kinds of  conceptual challenges.   Examples of experimentally established new phenomena are quantum teleportation \cite{BBC}, which we explain in detail below, and quantum key exchange \cite{Ekert91}, for which we refer the reader to \cite{QKD}. In particular, the field of quantum information has emerged from embracing `quantum weirdness',
\em not as a bug, but as a feature\,\em\mbox{!} 
\item 
Within this quantum informatic endeavour we are becoming increasingly conscious of how central the particular behaviour of \em compound systems \em is to quantum theory.  One nowadays refers to this as the existence of \em quantum entanglement\em.  It is when compound quantum systems are in these entangled states that the non-local correlations can occur. The first to point at the  key role of quantum  entanglement within quantum theory was Schr\"odinger in 1935 \cite{Schrodinger}.  Most of the new phenomena discovered in the  quantum information era crucially rely on quantum entanglement.  But this key role of quantum  entanglement is completely ignored within the proposed alternatives to the Hilbert space formalism to which we referred above.  The key concepts of those approaches solely apply to individual quantum systems, and, it is a recognised soft spot of these approaches that they weren't able to reproduce entanglement in a canonical manner. In hindsight, this is not that surprising.  Neither the physical evidence nor the appropriate mathematical tools were available (yet) to establish 
a new formalism for quantum theory in which quantum entanglement plays the leading role. 
\eit
But today, more than 70 years later, this situation has changed, which brings us to other important recent developments.  These did not take place in physics, but in other areas of science:
\bit
\item 
Firstly, not many might be aware of the enormous effort that has been made by the computer science community to understand the mathematical structure of general \em processes\em, and in particular, the way in which they \em interact\em, how different configurations of interacting processes might result in the same overall process, and similar fairly abstract questions.  An accurate description of how concurrent processes precisely interact turns out to be far more delicate than one would imagine at first.  Key to solving these problems are appropriate mathematical means for describing these processes, usually referred to as their \em semantics\em. The research area of \em computer science semantics \em has produced a vast amount 
of new mathematical structures which enable us to design high-level programming languages.  You may ask, why do we need these high-level programming languages?  Well,  because otherwise there wouldn't be internet, there wouldn't be operating systems for your Mac or PC, there wouldn't be mobile phone networks, and there wouldn't be secure electronic payment mechanisms, simply because these systems are so complicated that getting 
things right wouldn't be possible without relying on the programming paradigms present in high-level programming languages  such as \em abstraction\em, \em modularity\em, \em compositionality\em, \em computational types\em, and many others.  
\item 
These developments in computer science went hand-in-hand with developments in proof theory, that is, the study of the structure of mathematical proofs.  In fact, the study of interacting programs is in a certain sense  `isomorphic' to the study of interacting proofs -- what this `certain sense' is should become clear to the reader after reading the remainder of this paper.  The subject of proof theory encompasses the subject of logic: 
while logic aims to establish whether one can derive a conclusion given certain premises in `yes/no'-terms, in proof theory one is also interested in 
how one establishes that something is either true or false. In other words, the \em process \em of proving things becomes an explicit part of the subject, and of particular interest is how certain `ugly' proofs can be transformed in `nicer' ones.  
In the late 1980's proof theoreticians became interested in how many times one uses (they say `consume') a certain premise within  proofs.  
To obtain a clear view on this they needed to strip logic from: 
\bit
\item[-] \em its implicit ability to clone premises\em.  This implicit ability to clone premises was made explicit as a logical rule by Gentzen in 1934 \cite{SequentCalculus}.  Concretely, `clone $A$ within context $\Gamma$' translates symbolically as $A,\Gamma \vdash A,A,\Gamma$ where the symbol $\vdash$ stands for `entails'. 
\item[-] \em its implicit ability to delete premises\em, cf.~`delete $A$ within context $\Gamma$' means $A,\Gamma\vdash \Gamma$. 
\eit
Stripping logic from these two rules gave rise to Girard's \em linear logic \em \cite{Girard}. Now, in quantum information theory we also have a no-cloning principle and a no-deleting principle:
\bit
\item[-] The \em no-cloning theorem\em, discovered in 1982 \cite{Dieks,WZ}, states that there is no physical operation which produces a copy of an arbitrary unknown quantum state.  Explicitly, there is no physical operation $f$ such that for any $|\psi\rangle$ we have $f(|\psi\rangle\otimes|0\rangle)=|\psi\rangle\otimes|\psi\rangle$.  The the fact that $|\psi\rangle$ is unknown is crucial here, since otherwise we could just prepare a copy of $|\psi\rangle$.
Although this fact was only discovered 25 years ago, its proof is extremely simple \cite{NoCloning}.
\item[-]  The \em no-deleting theorem \em discovered in 2000 \cite{Pati} requires a slightly more subtle formulation.
\eit
One may wonder whether there is a connection between the logical and the physical no-cloning and no-deleting laws.  In particular, the above indicates that maybe   this new `linear logic' might be more of a `quantum logic' than the original `Birkhoff-von Neumann quantum logic' which according to most logicians wasn't even a `logic'.  Another important new feature 
of linear logic was the fact that it had a manifestly geometrical aspect to it, which translated in purely diagrammatic characterisations of linear logic proofs and of proof transformations \cite{DanosReignier}.  These \em proof diagrams \em look very similar to those that you will encounter in this paper \cite{RossThesis}.
\item 
There exists an algebraic structure which captures interacting computational processes as well as linear logic, namely \em monoidal categories\em. 
Monoidal categories are a particular kind of \em categories\em.
Initially,  categories  were introduced as a meta-theory for mathematical structures \cite{EilenbergMacLane}, which enables one to import results of one area of mathematics into another.  Its 
consequently highly abstract nature earned it the not all too flattering name `generalised abstract nonsense'.  Nonetheless, categories, and monoidal categories in particular,  meanwhile play an important role in several areas of mathematical physics, e.g.~in a variety of approaches to quantum field theory, in statistical physics, and in several proposals for a theory of quantum gravity.   Important mathematical areas such as knot theory are also naturally described in terms of monoidal categories. But for us their highly successful use in logic and computer science is more relevant.    In those areas category theory is very established e.g.~at Oxford University Computing Laboratory we offer it to our undergraduates. 
To pass from categories in computer science to categories in physics, the following substitution will start the ball rolling:
\begin{center}
`computational process' $\mapsto$ `physical process'.
\end{center}
Once we find ourselves in the world of monoidal categories, language becomes purely diagrammatical.
Structuralism becomes picturalism, ... It are the monoidal categories which underpin linear logic that provide it with its diagrammatic proof theory.  
Physicist friendly introductions to monoidal categories are \cite{Cats,Baez,CPaqII,LNPBS,LNPAT,Selinger2}. These are ordered by increasing level of technicality.  A very pedestrian introduction to category theory is Lawvere and Schanuel's \em Conceptual Mathematics \em \cite{LawvereSchanuel}.  
Standard textbooks on category theory such as \cite{MacLane} are unfortunately mostly directed at pure mathematicians, what makes them somewhat inappropriate for physicists.
\eit
All these developments together justify a new attempt for a `better' formalism for quantum theory, say quantum logic mark II.  We are not saying that there is something wrong with the (current) predictions of quantum theory, but that the way in which we obtain these isn't great.  

In Section \ref{sec:ContKeyDevelopments} at the end of this paper we discuss what the main  applications of this new formalism are, as well as an account of the development of the subject and its contributors.  Before that, our main goal is to convince you of the following:\footnote{This presentation was directly taken from the report of on of the referees, since I couldn't put it better myself.}
\bit
\item That there is an algebraic structure -- monoidal category theory -- which is very general and abstract, but has a clear physical significance.  It is concerned with the description of any kind of processes, including physical processes, but also computational processes, and even cookery.
\item  That there is a pictorial representation of this algebraic structure which can be used to analyze this theory at a detailed level, and almost trivializes  reasoning within monoidal categories.
\item That the algebra of linear maps of finite dimensional Hilbert spaces is particularly strongly related to this pictorial approach, in particular in relation to the Hilbert space tensor product which we use in quantum theory to describe compound quantum systems. So a pictorial representation of quantum mechanics is particularly appropriate.
\item That in order to use this approach for quantum mechanical reasoning, a set of pictorial elements and rules for combining them is needed. 
Describing these constitutes the bulk of the article.  
These pictorial rules are easier to work with than the standard rules of quantum mechanics, and so lead to useful results more quickly. We claim that using them, a child of eight could do better at reasoning about  quantum phenomena than a high-school physics teacher.
\eit

The kind of quantum physics we are concerned with in this paper, rather than energy spectra of systems, are the new quantum phenomena that have been exposed in the quantum information theory era.  Many of these phenomena are presented in the form of a \em protocol\em. Such a protocol usually involves a number of agents, and each of these is supposed to perform a number of operations in a certain order, including communication actions, to achieve a certain goal.  By  \em quantum protocols \em one refers to the fact that when relying on quantum systems one can perform tasks which could not have been achieved when only relying on classical systems.

While this survey is written towards a non-specialist audience, we did include some notes and remarks for the more specialized reader, so that also he may find this survey useful.  

\section{A higher level of structure}

What do we mean by `high-level'? We explain this concept with an example and a metaphor.  

\subsection{High-level methods for linear algebra 101}\label{sec:LA101}

In linear algebra, \em projectors \em are linear operators ${\rm P}:{\cal H}\to{\cal H}$ which are: 
\bit
\item[-] \em self-adjoint \em  i.e.~${\rm P}^\dagger={\rm P}$, where ${\rm P}^\dagger$ means that we both conjugate and transpose ${\rm P}'s$ matrix, 
\item[-] \em idempotent \em i.e.~${\rm P}\circ{\rm P}={\rm P}$, where $\circ$ is composition of linear operators ($\simeq$ matrix multiplication).  
\eit
These projectors play a very important role in quantum theory since what happens to a state in a quantum measurement is described by a projector. Indeed, in quantum theory measurements are represented by self-adjoint operators, and it can be shown that for each self-adjoint operator $M$ on a finite-dimensional Hilbert space there are projectors ${\rm P}_1, \ldots, {\rm P}_n$ such that
\[
M=\sum_{i=1}^{i=n} r_i \cdot {\rm P}_i\qquad\mbox{for \ some}\qquad r_i\in\R\,.
\]
If an orthonormal basis $\{|i\rangle\mid 1\leq i\leq n\}$ consists of eigenvectors of $M$ then we can set  
${\rm P}_i=|i\rangle\langle i|$\,. Performing the quantum measurement $M$ means that we obtain one of the values $r_i\in\R$ as the outcome, and that the initial state $|\psi\rangle$ of the quantum system that is measured undergoes a change and becomes ${\rm P}_i(|\psi\rangle)$.  We can represent this change of state as $|\psi\rangle\mapsto{\rm P}_i(|\psi\rangle)$.  Below a basis will always be orthonormal and we abreviate its notation to $\{|i\rangle\}_i$.

We now recall the definition and a key property of the Hilbert space tensor product. If $\{|i\rangle\}_i$ is a basis for Hilbert space ${\cal H}$ and $\{|j\rangle\}_j$ is a basis for Hilbert space ${\cal H}'$ then we have\footnote{The notation $x:=y$ stands for ``$x$ is defined to be $y$''.  One doesn't loose much if one reads this as an ordinary equality.}
\[
{\cal H}\otimes{\cal H}':=\left\{\sum_{ij}\omega_{ij}\cdot|i\rangle\otimes|j\rangle \Biggm| \forall i,j: \omega_{ij}\in\C\right\}\,,
\]
that is, in words, the \em tensor product \em of Hilbert spaces ${\cal H}$ and ${\cal H}'$ consists of every vector of the form $\sum_{ij}\omega_{ij}\cdot|ij\rangle$, with complex coefficients $\omega_{ij}\in\C$, where we abbreviated $|i\rangle\otimes|j\rangle$ to $|ij\rangle$.
Two such vectors are equal if and only if for all $i,j$ the coefficients $\omega_{ij}$ coincide.  Hence two vectors  $\sum_{ij}\omega_{ij}\cdot|ij\rangle$ and $\sum_{ij}\omega_{ij}'\cdot|ij\rangle$ are equal if and only if  the matrices $(\omega_{ij})_{ij}$ and $(\omega_{ij}')_{ij}$ are equal. 
But each  matrix $(\omega_{ij})_{ij}$ is the matrix of some linear operator $\omega:{\cal H}'\to{\cal H}$, namely the one for which we have
$\omega(|j\rangle)=\sum_{i} \omega_{ij}\cdot |i\rangle$. This implies that there is a bijective correspondence between linear operators $\omega:{\cal H}'\to {\cal H}$ and the vectors of ${\cal H}\otimes{\cal H}'$. 
That is, after taking the transpose of the matrices $(\omega_{ij})_{ij}$, a bijective correspondence between linear operators $\omega:{\cal H}\to {\cal H}'$ and the vectors of ${\cal H}\otimes{\cal H}'$.
We exploit this correspondence in the following exercise.

\bigskip\noindent{\textbf{{Exercise:}} 
Consider a special kind of projectors, namely those of the form ${\rm P}=|\Psi\rangle\langle\Psi|$ with 

\begin{wrapfigure}{r}{0.22\textwidth}
\vspace{-10pt}
  \begin{center}
    \epsfig{figure=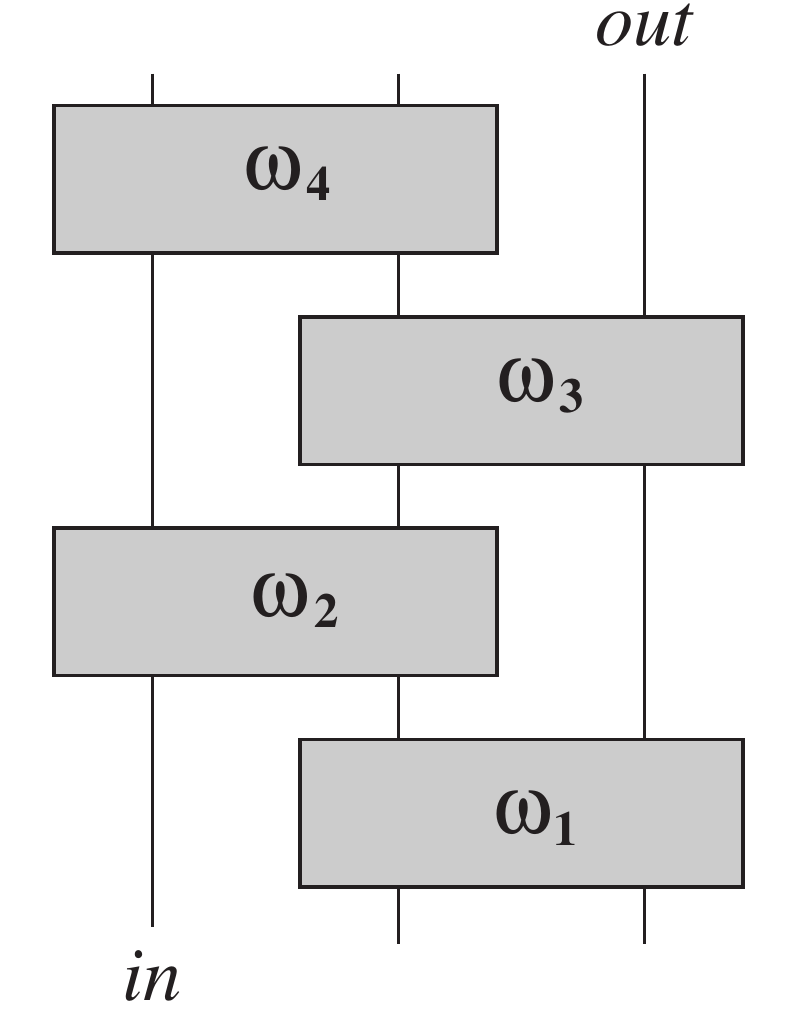,width=108pt}
  \end{center}
  \vspace{-10pt}
  \caption{Diagrammatic statement of the problem.  The boxes with labels $\omega_i$ represent the projectors ${\rm P}_i$. The reason why we take $\omega_i$ as labeling rather than labelling them ${\rm P}_i$ will become clear below.}
    \vspace{-10pt}
\end{wrapfigure}\noindent
\[
|\Psi\rangle:=\sum_{ij}\omega_{ij}\cdot|ij\rangle\in {\cal H}\otimes{\cal H}\,.
\]   
Hence these projectors act on the Hilbert space ${\cal H}\otimes{\cal H}$. As discussed above, we can think of the coefficients $\omega_{ij}$ on which they depend as the entries of the matrix of a linear operator $\omega:{\cal H}\to{\cal H}$.  The fact that ${\rm P}=|\Psi\rangle\langle\Psi|$ is a projector and hence idempotent imposes a normalization constraint on the operator $\omega$, but that won't be of any further importance here.  We will consider four such projectors ${\rm P}_1,{\rm P}_2,{\rm P}_3,{\rm P}_4$, respectively corresponding with linear operators $\omega_1,\omega_2,\omega_3,\omega_4$. Now,  consider a vector described in the tensor product of three Hilbert spaces, $\Phi\in{\cal H}_1\otimes{\cal H}_2\otimes{\cal H}_3$. Then,  first apply projector ${\rm P}_1$ to ${\cal H}_2\otimes{\cal H}_3$, then apply projector ${\rm P}_2$ to ${\cal H}_1\otimes{\cal H}_2$, then apply projector ${\rm P}_3$ to ${\cal H}_2\otimes{\cal H}_3$, and then apply  projector ${\rm P}_4$ to ${\cal H}_1\otimes{\cal H}_2$.  The question is: Given that $\Phi=\phi_{in}\otimes\Xi$ with $\phi_{in}\in{\cal H}_1$,  what is the resulting vector after applying all four projectors?  More specifically, since the resulting vector  will always be of the form $\Xi'\otimes \phi_{out}$ with $\phi_{out}\in{\cal H}_3$, something which follows from the fact that the last projector is applied to ${\cal H}_1\otimes{\cal H}_2$, what is the resulting vector $\phi_{out}$?  In short, can you write $\phi_{out}$ as a function of  $\phi_{in}$ given that
\[
({\rm P}_4\otimes 1_{{\cal H}_3})\circ
(1_{{\cal H}_1}\otimes {\rm P}_3)\circ
({\rm P}_2\otimes 1_{{\cal H}_3})\circ
(1_{{\cal H}_1}\otimes {\rm P}_1)\circ
(\phi_{in}\otimes\Xi)=\Xi'\otimes \phi_{out}\,?
\]
Since $\Phi\in{\cal H}_1\otimes{\cal H}_2\otimes{\cal H}_3$ describes the state of a tripartite quantum system, this situation could occur in physics when performing four bipartite measurements $M_1$, $M_2$, $M_3$ and $M_4$.

\bigskip
\noindent\textbf{Solution.}
However complicated the problem as stated might look, the solution is simple:
\[
\phi_{out}=(\omega_3\circ\bar{\omega}_4\circ\omega_2^T\circ\omega_3^\dagger\circ\omega_1\circ\bar{\omega}_2)(\phi_{in})
\]
where $\bar{\omega}_4$ is obtained by complex conjugating all matrix entries in the matrix of ${\omega}_4$, where $\omega_2^T$ is the transpose of $\omega_2$, and where $\omega_3^\dagger:=\bar{\omega}_3^T$ is the adjoint to $\omega_3$.  But what is more fascinating is that we can `read' this solution directly from the graphical representation -- see Figure \ref{fig:readsolution}.  We draw a line starting from `$in$' and whenever we enter a projector at one of its two inputs, we get \linebreak\vspace{-12pt}

\begin{wrapfigure}{r}{0.25\textwidth}
\vspace{-20pt}
  \begin{center}
    \epsfig{figure=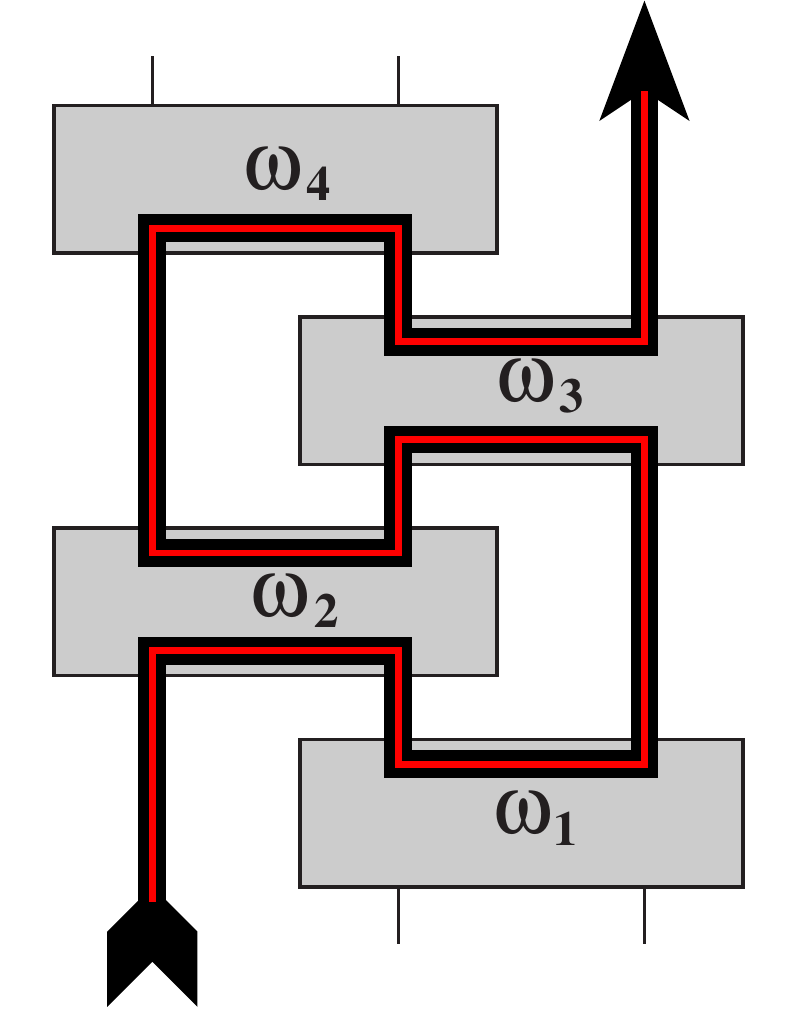,width=100pt}
  \end{center}
  \vspace{-10pt}
  \caption{`Reading' the solution of the exercise.}\label{fig:readsolution}
    \vspace{-10pt}
\end{wrapfigure}\noindent
out via the  other input,  and whenever we enter a projector at one of its two outputs, we get out via the other output. The expression 
\beq\label{ex:order}
\omega_3\circ\bar{\omega}_4\circ\omega_2^T\circ\omega_3^\dagger\circ\omega_1\circ\bar{\omega}_2
\eeq
is obtained by following this line and by composing all labels we encounter on our way, in the order we encounter them, and whenever we encounter it after entering from an input we moreover conjugate all matrix entries, and whenever we encounter it while going from right to left we also take the transpose.   

\bigskip
So  what at first looks like a pure `matrix hacking'-problem is governed by beautiful `hidden' geometry.  This principle is not specific to the above four-projector situation, but applies to any configuration of such projectors \cite{LE}. 

At first sight it might seem that the problem which we solved is totally artificial without any applications.  But it isn't, since  special cases of this exercise, depicted in Figures \ref{TelePicLE}, \ref{TeleBisPicLE} and  \ref{SwapPicLE}, constitute the structural core of the quantum teleportation protocol \cite{BBC}, the logic-gate teleportation protocol \cite{Gottesman}, and the entanglement swapping protocol \cite{BBC,Swap}, where missing labels stand for identities.  Let us here briefly sketch what these protocols are.

\begin{wrapfigure}{r}{0.24\textwidth}
\vspace{-26pt}
\begin{center}
\epsfig{figure=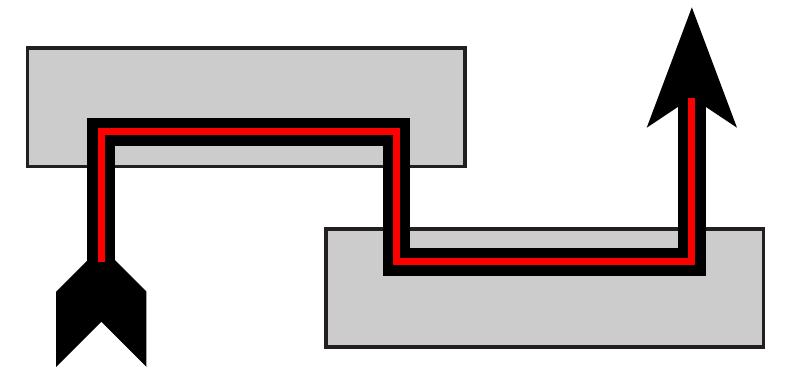,width=100pt}
\end{center}
\vspace{-10pt}
  \caption{The structural core of quantum teleportation.}\label{TelePicLE}
\vspace{-10pt}
\end{wrapfigure}
The \em quantum teleportation \em protocol involves two agents, usually named Alice and Bob. Alice possesses a \em qubit\em, that is, a quantum system described in a two-dimensional Hilbert space.  It is in an unknown state.  She and Bob also each possess one qubit of a pair of qubits in the entangled state $\sum_i{1\over\sqrt{2}}|ii\rangle$, called the \em Bell-state\em. Alice performs a \em Bell-base measurement \em on her two qubits, that is, a bipartite measurement which has the Bell-state as one of its eigenvectors.  If the measurement outcome is the eigenvalue corresponding to the 
Bell-state, then Bob's qubit turns out to be in the unknown state Alice's qubit initially was.  Figure \ref{TelePicLE} captures this situation as follows: the incoming arrow represents Alice's initial unknown qubit, the box besides it represents the shared Bell-state, the box above it represents the measurement that Alice makes, and the outgoing arrow represents Bob's resulting qubit.  The fact that the line now connects Alice's initial qubit with Bob's resulting qubit, and that the only labels it encounters are identities, implies that Bob's qubit is now in the state Alice's qubit was, formally $\phi_{out}=\phi_{in}$.  The full-blown quantum teleportation protocol goes a bit further, also accounting for when the measurement outcome does not have the Bell-state as one of its eigenvectors.  This can also be easily accounted for diagrammatically, as we show in Section \ref{sec:Uturns}.
 
\begin{wrapfigure}{r}{0.24\textwidth}
\vspace{-12pt}
\begin{center}
\epsfig{figure=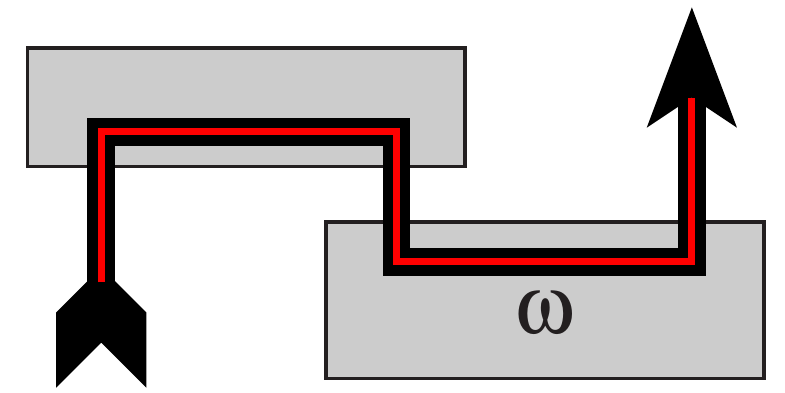,width=100pt}
\end{center}
\vspace{-10pt}
  \caption{The structural core of logic-gate teleportation.}\label{TeleBisPicLE}
\vspace{-10pt}
\end{wrapfigure}
The \em logic gate teleportation \em protocol is an elaboration on the quantum teleportation protocol which not only transfers a state from one agent to another, but at the same time applies a linear operator $\omega$ to it.  It should be clear from the above that Figure \ref{TeleBisPicLE} indeed realises this.  This logic-gate teleportation protocol is important since it enables one to \em process \em quantum information in a robust manner.

\begin{wrapfigure}{r}{0.27\textwidth}
\vspace{-16pt}
\begin{center}
\epsfig{figure=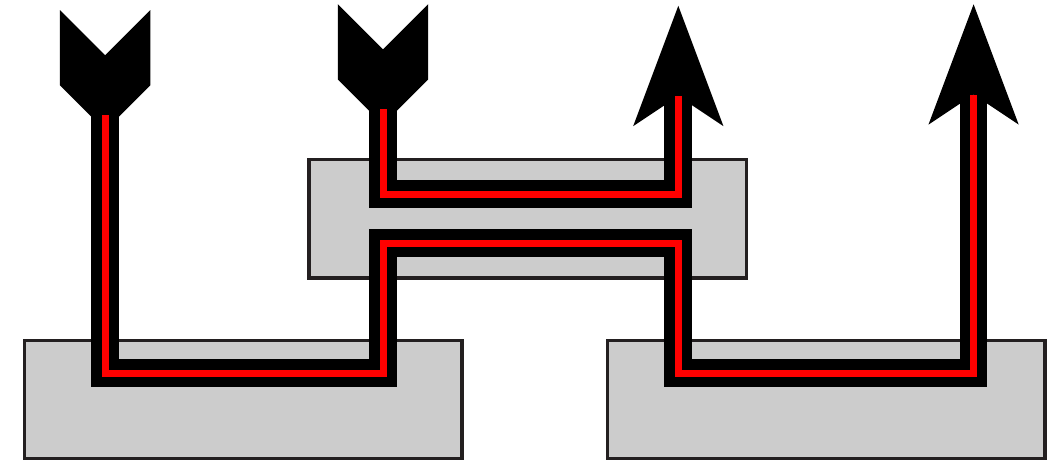,width=135pt}
\end{center}
\vspace{-10pt}
  \caption{The structural core of entanglement swapping protocol.}\label{SwapPicLE}
  \vspace{2pt}
\end{wrapfigure}
The \em entanglement swapping protocol \em is another variation on the same theme.  We start with two Bell-states, then apply a Bell-base measurement to one qubit in each pair.  The result of this is that not only the two qubits we measured are in a Bell-state, but also the two ones we didn't touch.
We can read this from Figure \ref{SwapPicLE} since these qubits are now connected by a line.

For a full derivation of these protocols in terms of the geometric reading of projectors, consult \cite{LE}.  

Note also that the resulting order of these labels $\omega_1,\ldots,\omega_4$ in expression (\ref{ex:order}) seems to ignore the order in which we applied the corresponding projectors ${\rm P}_1,\ldots,{\rm P}_4$. 
Here we won't discuss the physical interpretation of this `line', but just mention that  this `seemingly acausal' flow of information in this diagram has been a source of serious confusion, e.g.~\cite{Laforest}.

\bigskip
The above example shows that pictures can do more than merely provide an illustration or a convenient representation: they can provide reasoning mechanisms, i.e.~\em logic\em.  We now show that they have the capability to comprise \em equational content\em.  The representation of linear operators as pictures on which we implicitly relied in the previous exercise went as follows:
\vspace{-1.5mm}\beq\label{pics:defs1}
f \  \equiv \ \raisebox{-0.52cm}{\epsfig{figure=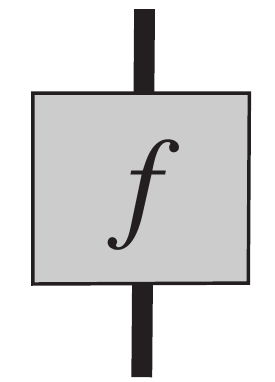,width=24pt}}
\qquad\qquad 
 1_A \  \equiv \ \raisebox{-0.52cm}{\epsfig{figure=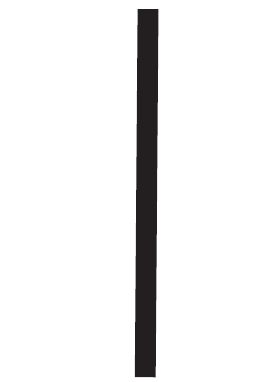,width=24pt}}
\quad\qquad
g\circ f \  \equiv \ \raisebox{-0.90cm}{\epsfig{figure=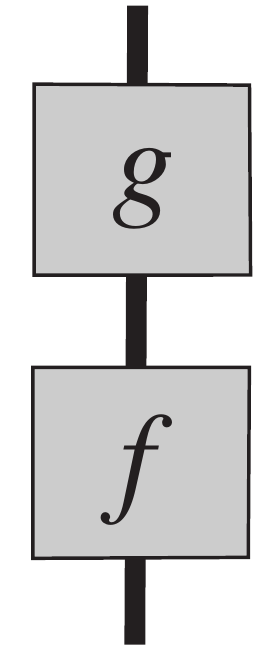,width=24pt}}
\quad\qquad 
f\otimes g \  \equiv \    \raisebox{-0.52cm}{\epsfig{figure=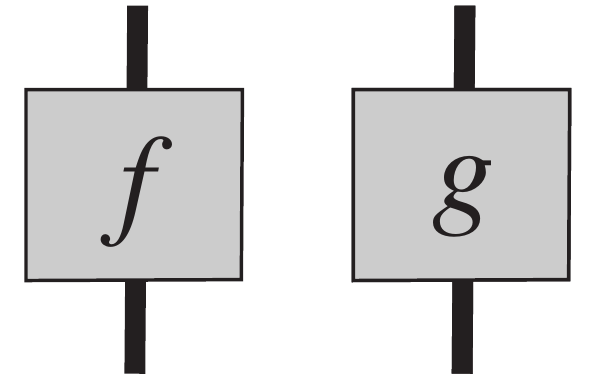,width=50pt}}\vspace{-1.5mm}
\eeq
So operators are represented by boxes with an input and an output wire.  In fact, we will also allow for more than one wire, or none. Identities are represented by wires, composition by connecting input wires to output wires, and the tensor by putting boxes side-by-side.  

You may or you may not know that any four linear operators satisfy the equation:
\beq\label{eq:bifunct}
( g\circ f)\otimes( k\circ h)
=
( g\otimes k)\circ( f\otimes h)
\eeq
It is an easy although somewhat tedious  exercise to verify this equation.  How does this equation, which only involves composition and tensor, translates into pictures?  We have:
\begin{center}
\quad\
$g\circ f$\ $\equiv$\ \raisebox{-0.87cm}{\epsfig{figure=Ober1.pdf,width=23pt}} 
\qquad\ \  and \qquad\ \ 
$k\circ h$\ $\equiv$\ \raisebox{-0.93cm}{\epsfig{figure=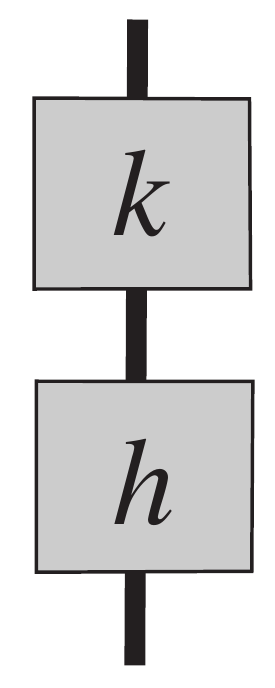,width=23pt}} 
\qquad\ \   so \quad
$( g\circ f)\otimes( k\circ h)$
\ $\equiv$\ \raisebox{-0.95cm}{\epsfig{figure=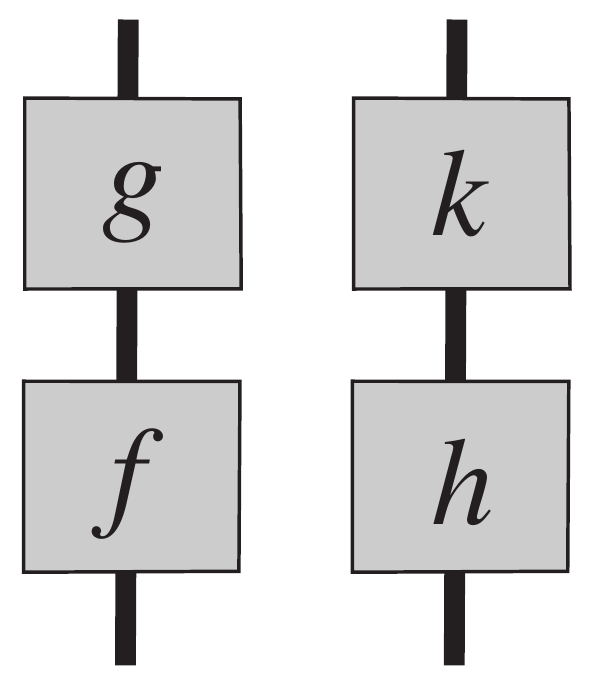,width=50pt}}
\end{center}
On the other hand we have:
\begin{center}
$f\otimes h$\ $\equiv$\    \raisebox{-0.50cm}{\epsfig{figure=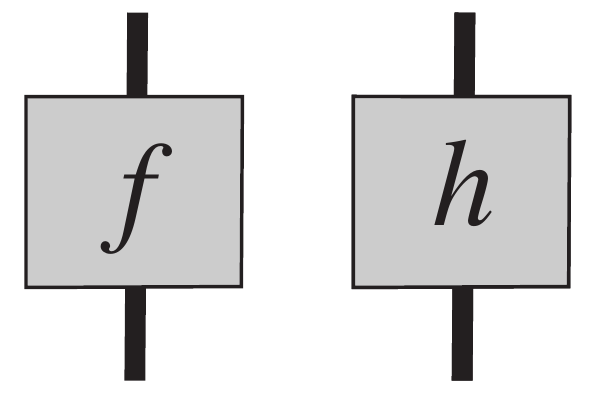,width=50pt}}
\quad and \quad
$g\otimes k$\ $\equiv$\    \raisebox{-0.51cm}{\epsfig{figure=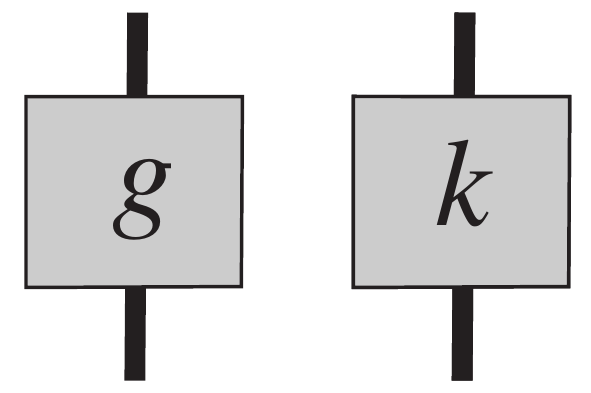,width=50pt}}
\quad  so \quad
$( g\otimes k)\circ( f\otimes h)$
\ $\equiv$\ \raisebox{-0.95cm}{\epsfig{figure=Ober1tris.pdf,width=50pt}}
\end{center}
So we obtain a \em tautology\,\em!  This means that the so innocent looking way in which we represented composition and tensor of linear operators as pictures already implies validity of eq.(\ref{eq:bifunct}).  Hence these simple pictures already carry non-trivial equational content.

\subsection{A metaphor: what do we look at when watching television?} 

So we just saw that there is more to linear algebra than `hacking with matrices'.  Other features, namely the role played by the line  in the above exercise, and  the tautological nature of eq.(\ref{eq:bifunct}), show that there are structures which emerge from the underlying matrix manipulations.   

Similar situations also occur in everyday life.  When watching television, we don't observe the `low-level' matrices of tiny pixels the screen is made up from, but rather the `high-level' gestalts of each of the figuring entities (people, animals, furniture, ...) which make up the story that the images convey. These entities and their story is the essence of the images, while the matrix of pixels is just a technologically convenient representation, something which can be send as a stream of data from broadcaster to living room.  What is special about this representation is that, provided the pixels are small enough, they are able to capture any image.  

A different representation consists of a library which includes images of all figuring entities, to which we attribute coordinates. This is done in computer games.  This representation is closer to the actual content of the images, but would be unfeasible unfeasible for television images.  

In modern computer programming, one does not `speak' in terms of arrays of 0s and 1s, although that's truly the data stored within the computer, but rather relies on high-level concepts about information flow.  A typical example are the flow charts which are purely diagrammatic.  

We sense an analogy of all of this with the status of the current quantum mechanical formalism.  The way we nowadays reason about quantum theory is still very `low-level', in terms of  arrays of complex numbers and matrices which transform these arrays. Just like the pixels of the television screen, the arrays of complex numbers have the special property that they allow to represent all entities of the quantum story.   So while we do obtain accurate representations of physical reality, it might not be the best way to understand it, and in particular, to reason about it.   


\section{General compositional theories}

Groups and vector spaces are examples of \em algebraic structures \em that are well-known to physicists. Obviously there are many other kinds of algebraic structures.  In fact, there exists an algebraic structure which is such that `something is provable from the axioms of this algebraic structure' if and only if `something can be derived within the above sketched diagrammatic language'.  

Let us  make this more precise.   An algebraic structure typically consists of: (i) some \em elements \em $a,b,c, ...$; (ii) some \em operations \em such as multiplying, taking the inverse,  and these operations also include special elements such as the unit;  (iii) some \em axioms \em (or otherwise put, \em laws\em).  For example, for a group the operations are a binary operation $-\cdot-$ which assigns to each pair of elements $a, b$ another element $a\cdot b$, a unitary operation $(-)^{-1}$ which assigns to each element $a$ another element $a^{-1}$, and  a special element $e$.   The axioms for a group are $x \cdot (y\cdot z)=(x \cdot y)\cdot z$, $x\cdot e=e\cdot x=x$, and $x^{-1}\cdot x=x\cdot x^{-1}=e$, where $x,y,z$ are now \em variables \em that range over all elements of the group. These axioms tell us that the operation $-\cdot-$ is associative and has $e$ as its unit, and that the operation $(-)^{-1}$ assigns the inverse to each element. The case of a vector space is a bit more complicated as it involves two sets of elements, namely the elements of the underlying field, as well as the vectors themselves, but the idea is again more or less the same. 

Let us be a bit more precise of what we mean by an axiom.  By a \em formal expression \em we mean an expression involving both elements and operations, and typically the elements are variables.  For example, in the case of a group $x\cdot (y\cdot z)^{-1}$ is such a formal expression.  An axiom is an equation between two formal expressions which holds as part of the definition of the algebraic structure.  But in general there are of course many other  equations between two formal expressions
that hold for that algebraic structure, e.g.~$x\cdot (y\cdot z)^{-1}=(x\cdot z^{-1})\cdot y^{-1}$ for groups.  

What we claim is that there is a certain algebraic structure defined in terms of elements, operations and axioms, such that for the picture calculus the following holds:
\ben
\item to each picture we can associate a formal expression of that algebraic structure\,;  
\item conversely, to each algebraic formal expression we can associate  a picture\,;
\item most surpringly, any equation between two pictures is derivable from the intuitive rules in the diagrammatic calculus, if and only if the corresponding  formal expressions are derivable from the axioms of the algebraic structure.
\een
In other words, the picture calculus and the algebraic structure are essentially the same, despite the fact that at first sight they look very different. But rather than just formally defining this algebraic structure, we want to provide the reader first with an intuitive feel for it, as it is quite different from the algebraic structures physicists are used to manipulate.  

Previous experiences have, somewhat surprisingly, indicated the nature of this structure, and its generality, is best conveyed without making reference to physics.  Therefore we present, ...

\subsection{The algebra of  cooking}

Let  \A~be a \bB raw potato\e.  \A~admits many \bR\underbar{\em\bBl states\e\em} \e e.g.~\bR dirty\e, \bR clean\e,  \bR skinned\e, ...  We want to \bR\underbar{\em\bBl process\e\em} \e \A~into \bB cooked potato \B\e.  Also  \B~admits many \bR\underbar{\em\bBl states\e\em} \e e.g.~\bR boiled\e, \bR fried\e,  \bR deep fried\e,  \bR baked with skin\e, \bR baked without skin\e, ... 
Correspondingly, there are several ways to turn  \A~into  \B~e.g.~\bR boiling\e, \bR frying\e, \bR baking\e, respectively referred to as $\bR f\e$,
$\bR f'\e$ and $\bR f''\e$.  We make the fact that these cooking \bR\underbar{\em\bBl process\e\em} \e apply to
\A~and  produce \B~explicit within the notation of these \bR\underbar{\em\bBl processes\e\em}\e:\vspace{-4pt}
\[ 
\bR \Aa\rTo^f \Bb\qquad\qquad \Aa\rTo^{f'} \Bb\qquad\qquad \Aa\rTo^{f''} \Bb\e\,.\vspace{-8pt}
\]
Our use of colours already indicated that \bR\underbar{\em\bBl states\e\em} \e are themselves \bR\underbar{\em\bBl processes\e\em} \e too:\vspace{-4pt}
\[ 
\bB \II\e\bR\rTo^\psi \Aa\e\,,\vspace{-8pt}
\]
where $\bB \II\e$ stands for \bB unspecified \e or \bB unknown\e, i.e.~we don't need to know from what \bB\underbar{\em\bBl system\e\em} \e \A\ has been produced, just that it is  in \bR\underbar{\em\bBl state\e\em} \e $\bR\psi\e$ and available for  \bR\underbar{\em\bBl processing\e\em}\e.  Let\vspace{-4pt}
\[
\bR \Aa\rTo^{f} \Bb\rTo^{g} \Cc \e \ = \ \bR \Aa\rTo^{g\,\Gcirc\, f} \Cc\e\vspace{-8pt}
\]
be the \bG\underbar{\em\bBl composite\e\em} \e \bR\underbar{\em\bBl process\e\em} \e of
first \bR boiling $\bBl=\e\Aa\rTo^f \Bb$ \e and then \bR salting $\bBl=\e\Bb\rTo^g \Cc$\e\,, and let\vspace{-4pt}
\[ 
\Xx\bR\rTo^{{\bf\bR 1\e}_{\Xx}}\e\Xx \vspace{-8pt}
\] 
be \bR doing nothing \e to $\Xx$. Clearly we have $\bR {\bf\bR 1\e}_{\Yy}\Gcirc \xi\e=\bR\xi\Gcirc {\bf\bR 1\e}_{\Xx}\e= \bR\xi\e$ for all  \bR\underbar{\em\bBl processes\e\em} \e  $\bR\Xx\rTo^{\xi}\Yy\e$.
Let $\bB A\bG\otimes\e D\e$ be \bB potato $A$ \bG and \e carrot $D$\e\,, and let\vspace{-4pt}
\[ 
\Aa\bG\otimes\e \Dd\bR\rTo^{f\bG \otimes\e h}\e\Bb\bG\otimes\e\Ee 
\qquad\mbox{and}\qquad \Cc\bG\otimes\e\Ff\bR\rTo^{x}\e \Mm\vspace{-8pt}
\] 
respectively be \bR boiling \e \bB potato \A \e\ \bG while \e \bR frying \e \bB carrot \Dd\e, and,  \bR mashing \e  \bB spiced cooked potato $\Cc$\ \e \bG and \e \bB spiced cooked carrot $\Ff$\e.  The whole  \bR\underbar{\em\bBl process\e\em} \e from \bB raw ingredients \e $\Aa$ and $\Dd$ to  \bB meal \e $\Mm$ is:
 \[ 
\bR\Aa\bG\otimes\e \Dd\rTo^{f\bG\otimes\e h}\Bb\bG\otimes\e \Ee\rTo^{g\bG\otimes\e k}\Cc\bG\otimes\e
\Ff\rTo^{x} \Mm\ \ \e {\bf=}\bR\ \Aa\bG\otimes\e \Dd\rTo^{x\bG\circ\e(g\bG\otimes\e k)\bG\circ\e (f\bG\otimes\e h)}\Mm\e.
\]
A \bM\underbar{\em\bBl recipe\e\em} \e is the sequence of consecutive \bR\underbar{\em\bBl processes\e\em} \e which we apply:
\[
\bM\left(\bR\Aa\bG\otimes\e \Dd\rTo^{f\bG\otimes\e h}\Bb\bG\otimes\e \Ee\e\mbox{\,\large\bf,\ }
\bR \Bb\bG\otimes\e \Ee\rTo^{g\bG\otimes\e k}\Cc\bG\otimes\e\Ff\e\mbox{\,\large\bf,\ }
\bR \Cc\bG\otimes\e \Ff\rTo^{x} \Mm\e
 \right)\e\,.
\]
Of course, many \bM\underbar{\em\bBl recipes\e\em} \e might actually result in the same \bR\underbar{\em\bBl process\e\em} \e -- cf.~in a group it is possible that while $x\not=x'$ and $y\not=y'$, and hence $(x,y)\not=(x',y')$, we have $x\cdot y=x'\cdot y'$.  Some equational statements may only apply to specific  \bM\underbar{\em\bBl recipes\e\em} \e while others apply at the level of formal expressions, and we refer to the latter as \bM\underbar{\em\bBl laws govering recipes\e\em}\e.  Here is one such \bM\underbar{\em\bBl law governing recipes\e\em}\e:
\[
({\bf\bR 1\e}_{\Yy}\bG\otimes\e \bR \zeta\e)\bG\circ\e(\bR \xi\e\bG\otimes\e {\bf\bR 1\e}_{\Zz}){\bf\ =\ }(\bR \xi\e\bG\otimes\e {\bf\bR 1\e}_{\Uu})\bG\circ\e({\bf\bR 1\e}_{\Xx}\bG\otimes\e \bR \zeta\e)\,.
\]
For example, for $\Xx:=\Aa$, $\Yy:=\Bb$, $\Zz:=\Cc$, $\Uu:=\Dd$, $\bR \xi\e:=\bR f\e$ and $\bR \zeta\e:=\bR g\e$ we have:
\begin{center}
\bR boil \bB potato \e\bG then \e fry \bB carrot\e \e
\ \  = \ \  \bR fry \bB carrot \e \bG then \e boil \bB potato\e\e\,.
\end{center}
This law is only an instance of a more general law on recipes, namely 
\[ 
(\bR \zeta\e\bG\circ\e \bR \xi\e)\bG\otimes\e(\bR \kappa\e\bG\circ\e \bR \omega\e){\bf\ =\ }(\bR \zeta\e\bG\otimes\e \bR \kappa\e)\bG\circ\e(\bR \xi\e\bG\otimes\e \bR \omega\e)\,,
\] 
which in the particular case of $\bR \xi\e:=\bR f\e$, $\bR \zeta\e:=\bR g\e$, $\bR \kappa\e:=\bR k\e$ and $\bR \omega\e:=\bR h\e$ reads as:\vspace{1.5mm}
\begin{center}
\bR boil \bB potato \e \bG then \e salt \bB potato\e\bBl,\ \e \bG while\e\bBl,\ \e fry \bB carrot \e \bG then \e pepper \bB carrot \e\e\vspace{-1mm}
\end{center}
\centerline{$|\hspace{0.05mm}|$}\vspace{-2mm}
\begin{center}
\bR boil \bB potato \e \bG while \e fry \bB carrot\e\bBl,\ \e \bG then\e\bBl,\ \e salt \bB potato \e \bG while \e pepper \bB carrot\e\e\,.\vspace{1.5mm}
\end{center}
Note in particular that we rediscover eq.(\ref{eq:bifunct}) of the previous section, which was then a tautology within the picture calculus, and is now a general law on cooking processes. 

It should be clear to the reader that in the above we could easily have replaced cooking processes, by either biological or chemical processes, or mathematical proofs or computer programs, or, obviously, \em physical processes\em.  So eq.(\ref{eq:bifunct}) is a general principle that applies whenever we are dealing with any kind of \bB\underbar{\em\bBl systems\e\em} \e and \bR\underbar{\em\bBl processes\e\em} \e thereon.  The mathematical structure of these is a bit more involved than that of a group. While for a group we had elements, operations, and laws i.e.~equations between formal expressions, here:
\bit
\item[(C1)] Rather than an underlying set of elements, as in the case of a group, we have two sorts of things, one to which we referred as \bB\underbar{\em\bBl systems\e\em}\e, and the other to which we referred as \bR\underbar{\em\bBl processes\e\em}\e.
\item[(C2)] There is an operation $-\bG\otimes\e-$ on \bB\underbar{\em\bBl systems\e\em} \e as well as an operation $-\bG\otimes\e-$ on \bB\underbar{\em\bBl processes\e\em}\e, with respective units $\bB\II\e$ and ${\bR 1\e}_{\bB\II\e}$.  Both of these are very similar to the multiplication of the group.  In addition to this operation, there is also an operation $-\bG\circ\e-$ on \bB\underbar{\em\bBl processes\e\em}\e, but for two \bB\underbar{\em\bBl processes\e\em} \e 
$\bR \Aa\rTo^f \Bb\e$ and $\bR \Cc\rTo^g \Dd\e$, their composite $\bR g\e\bG\circ\e\bR f\e$ exists if and only if we have $\Bb=\Cc$.
\item[(C3)] The way in which $-\bG\otimes\e-$ and $-\bG\circ\e-$ interact with each other is given by the laws:
\[ 
(\bR g\e\bG\circ\e \bR f\e)\bG\otimes\e(\bR k\e\bG\circ\e \bR h\e){\bf\ =\ }(\bR g\e\bG\otimes\e \bR k\e)\bG\circ\e(\bR f\e\bG\otimes\e \bR h\e)\qquad \mbox{and}\qquad
{\bR 1\e}_{\bB A \bG\otimes\e B\e}={\bR 1\e}_{\bB A\e}\bG\otimes\e{\bR 1\e}_{\bB B\e}\,.
\] 
\eit
The items (C1), (C2) and (C3), up to some subtleties for which we refer the reader to \cite{Cats,CPaqII,LNPAT}, define what it means to be a \em monoidal category\em, a mathematical structure which has been around now for some 45 years \cite{Benabou}. It has become prominent in computer science, and is gaining prominence in physics. \bB\underbar{\em\bBl Systems\e\em} \e are typically referred to as \bB\underbar{\em\bBl objects\e\em}\e, \bR\underbar{\em\bBl processes\e\em} \e are referred to as \bR\underbar{\em\bBl morphisms\e\em}\e, the operation $-\bG\circ\e-$ as \bG\underbar{\em\bBl composition\e\em}\e, and the  operation $-\bG\otimes\e-$ as the \bG\underbar{\em\bBl tensor\e\em}\e.  

The words \bG then \e  and \bG while \e we used to refer to
$-\bG\circ\e-$ and $-\bG\otimes\e-$ are clearly connected to the `time-like' and `space-like' separation one has in relativistic spatio-temporal causal structure.  Put differently, we can compose processes both `sequentially' and `in parallel'.  

\begin{definition}
A scientific theory of systems and processes thereon, in which we have two interacting modes of composing systems/processes, and such that mathematically it is described by a monoidal category in the above sense, will be called a \em compositional theory\em.  
\end{definition}

 \noindent {\bf Remark:}
The notion of system used above does not straightforwardly extrapolate to all physical theories, e.g.~quantum field theories, due to creation/annihilation of particles.  
 A more elaborate notion of system, which is very much the same as the one used in algebraic quantum field theory \cite{Haag}, admits the same algebraic description, and does apply to quantum fields. 
 

\subsection{Another metaphor: why does a tiger have stripes and a lion doesn't?}
 
A particle physicist might expect that the explanation is written within the fundamental building
blocks which these animals are made up from, so he would dissect the tiger and the lion.  One finds intestines, but these are the same for both animals.  When looking for even smaller building blocks, one discovers cells.  Again, these are very much the same for both animals.  When going even smaller one discovers DNA, and now there is a difference.  But does one now have a satisfactory \em explanation \em for the fact that tigers have stripes and  lions don't?   Your favorite nature channel would probably disagree. It would tell you that the explanation is given by the \bR\underbar{\em\bBl process\e\em}\e\vspace{-2mm}
\[ 
\bR\begin{diagram}
\hspace{-5.5cm}\mbox{\bB prey \bG$\otimes$ \e predator \bG$\otimes$ \e environment\e}&
\rTo^{\mbox{ hunt}}&
\mbox{\bB dead  prey \bG$\otimes$ \e eating predator\e}\,{\bBl.\e}\hspace{-5.2cm}\vspace{-2mm}
\end{diagram}\e
\] 
It represents the successful challenge of  a predator, operating within some habitat, on some prey. Key to  the success of such a  challenge is the predator's camouflage. Sandy savanna is the lion's habitat while forests constitute the tiger's habitat, so their respective  coat blends them within their natural habitat.  Darwinist biologists would claim that the fact that this is encoded in the animal's DNA is not a cause, but rather a consequence, via the process of natural selection. 

This example clearly illustrates that there are different levels of structural description that apply to a certain situation, and that some of these might be more relevant than others.  Rather than looking at the individual structure of systems, and their constituents, above we looked at how systems interact with others.  This is exactly what monoidal category theory enables one to describe, contra the traditional intrinsically monolithic mathematical structures. 

More philosophically put, this passage enables us to (at least to some extent) consider other perspectives than a purely reductionist one.  In particular, for quantum theory, it enables us to put more emphasis on the way in which quantum systems interact. This leads to new modes for explaining physical phenomena.  These modes are our subject of study here.

\subsection{Compositional theories $\equiv$ picture calculi}

\begin{wrapfigure}{r}{0.18\textwidth}
\vspace{-20pt}
  \begin{center}
    \epsfig{figure=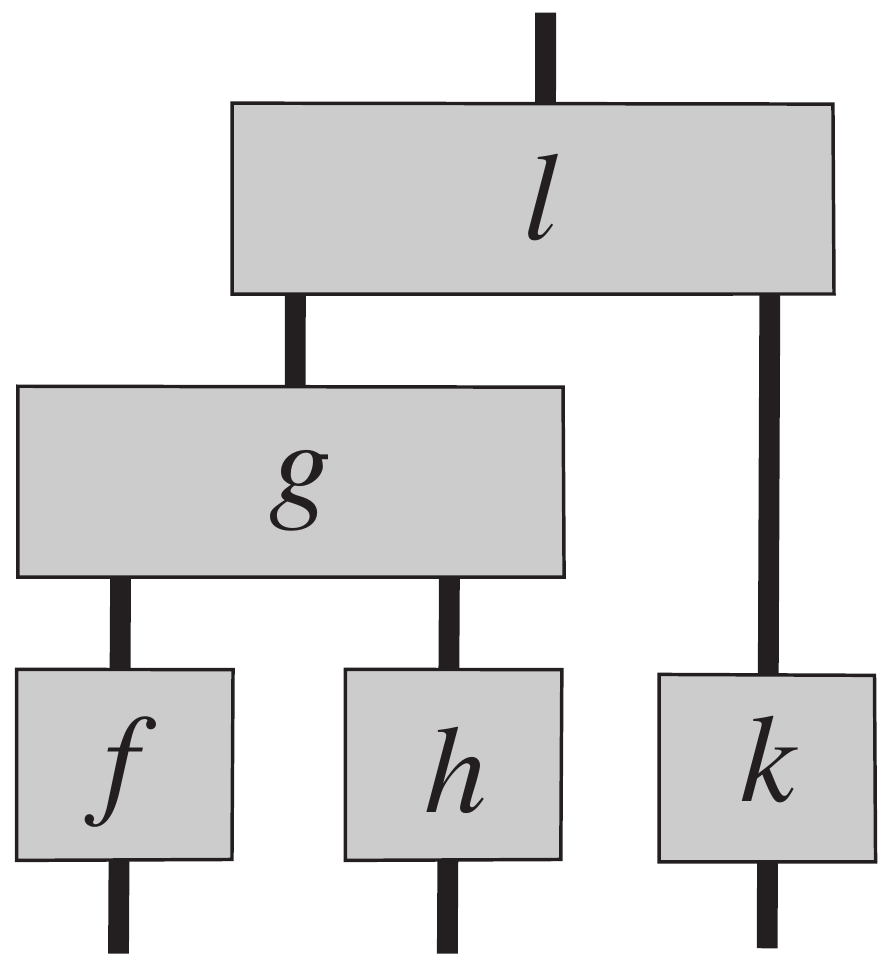,width=80pt}
  \end{center}
  \vspace{-10pt}
  \caption{Compound processes as pictures}
    \vspace{-10pt}
\end{wrapfigure}
We already introduced some basics of the diagrammatic language in eqs.(\ref{pics:defs1}).
For example, on the right is the diagrammatic representation of 
\[
l\circ(g\otimes 1)\circ(f\otimes h\otimes k)\,,
\]
or, by applying eq.(\ref{eq:bifunct}), it is also the  diagrammatic representation of 
\[
l\circ ((g\circ(f\otimes h))\otimes k)\,, 
\]
where we relied on $1\circ k=k$.  We represent the `unspecified' system ${\rm I}$  by `nothing', that is, no wire.  We represent \em states \em (cf.~kets), \em effects \em (cf.~bras), and \em numbers \em (e.g.~bra-kets) by:\vspace{-3mm}
\[
{\rm I}\rTo^{\psi} A  \  \equiv \ \raisebox{-0.52cm}{\epsfig{figure=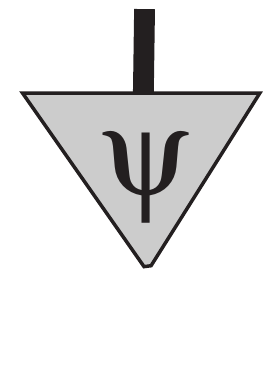,width=24pt}}
\qquad\qquad
A{\rTo^{\pi}} {\rm I}  \  \equiv \ \raisebox{-0.52cm}{\epsfig{figure=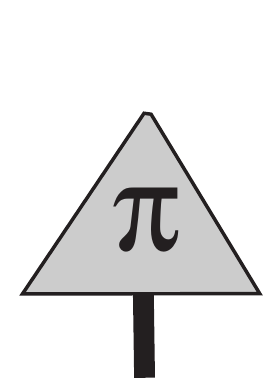,width=24pt}}
\qquad\qquad
{\rm I}{\rTo^{\pi}} {\rm I}  \  \equiv \ \raisebox{-0.52cm}{\epsfig{figure=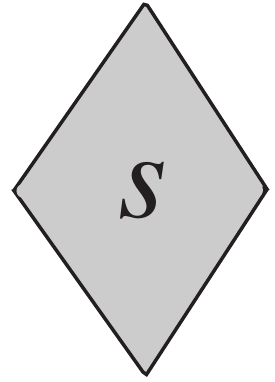,width=24pt}}\,.\vspace{-3mm}
\]
Note how these triangles and diamonds are essentially the same as Dirac notation:\footnote{That we have to rotate the ket's and bra's is merely a consequence of our convention to read the pictures from bottom to top with respect to composition. In other words, in our pictures  time flows upward.}
  \begin{center}
    \epsfig{figure=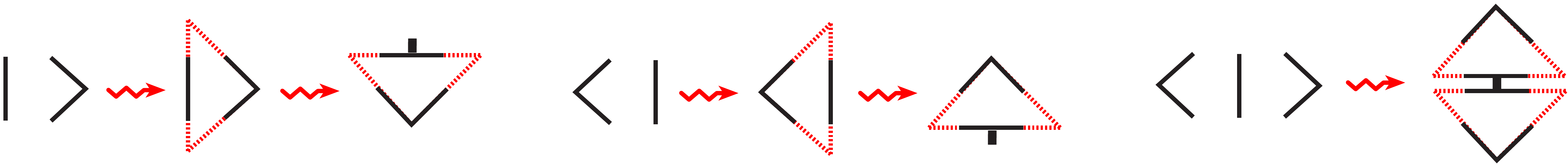,width=340pt}
  \end{center}
 Hence the graphical language builds further on something physicists already know very well.  
  Within the mathematical definition of a monoidal category, these special morphisms state, effect and number are subject to some equational constraints, but in the graphical calculus this is completely accounted for by the fact that ${\rm I}$ corresponds to `no wire'.

\begin{wrapfigure}{r}{0.48\textwidth}
\vspace{-8pt}
  \begin{center}
    \epsfig{figure=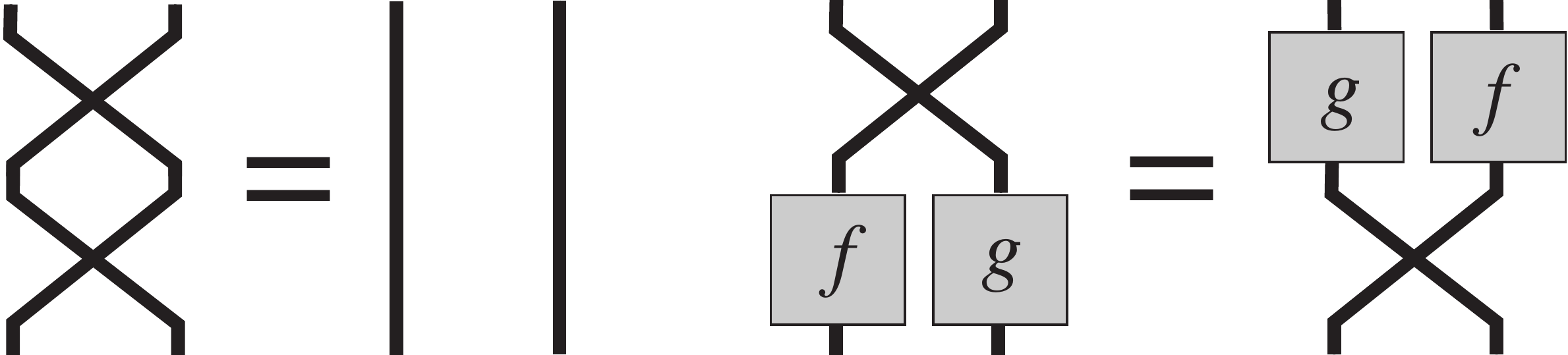,width=220pt}
  \end{center}
  \vspace{-10pt}
  \caption{Laws on `swapping systems'.}\label{Fig:swapping}
    \vspace{-20pt}
\end{wrapfigure}
Sometimes one wishes to have a process 
\[
A\otimes B\rTo^{\sigma_{A,B}}B\otimes A  \  \equiv \ \raisebox{-0.44cm}{\epsfig{figure=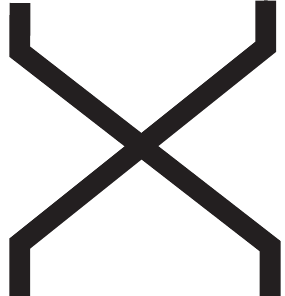,width=32pt}}
\] 
that \em swaps \em systems  in compositional theories.  Again this can be made mathematically precise, and is captured by the mathematical notion of \em symmetric monoidal category\em.
This involves substantially more equational requirements but each of these is again intuitively evident in diagrammatic terms, e.g.~in Figure \ref{Fig:swapping} we depicted:
\[
\sigma_{B,A}\circ\sigma_{A,B}=1_{A,B}\qquad\quad \mbox{and} \qquad\quad
\sigma_{A,B}\circ(f\otimes g)=(g\otimes f)\circ\sigma_{A,B}\,.
\]

\begin{theorem}\label{thm:JS}{\rm\cite{JoyalStreet}}
The graphical calculi for monoidal categories and symmetric monoidal categories is such that an equational statement between formal expressions in the language of (symmetric) monoidal categories holds if and only if it is derivable in the graphical calculus.
 \end{theorem}

The theory of graphical languages for a variety of different species of monoidal categories, including so-called \em braided \em ones,  is surveyed in a recent paper by Selinger  \cite{Selinger2}.
 
\section{Picture calculus for quantum theory I: lots from little}

In quantum theory systems are described by Hilbert spaces, and processes by linear maps. Therefore the symmetric monoidal category which has Hilbert spaces as objects, (bounded) linear maps as morphisms, and the tensor product as the tensor, plays an important role in this paper.  We denote this category by ${\bf Hilb}$.  When restricting to finite dimensional Hilbert spaces we write ${\bf FHilb}$ instead.  A detailed physicist-friendly description of ${\bf FHilb}$ is in \cite{CPaqII}.\footnote{There are of course several important subtleties when thinking of ${\bf Hilb}$ or ${\bf FHilb}$ as modeling quantum processes.  For example, the states of a quantum system are not described by vectors in a Hilbert space but rather by one-dimensional subspaces.  The categorical formalism can easily handle this \cite{deLL}, but a detailed discussion is beyond the scope of this paper.}


In  ${\bf Hilb}$ we have ${\rm I}:=\C$, since for any Hilbert space ${\cal H}$ we have that ${\cal H}\otimes\C\simeq{\cal H}$, where we conceive $\C$ itself as a one-dimensional Hilbert space.  Consequently,  states are linear maps $\psi:\C\to{\cal H}$. 
How do these relate to the states of quantum theory, that is, vectors $|\psi\rangle\in{\cal H}$?  It turns out that these two mathematical concepts are essentially one and the same thing.  Indeed, each $|\psi\rangle\in{\cal H}$ defines a unique linear map 
\[
\psi:\C\to{\cal H}::1\mapsto |\psi\rangle\,, 
\]
since by setting $\psi(1)=|\psi\rangle$, the map $\psi$ is completely determined due to linearity.\footnote{So the syntax $f: X\to Y:: x\mapsto y$ that we use to denote functions consists of two parts.  The part $X\to Y$ tells us that $X$ is the set of arguments and that the function takes values in $Y$.  The part $x\mapsto y$ tells us that $f(x):= y$.}  Conversely, such a linear map defines a unique state by setting $|\psi\rangle:=f(1)$. Hence these linear maps $\psi:\C\to{\cal H}$ and vectors $|\psi\rangle\in{\cal H}$ are in bijective correspondence.  Similarly one shows that the linear maps $s:\C\to\C$ are in bijective correspondence with the complex numbers $s\in\C$. 
\begin{center} 
\begin{tabular}{r|c|c|c|c|} 
\hline
\bW\fbox{\bBl\bf pics:\e}\e       &  \bW\fbox{\bBl\raisebox{-0.5cm}{\epsfig{figure=No2.pdf,width=24pt}}\e}\e  & \bW\fbox{\bBl\raisebox{-0.5cm}{\epsfig{figure=No1.pdf,width=24pt}}\e}\e  & & \bW\fbox{\bBl\raisebox{-0.5cm}{\epsfig{figure=State.pdf,width=24pt}}\e}\e 
 \\ \hline
\bW\fbox{\bBl\bf cats:\e}\e  & object $A$                          & 
morphism  \bW$\underline{\overline{\bBl A\rTo^f B\e}}$\e                  
                                    & \ \ ${\rm I}$  \ \          & 
                                    ${\rm I}\rTo^\psi B$                  
\\ \hline
\bW\fbox{\bBl\bf ${\bf Hilb}$:\e}\e              & Hilbert space ${\cal H}$ & 
linear map \bW\fbox{\bBl $f:{\cal H}\to {\cal H}'$\e}\e
                                    & $\C$ & 
                                    $|\psi\rangle\in{\cal H}$
\\ \hline                                    
\end{tabular}
\end{center} 
So ${\bf Hilb}$ is (a yet somewhat naive version of) quantum theory recast as a compositional theory, but still with explicit reference  to Hilbert spaces.  What we truly would like to do is to describe quantum theory in purely diagrammatic terms, without reference to Hilbert space.  In the remainder of this section we will adjoin two intuitively natural features to the graphical language which bring us  substantially closer to the fundamental concepts of the quantum realm, and will already allow for some protocol derivation.  This is taken from a joint paper with Abramsky \cite{AC1}.

\subsection{Concepts derivable from flipping boxes upside-down}

\begin{wrapfigure}{r}{0.43\textwidth}
\vspace{-20pt}
 \begin{center} 
\begin{tabular}{r|c|c|} 
\hline
\bW\fbox{\bBl\bf pics:\e}\e       &  \bW\fbox{\bBl\raisebox{-0.5cm}{\epsfig{figure=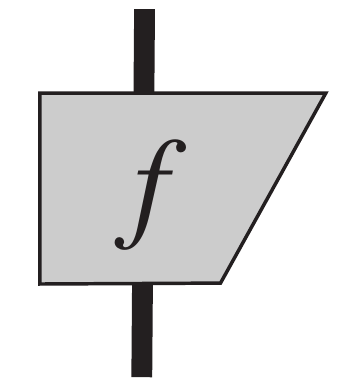,width=30pt}}\e}\e
                          &  \bW\fbox{\bBl\raisebox{-0.5cm}{\epsfig{figure=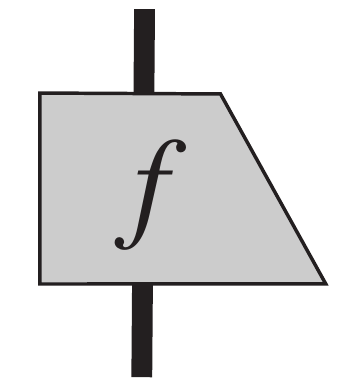,width=30pt}}\e}\e
\\ \hline
\bW\fbox{\bBl\bf cats:\e}\e    & \bW\fbox{\bBl\bf$A\rTo^f B$\e}\e &  \bW\fbox{\bBl\bf$B\rTo^{f^\dagger}A$\e}\e
\\ \hline 
\bW\fbox{\bBl\bf${\bf Hilb}$:\e}\e    & \bW\fbox{\bBl linear map\e}\e &  \bW\fbox{\bBl its adjoint\e}\e
\\ \hline 
\end{tabular}
\end{center} 
    \vspace{-10pt}
\end{wrapfigure}
Assume that for each graphical element there is a corresponding one obtained by flipping it upside-down.  To make this visible in the graphical calculus we introduce asymmetry.  In the case of ${\bf Hilb}$ we can interpret this `flipping' in terms of the linear-algebraic adjoint, obtained by transposing a matrix and conjugating its entries.  Therefore we also denote such a `flipping' operation by $\dagger$ in arbitrary monoidal categories.  We call a monoidal category with such a flipping operation a \em dagger monoidal category\em.  

Again, while in the graphical language we can simply define this operation by saying that we flip things upside-down, in category-theoretic terms we have to specify several equational requirements, for example, $(f\otimes g)^\dagger=f^\dagger\otimes g^\dagger$, $(g\circ f)^\dagger=f^\dagger\circ g^\dagger$ and $1_A^\dagger=1_A$.  

So what do adjoints buy us?  They let us define the following in any dagger monoidal category:

\begin{definition}
The \em inner product \em of two states ${\rm I}\rTo^\psi A$ and ${\rm I}\rTo^\phi A$ is the number
${\rm I}{\rTo^{\phi^\dagger\circ\psi}} {\rm I}$.  A morphism $A\rTo^f B$ is \em unitary \em if and only if $f^\dagger=f^{-1}$, where $B\rTo^{f^{-1}}A$ is the \em inverse \em to $f$. 
Such an \em inverse\em, if it exists, is defined in terms of the equations  $f\circ f^{-1}=1_B$ and $f^{-1}\circ f=1_A$.  A morphism $A\rTo^f A$ is \em self-adjoint \em iff $f=f^\dagger$, and it is a \em projector \em if moreover  $f\circ f= f$.
\end{definition}

The names of these concepts are justified by the fact that in ${\bf Hilb}$ they coincide with the usual notions \cite{CPaqII}. 
For the case of the inner product this can easily be seen when writing the linear maps $\psi$ and $\phi$ in terms of their respective matrices:\vspace{-1.5mm}
\[
\phi=\left(\begin{array}{c}
\phi_1\\ \vdots \\ \phi_n
\end{array}\right)
\qquad\ \ 
\psi=\left(\begin{array}{c}
\psi_1\\ \vdots \\ \psi_n
\end{array}\right)
\qquad\ \ 
\phi^\dagger\circ\psi = 
\left(\begin{array}{ccc}
\bar{\phi}_1& \ldots & \bar{\phi}_n
\end{array}\right)
\left(\begin{array}{c}
\psi_1\\ \vdots \\ \psi_n
\end{array}\right)
=\bar{\phi}_1\psi_1+\ldots+\bar{\phi}_n\psi_n\,.\vspace{-1.5mm}
\]

Note that self-adjointness of a linear operator translates in diagrammatic terms as `invariance under flipping it upside-down'.  Also, in any dagger monoidal category we can derive the more usual definition of unitarity in terms of preservation of the inner-product:

\begin{proposition}
Unitary morphisms preserve inner-products.
\end{proposition}

\begin{wrapfigure}{r}{0.75\textwidth}
\vspace{-16pt}
 \begin{center} 
\begin{tabular}{r|c|c|} 
\hline
 {\bf pics:}       & \raisebox{-0.85cm}{\epsfig{figure=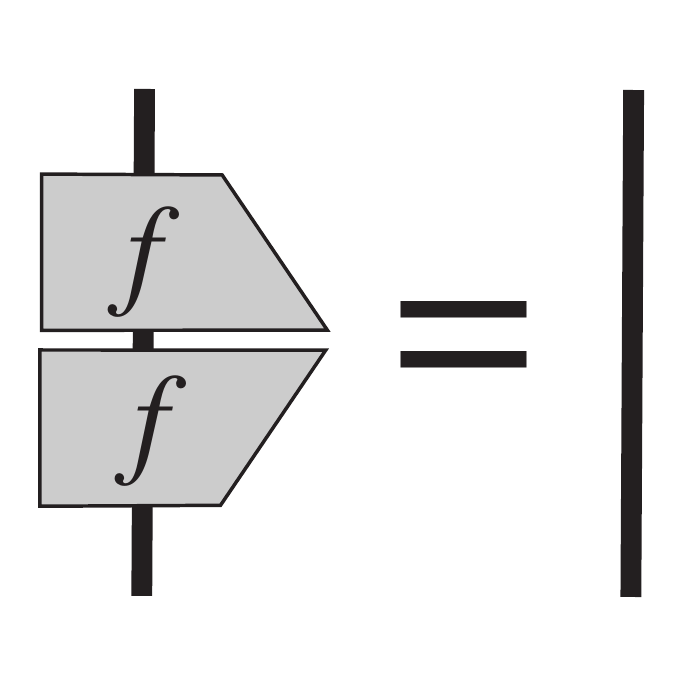,width=60pt}}
                          &  \bW\fbox{\bBl\raisebox{-0.85cm}{\epsfig{figure=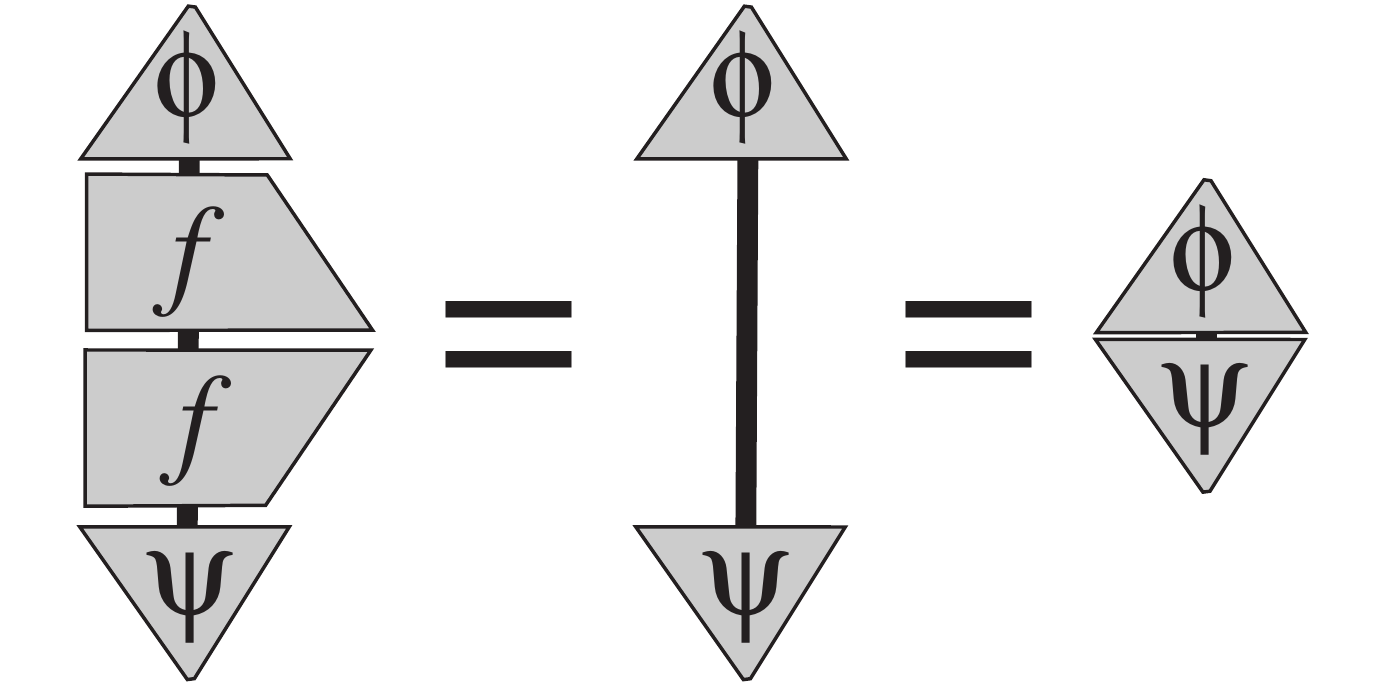,width=120pt}}\e}\e
\\ \hline
\bW\fbox{\bBl\bf cats:\e}\e    
                          & \bW\fbox{\bBl\bf$f^\dagger\!\circ f = 1_B$\e}\e 
                          &  \!\!\!\bW\fbox{\bBl\bf$(f\circ\phi)^\dagger\!\circ (f\circ\psi) =\phi^\dagger\circ (f^\dagger\!\circ f)\circ\psi=\phi^\dagger\!\circ\psi $\e}\e\!\!\!
\\ \hline 
{\bf${\bf FHilb}$:}
                          &\!\!\!\bW\fbox{\bBl$f$\! is\! an\! isometry\e}\e \!\!\!
                          &  \bW\fbox{\bBl$\langle f(\phi)|f(\psi)\rangle=\langle \phi|(f^\dagger\!\circ f)(\psi)\rangle=\langle \phi|\psi\rangle$\e}\e
\\ \hline 
\end{tabular}
\end{center} 
    \vspace{-10pt}
\end{wrapfigure}
The proof is depicted in the table on the right.  Recall here that $f$ is unitary if both $f$ and $f^\dagger$ are isometries, and that a linear map $f:{\cal H}\to {\cal H}'$ is an isometry whenever $f^\dagger\circ f=1_{\cal H}$.  Also the notion of \em positivity \em generalises to dagger monoidal categories, but more interesting is the notion of \em complete positivity\em. In standard quantum theory completely positive maps, roughly speaking,  assign to each density matrix another density matrix in such a way that mixtures of pure states are preserved. They are of key importance to describing noisy processes, open systems, and decohence.  It turns out that they can already be defined at the general level of dagger symmetric monoidal categories, such that in the case of ${\bf Hilb}$ we obtain the usual notion.  We only mention this result here, and  refer the reader to \cite{Selinger,CPMbob} for a detailed discussion. 

\subsection{Concepts derivable from  {\sf U}-turns}\label{sec:Uturns}

We adjoin new graphical elements to the calculus, namely a $\cup$-shaped  and a $\cap$-shape wire.\footnote{Depending on one's taste one can depict these either as \raisebox{-0.15cm}{\epsfig{figure=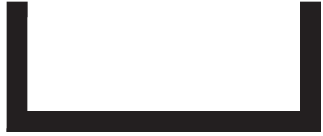,width=30pt}} or as \raisebox{-0.15cm}{\epsfig{figure=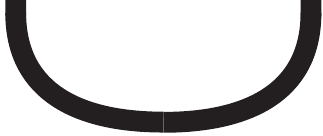,width=30pt}}\,; here we picked the latter.}
 
 \begin{center} 
\begin{tabular}{r|c|c|c|} 
 & {\bf element 1} & {\bf element 2} & {\bf rule}
\\ \hline
{\bf pics:}       & \bW\fbox{\bBl\raisebox{-0.15cm}{\epsfig{figure=cup.pdf,width=30pt}}\e}\e
                          & \bW\fbox{\bBl\raisebox{-0.15cm}{\epsfig{figure=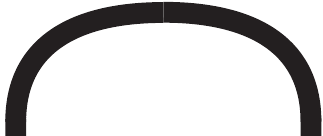,width=30pt}}\e}\e
                          & \bW\fbox{\bBl\raisebox{-0.40cm}{\epsfig{figure=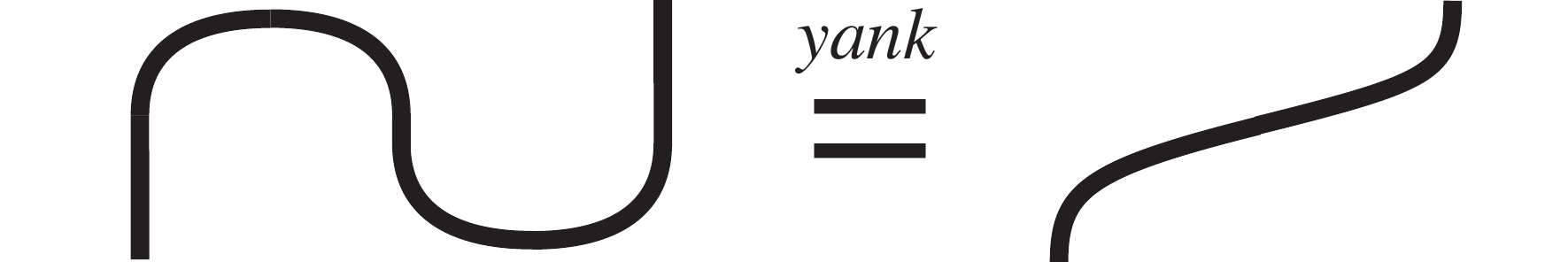,width=162pt}}\e}\e
\\ \hline
 \raisebox{0.08cm}{\bW\fbox{\bBl\bf cats:\hspace{-3pt}\e}\e}    
                          & \bW\fbox{\bBl\bf${\rm I}\rTo^{\eta_A} A\otimes A$\e}\e 
                          & \bW\fbox{\bBl\bf$A\otimes A{\rTo^{\epsilon_A}}{\rm I}$\e}\e
                          &  \raisebox{0.08cm}{\bW\fbox{\bBl$(\epsilon_A\otimes 1_A)\circ(1_A\otimes\eta_A)=1_A$\e}\e}
\\ \hline 
{\bf${\bf FHilb}$:}
                          &\bW\fbox{\bBl$\sum_i|ii\rangle$\e}\e 
                          &\bW\fbox{\bBl$\sum_i\langle ii|$\e}\e
                          &\bW\fbox{\bBl
                          $\Bigl(\sum_i\langle ii|\otimes 1_{\cal H}\Bigr)\Bigl(1_{\cal H}\otimes\sum_i|ii\rangle\Bigr)=1_{\cal H}$
                          \e}\e
\\ \hline 
\end{tabular}\vspace{-1.5mm}
\end{center} 
We refer to $\cup$'s as \em Bell-states \em and to $\cap$'s as \em Bell-effects\em.   These $\cup$'s and $\cap$'s are required to obey an intuitive graphical rule, which is depicted in the table above. While sym-
\linebreak\vspace{-12pt}

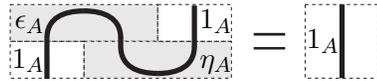
\begin{wrapfigure}{l}{0.32\textwidth}
\vspace{-10pt}
  \begin{center}
\begin{minipage}[b]{1\linewidth} 
\begin{picture}(142,30)
\centering\epsfig{figure=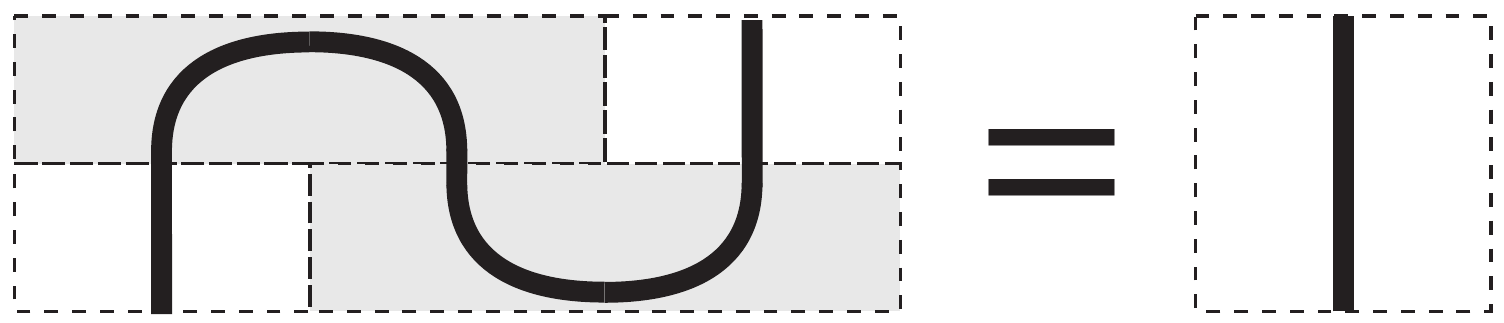,width=142pt}\hspace{-142pt}
\put(73,5){$\eta_A$}
\put(73,20){$1_A$}
\put(3,20){$\epsilon_A$}
\put(2,5){$1_A$}
\put(113.5,12.5){$1_A$}
\end{picture}
\end{minipage}
  \end{center}
  \vspace{-10pt}
  \caption{Comparison of the diagrammatic and the category-theoretic description of `straightening/yanking'.}\label{fig:analyanking}
    \vspace{-0pt}
\end{wrapfigure}\noindent
bolically this rule is quite a mouthful, graphically it is so simple that it looks somewhat silly: a line involving $\cup$'s and $\cap$'s can always been `straightened' or `yanked'.  
Figure \ref{fig:analyanking} explains how the diagrammatic and the symbolic descriptions of this rule relate. 
The reason for depicting the identity as \raisebox{-0.18cm}{\epsfig{figure=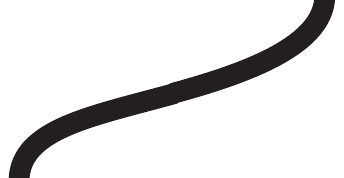,width=30pt}} in the table will become clear in later uses of this rule.    Since a $\cup$-shaped wire has no input and two outputs it corresponds to a morphism ${\rm I}\rTo^{\eta_A} A\otimes A$ in a monoidal category, so in ${\bf FHilb}$ it corresponds to some linear map $\eta_{\cal H}:\C\to {\cal H}\otimes{\cal H}$.  As explained above, to specify which linear map, it suffices to say what 
the state $\eta_{\cal H}(1)$ is.  States and  effects  satisfying this property do exist in ${\bf FHilb}$, and the Bell-state $\sum_i|ii\rangle$ and the Bell-effect $\sum_i\langle ii|$ are indeed examples.   We have:     
\[
\Bigl(1_{\cal H}\otimes\sum_i|ii\rangle\Bigr)|k\rangle=|k\rangle\otimes\sum_i|ii\rangle=\sum_i|kii\rangle\vspace{-5mm}
\]
so
\[
\Bigl(\sum_j\langle jj|\otimes 1_{\cal H}\Bigr)\Bigl(1_{\cal H}\otimes\sum_i|ii\rangle\Bigr)|k\rangle=
\Bigl(\sum_j\langle jj|\otimes 1_{\cal H}\Bigr)\sum_i|kii\rangle=
\sum_{ij}\langle jj|ki\rangle |i\rangle=\sum_{ij}\delta_{jk}\delta_{ji}|i\rangle=|k\rangle\,.\vspace{-2mm}
\]
Hence each basis vector is mapped on itself, so we indeed obtain the identity.


\begin{wrapfigure}{r}{0.67\textwidth}
\vspace{-14pt}
 \begin{center} 
\begin{tabular}{r|c|c|} 
 \hline
\bW\fbox{\bBl\bf pics:\e}\e 
 	&  \bW\fbox{\bBl\raisebox{-0.5cm}{\epsfig{figure=dag1.pdf,width=30pt}}\e}\e
          & \bW\fbox{\bBl\raisebox{-0.5cm}{\epsfig{figure=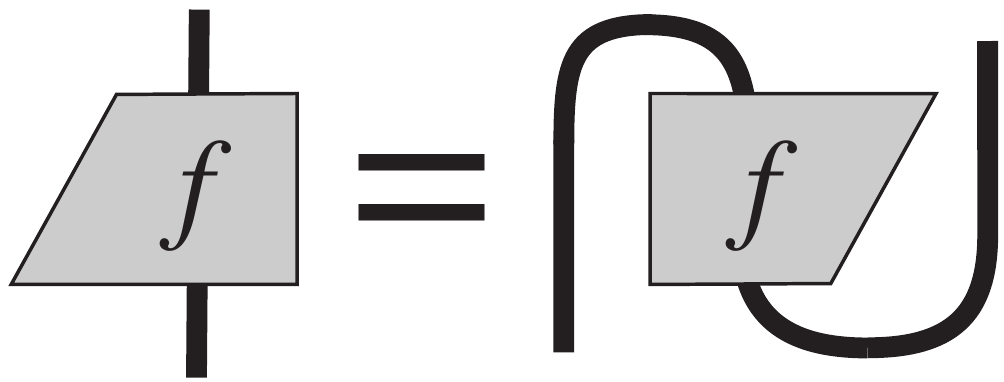,width=88pt}}\e}\e
\\ \hline
\bW\fbox{\bBl\bf cats:\e}\e   
	& \bW\fbox{\bBl\bf$\!A\rTo^f B\!$\e}\e 
	&  \bW\fbox{\bBl\bf$\!\!f^T=B\rTo^{(\epsilon_B\otimes 1_A)\circ(1_B\otimes f \otimes 1_A)\circ(1_B\otimes\eta_A)}A\!\!$\e}\e
\\ \hline 
\bW\fbox{\bBl\bf ${\bf FHilb}$:\e}\e 
	& linear map
	& its transpose
\\ \hline 
\end{tabular}
\end{center} 
    \vspace{-10pt}
\end{wrapfigure}
These $\cup$'s and $\cap$'s capture a surprising amount of linear-algebraic structure.
They for example allow one to generalise the linear-algebraic notion of \em transpose \em to arbitrary compositional theories.  The table on the right  shows how the transpose can be expresses in terms of $\cup$'s and $\cap$'s.  In ${\bf FHilb}$ this concept coincides with the usual notion.  The reader can verify  that 
\[
\Bigl(\sum_i\langle ii|\otimes 1_{\cal H}\Bigr)\bigl(1\otimes f\otimes 1\bigr) \Bigl(1_{{\cal H}'}\otimes\sum_i|ii\rangle\Bigr)=f^T:{\cal H}'\to{\cal H}\vspace{-2mm}
\]
indeed holds for any linear map $f:{\cal H}\to{\cal H}'$. The computation proceeds very much in the same manner as our verification of  the yanking rule for Bell states and Bell effects.

As is also illustrated in the table above, graphically we denote this adjoint by rotating 
the box representing the morphism by 180 degrees.  This choice is not at all arbitrary.  As shown in Figure \ref{fig:sliding}, the  definition of the transpose together with the yanking axiom for the $\cup$'s and 
$\cap$'s allows us to prove that we can `slide' boxes along these $\cup$'s and $\cap$'s, which indeed exactly corresponds to 
\linebreak\vspace{-12pt}

\begin{wrapfigure}{l}{0.525\textwidth}
\vspace{-18pt}
  \begin{center}
    \epsfig{figure=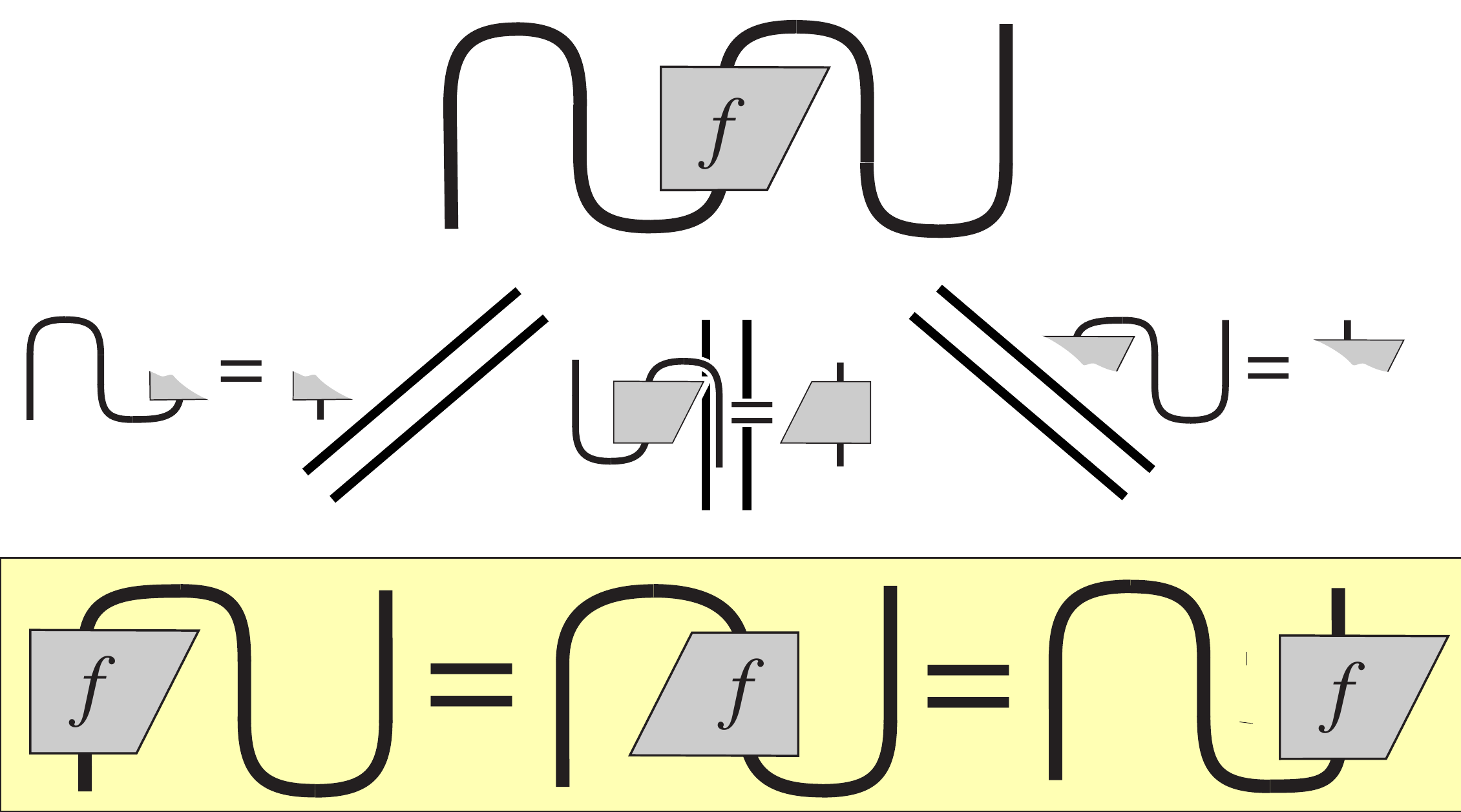,width=240pt}
  \end{center}
  \vspace{-10pt}
  \caption{Proof of the sliding rule. We apply yanking to the picture at the top to obtain bottom-left and bottom right.  The bottom-middle picture follows by the definition of the transpose.}\label{fig:sliding}
    \vspace{-6pt}
\end{wrapfigure}\noindent
rotating the box 180 degrees.  We'll see further how this principle, and nothing but this principle,  will allow us to derive several quantum informatic protocols.  

At the beginning of Section \ref{sec:LA101} we discussed `map-state' duality,
that is, to a linear map $f:{\cal H}\to{\cal H}'$ with matrix $(\omega_{ji})_{ji}$ in basis $\{|i\rangle\}_i$ of ${\cal H}$ and basis $\{|j\rangle\}_j$ of ${\cal H}'$  we can always associate a bipartite vector $\Psi_f:=\sum_{ji}\omega_{ji}\cdot|ij\rangle\in{\cal H}\otimes{\cal H}'$. This correspondence between linear maps from ${\cal H}$ to ${\cal H}'$ and vectors in ${\cal H}\otimes{\cal H}'$ is a bijective one.  The $\cup$'s and $\cap$'s  generalize this `map-state' duality to arbitrary compositional theories.  First, note that we can write the bipartite state $\Psi_f$ in terms of $f$ itself and a Bell-state, namely as $\Psi_f=(1_{\cal H}\otimes f)\sum_{i}|ii\rangle$.
More generally, in the graphical calculus the bijective correspondence between morphisms and bipartite states is:
\beq\label{eq:mapstate}
f  \  \equiv \ \raisebox{-7mm}{\epsfig{figure=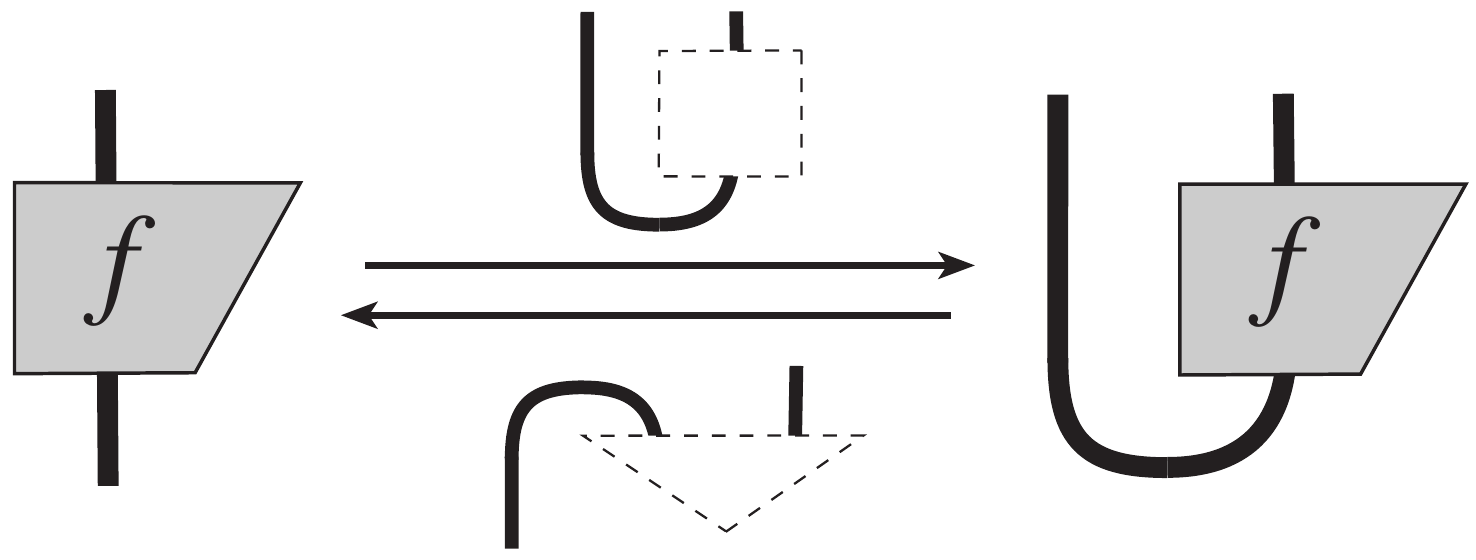,width=120pt}} 
 \  \equiv \ (1_B\otimes f)\circ\eta_A\,.
\eeq
So going from the linear map to the bipartite state consists of `plugging' it on the second output-wire of  a $\cup$, and to convert a bipartite state back to a linear map we have to `plug' its first 
\linebreak\vspace{-12pt}

\begin{wrapfigure}{r}{0.42\textwidth}
\vspace{-10pt}
 \begin{center} 
\begin{tabular}{r|c|} 
\hline
\bW\fbox{\bBl\bf pics:\e}\e 
          & \bW\fbox{\bBl\raisebox{-0.5cm}{\epsfig{figure=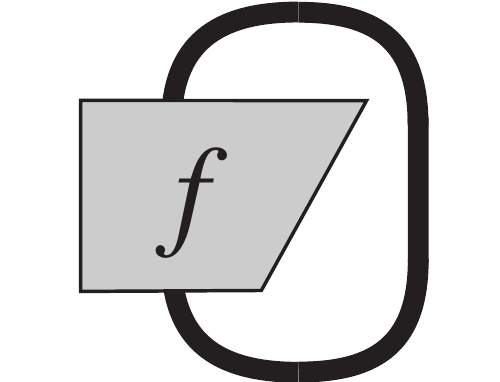,width=44pt}}\e}\e
\\ \hline
\bW\fbox{\bBl\bf cats:\e}\e   
	&  \bW\fbox{\bBl\bf$tr(f)={\rm I}{\rTo^{\eta_A\circ(f\otimes 1_A)\circ\epsilon_A}}{\rm I}$\e}\e
\\ \hline 
\bW\fbox{\bBl\bf ${\bf FHilb}$:\e}\e 
	& \bW\fbox{\bBl trace i.e.~$\sum_i m_{ii}$\e}\e
\\ \hline 
\end{tabular}
\end{center} 
   \vspace{-10pt}
  \caption{Also the trace allows a diagrammatic presentation in terms of $\cup$'s and $\cap$'s.  It's abstract category-theoretic axiomatisation is in \cite{JSV}.}
    \vspace{-16pt}
\end{wrapfigure}\noindent 
output-wire into the second input-wire of a $\cap$.
The yanking rule guarantees that we recover the linear map we started from. This bijective correspondence lifts to completely positive maps, yielding a generalized \em Choi-Jamiolkowski isomorphism\em.

  Other  concepts of linear algebra which can be expressed in terms of $\cup$'s and $\cap$'s are the \em trace\em, which is depicted on the right, the \em partial trace\em, and the \em partial transpose\em, which all play an important role in quantum theory.

 \bigskip\noindent {\bf Remark:}
Rather than  defining $\cup$'s as morphisms $A\otimes A{\rTo^{\epsilon_A}}{\rm I}$\,, like we did above, there are good reasons to define $\cup$'s as morphisms $A^*\otimes A{\rTo^{\epsilon_A}}{\rm I}$\,,  where $A^*$ is referred to as the \em dual\em.  For example, when we take ${\cal H}^*$ to be the dual Hilbert space of a Hilbert space ${\cal H}$ (i.e.~the space of functionals) then the Bell-states, trace and transpose are basis independent \cite{AC1,Coecke-Paquette-Perdrix}.

\subsection{$2 \times 2 = 4$}

\begin{wrapfigure}{r}{0.70\textwidth}
\vspace{-22pt}
 \begin{center} 
\begin{tabular}{r|c|c|} 
\hline
\bW\fbox{\bBl\bf pics:\e}\e 
 	&  \bW\fbox{\bBl\raisebox{-0.5cm}{\epsfig{figure=dag1.pdf,width=30pt}}\e}\e
          & \bW\fbox{\bBl\raisebox{-0.5cm}{\epsfig{figure=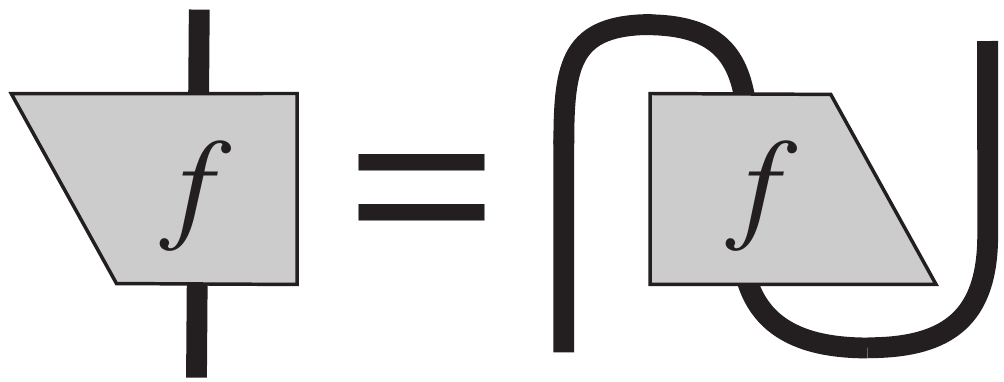,width=88pt}}\e}\e
\\ \hline
\bW\fbox{\bBl\bf cats:\e}\e   
	& \bW\fbox{\bBl\bf$A\rTo^f B$\e}\e 
	&  \bW\fbox{\bBl\bf$f^\sharp=A\rTo^{(\epsilon_A\otimes 1_B)\circ(1_A\otimes f^\dagger \otimes 1_B)\circ(1_A\otimes\eta_B)}B$\e}\e
\\ \hline 
\bW\fbox{\bBl\bf ${\bf FHilb}$:\e}\e 
	& linear map
	& its conjugate
\\ \hline 
\end{tabular}
\end{center} 
    \vspace{-14pt}
\end{wrapfigure}
If we combine the structures introduced in the previous two sections, we can construct the transpose of the adjoint, or equally, as is obvious from the graphical calculus, the adjoint of the transpose.  
In ${\bf FHilb}$ this corresponds to conjugating matrix 
\linebreak\vspace{-12pt}

\begin{wrapfigure}{l}{0.30\textwidth}
\vspace{-5pt}
  \begin{center}
    \epsfig{figure=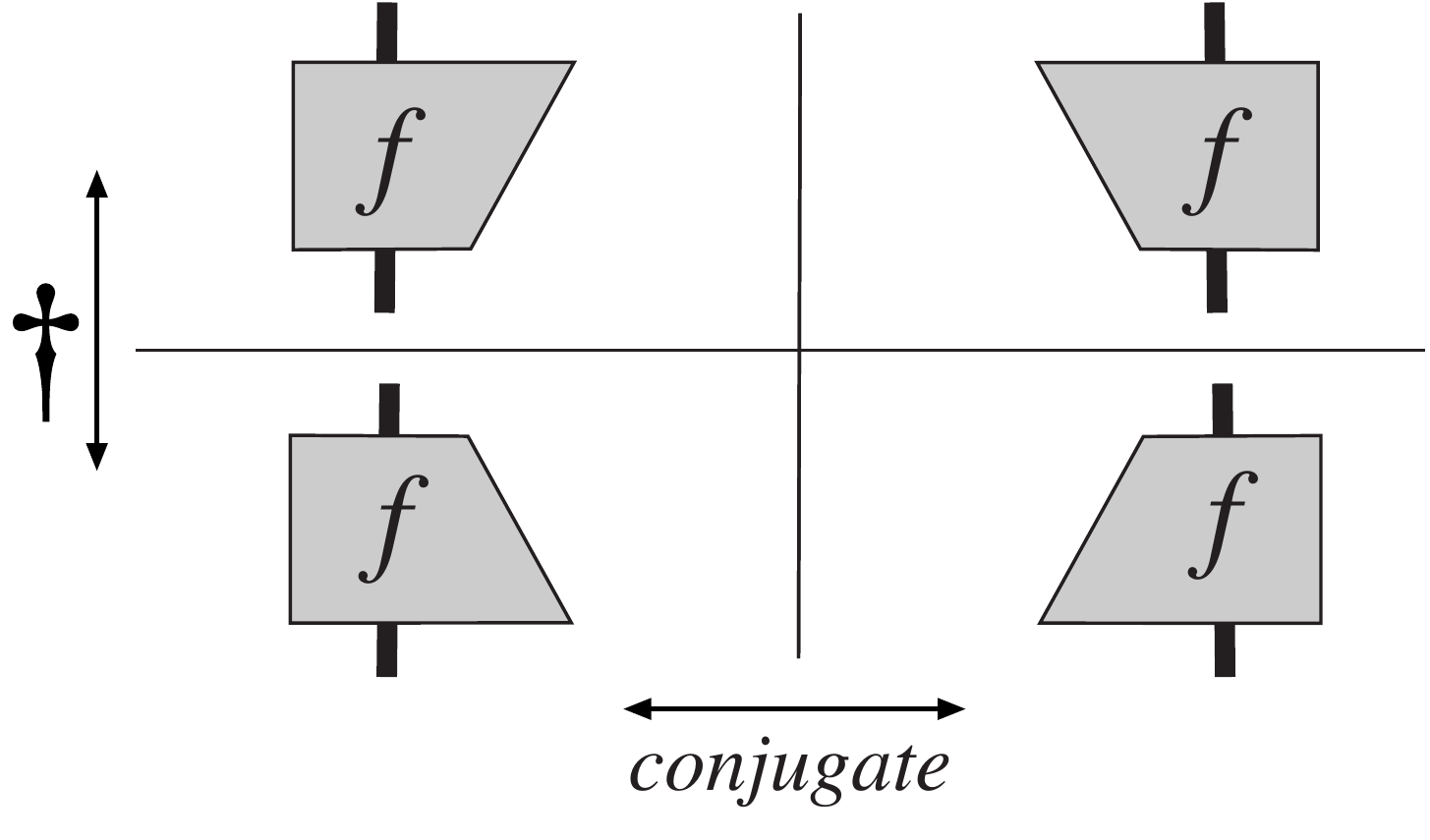,width=134pt}
  \end{center}
    \vspace{-20pt}
\end{wrapfigure}\noindent
entries.  On the left we summarise the graphical representation of the adjoint, the conjugate and the transpose, and the ways they relate to each other.   This is the setting in which things start to become interesting, and we 
can start our explorations in the area quantum informatic protocols.  All the results that we will derive apply to arbitrary compositional theories in which we can flip boxes upside-down and have $\cup$'s and $\cap$'s, so in particular, to ${\bf FHilb}$.  

First we derive the quantum teleportation protocol.  Assume that $f$ is a \em unitrary morphism \em i.e.~its adjoint is equal to its inverse.  Physically it represents a reversible operation. We have:  \vspace{2mm}
  \begin{center}
    \epsfig{figure=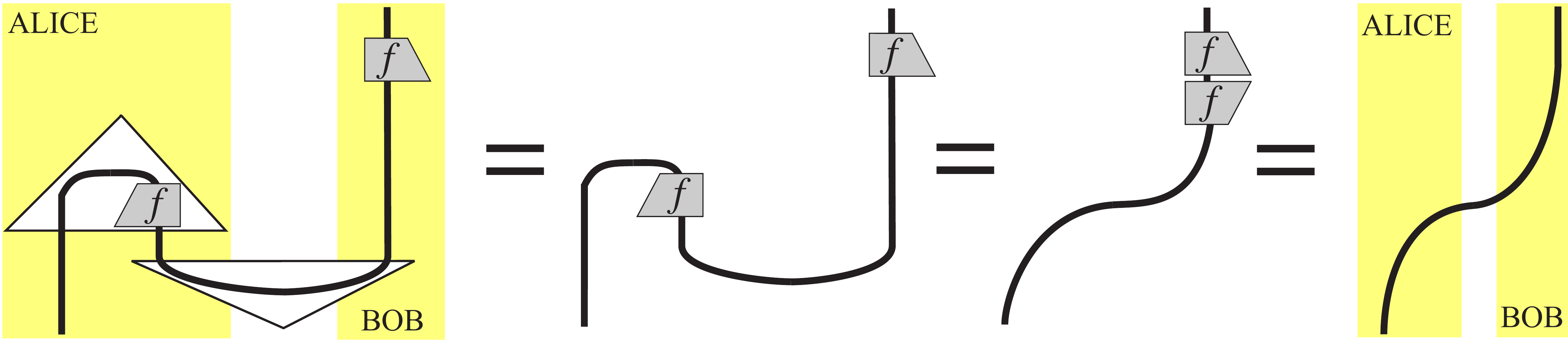,width=456pt}\vspace{2mm}
  \end{center}
The picture on the left describes the setup. Alice and Bob share a Bell-state (= the white triangle at the bottom). Alice also possesses another qubit in an unknown state (= the leftmost black wire at the bottom). She performs a bipartite measurement on her two qubits for which the resulting corresponding effect is the remaining triangle, that is,  $(\Psi_{f^T})^\dagger$ in the notation of the previous\linebreak\vspace{-12pt}

\begin{wrapfigure}{5}{0.460\textwidth}
\vspace{-12pt}
  \begin{center}
     \epsfig{figure=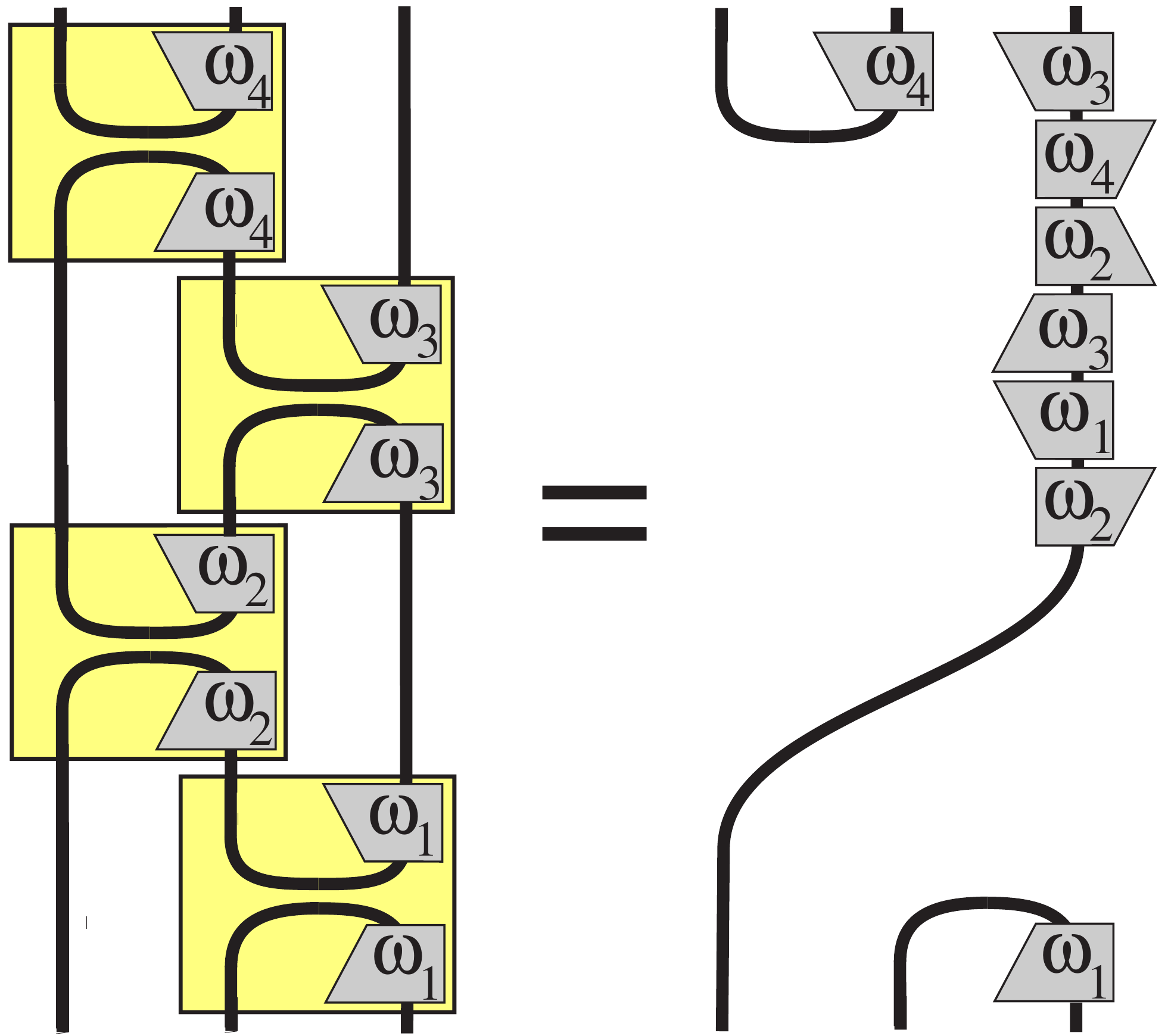,width=210pt}
  \end{center}
  \vspace{-12pt}
\end{wrapfigure}\noindent
 section. By map-state duality we know that any bipartite-effect can be represented in this manner for some $f$.  The fact that $f$ is here  unitary guarantees that the effect is maximally entangled.  Finally Bob performs the adjoint to $f$ on his qubit.  The picture on the right shows that the overall result of doing all of this is that Alice's qubit ends up with Bob. Importantly, the fact that Alice's measurement  and Bob's operation are labelled by the same symbol $f$ implies that Alice needs to communicate what her $f$ is (i.e.~her measurement outcome) to Bob. As a consequence this protocol does not violate no-faster-than-light-communication, and hence it is in perfect harmony with special relativity.

On the right you find the solution to the exercise we presented in Section \ref{sec:LA101}. Indeed, that's all there is to it. Since ${\bf FHilb}$ is an example of a compositional theory, this general proof implies the result for the specific case of linear algebra.

We also derive the entanglement swapping protocol:
  \begin{center}
    \epsfig{figure=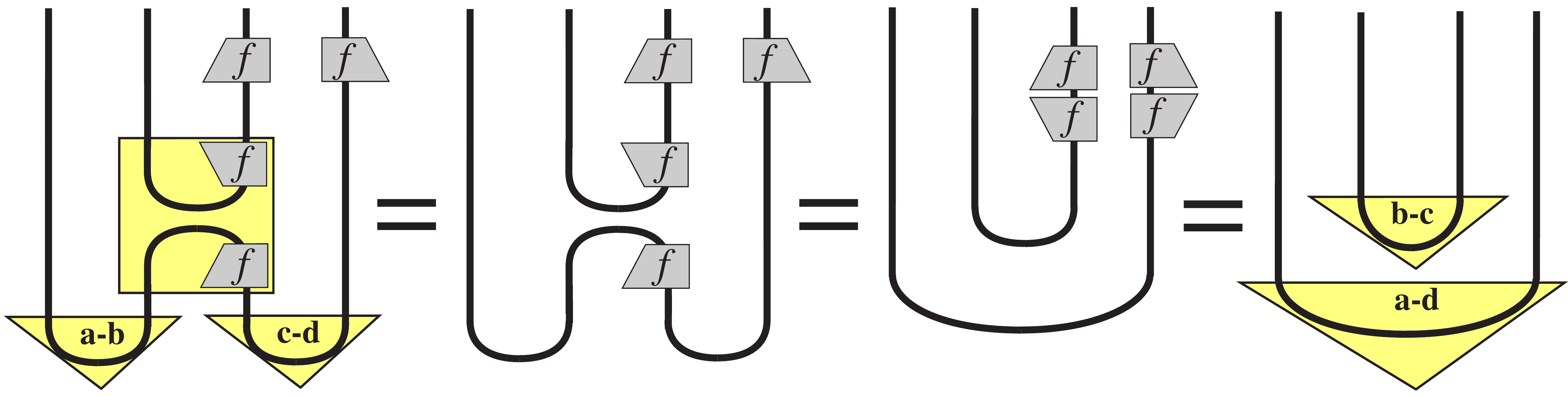,width=456pt}
  \end{center}
The four qubits involved, $a,b,c,d$, are initially in two Bell-states, $a\mbox{-}b$ and $c\mbox{-}d$.  By performing
a (non-destructive) measurement on $b$ and $c$ (= yellow square), and by then performing the corresponding unitaries on $c$ and $d$, we get a situation where the Bell-states are now $a\mbox{-}d$ and $b\mbox{-}c$.

Dagger symmetric monoidal categories in which each object comes with a $\cup$ and a $\cap$, subject to certain conditions which make all of these live happily together, are \em dagger compact categories\em.  Theorem \ref{thm:JS} extends to dagger compact categories.

\begin{theorem}{\rm\cite{KellyLaplaza,Selinger}}
The graphical calculus for dagger compact categories is such that an equational statement between formal expressions in the language of dagger compact categories holds if and only if it is derivable in the graphical calculus.
 \end{theorem}

But in fact, now there is even more.  As mentioned before, ${\bf FHilb}$ is an example of a dagger compact category, but of course there are also many other ones.  To give two examples: 
\bit
\item Taking  sets as objects, relations as morphisms, the cartesian product as tensor, and relational converse as the dagger, results in a dagger compact category ${\bf Rel}$  \cite{CPaqII}.
\item Taking closed $n-1$-dimensional manifolds as objects, $n$-dimensional manifolds connecting these as morphisms (= \em cobordisms\em), the disjoint union of these manifolds as the tensor, and reversal of the manifold as the dagger, results in a dagger compact category ${\bf nCob}$ \cite{Baez}.\footnote{Following Atiyah in \cite{Atiyah}, \em topological quantum field theories \em can be succinctly defined as \em monoidal functors \em from ${\bf nCob}$ into ${\bf FHilb}$, where a functor is a map both on objects and on morphisms which preserves composition and tensor~\cite{Kock,Baez,CPaqII}.}
\eit

 \medskip
The dagger compact categories ${\bf Rel}$ and ${\bf nCob}$, in particular the latter, are radically different from  ${\bf FHilb}$. This would make one think that there is nothing special about ${\bf FHilb}$ within the context of dagger compact categories.  But in fact, ${\bf FHilb}$ is very special as a dagger compact category, as the following beautiful result due to Selinger demonstrates.  

\begin{theorem}\label{ThmCompletenessSelinger}{\rm\cite{Plotkin,Selinger3}}
An equational statement between formal expressions in the language of dagger compact categories holds if and only if it holds in the dagger compact category ${\bf FHilb}$.
\end{theorem}

Let us spell out what this exactly means. Obviously, any statement provable for dagger compact categories carries over to ${\bf FHilb}$ since the latter is an example of a dagger compact category.  So anything that we prove in the graphical calculus automatically applies to Hilbert spaces and linear maps.  But this theorem now tells us that the converse is also true, that is, if some equational statement happens to hold  for Hilbert spaces and linear maps, which is expressible in the language of dagger compact categories, then we can always derive it in the graphical language.  This of course does not mean that all that we can prove about quantum theory can be proven diagrammatically. But all those statements involving  identities, adjoints, (partial) transposes, conjugates, (partial) traces, composition, tensor products, Bell-states and Bell-effects, and with Hilbert spaces, numbers, states and linear maps as variables, can be proven diagrammatically.  For dagger compact categories such as ${\bf Rel}$ and ${\bf nCob}$ there does not exist an analogous result.

A current challenge is to extend this so-called \em completeness \em theorem to richer graphical languages, e.g.~the one presented in the next section of this paper, which capture even more of the Hilbert space structure.  The ultimate challenge would be to find  a graphical language which captures the complete Hilbert space structure, if that is even possible of course.  

Obviously most of the results in quantum informatics use a much richer language than that of dagger compact categories.  But that doesn't necessarily mean that it could not be formulated merely in this restrictive language.  An example is the no-cloning theorem.  While usually stated in linear-algebraic terms, the no-cloning theorem can in fact already be proven for arbitrary dagger compact categories, a result due to Abramsky that reads as follows:

\begin{theorem}{\rm\cite{AbrClone}}
If in a dagger compact category there exists a universal cloning morphism then this dagger compact category must be a trivial one.  In other words, there are no non-trivial dagger compact categories which admit a universal cloning morphism.
\end{theorem}

We have to explain what exactly we mean by trivial here, since many different notions of trivial could apply.  The most trivial notion of trivial is of course that in the whole category there is only one object and one morphism, namely the object's identity.  The notion of triviality that applies to the above is that each $A\rTo^{f}  A$ is equal to the identity $A\rTo^{1_A} A$ up to a number ${\rm I}{\rTo^s}{\rm I}$.  This means that the state ${\rm I}\rTo^{\psi}  A$ of any system $A$ can never change, and hence we indeed have a very useless and hence very trivial compositional theory.

\section{Picture calculus for quantum theory II: observables, complementarity, and phases}

The aim is now to further refine our graphical language to the extent that we can describe arbitrary linear maps within it, hence the whole of quantum theory.  This will enable us to perform more sophisticated calculations diagrammatically, and study important quantum phenomena such as non-locality in a high-level manner.
This requires few additional concepts.  


\subsection{Observables as pictures}

The following are not expressible in the graphical language of dagger compact categories:\vspace{-12pt}

\begin{wrapfigure}{r}{0.30\textwidth}
\vspace{-0pt}
  \begin{center}
    \epsfig{figure=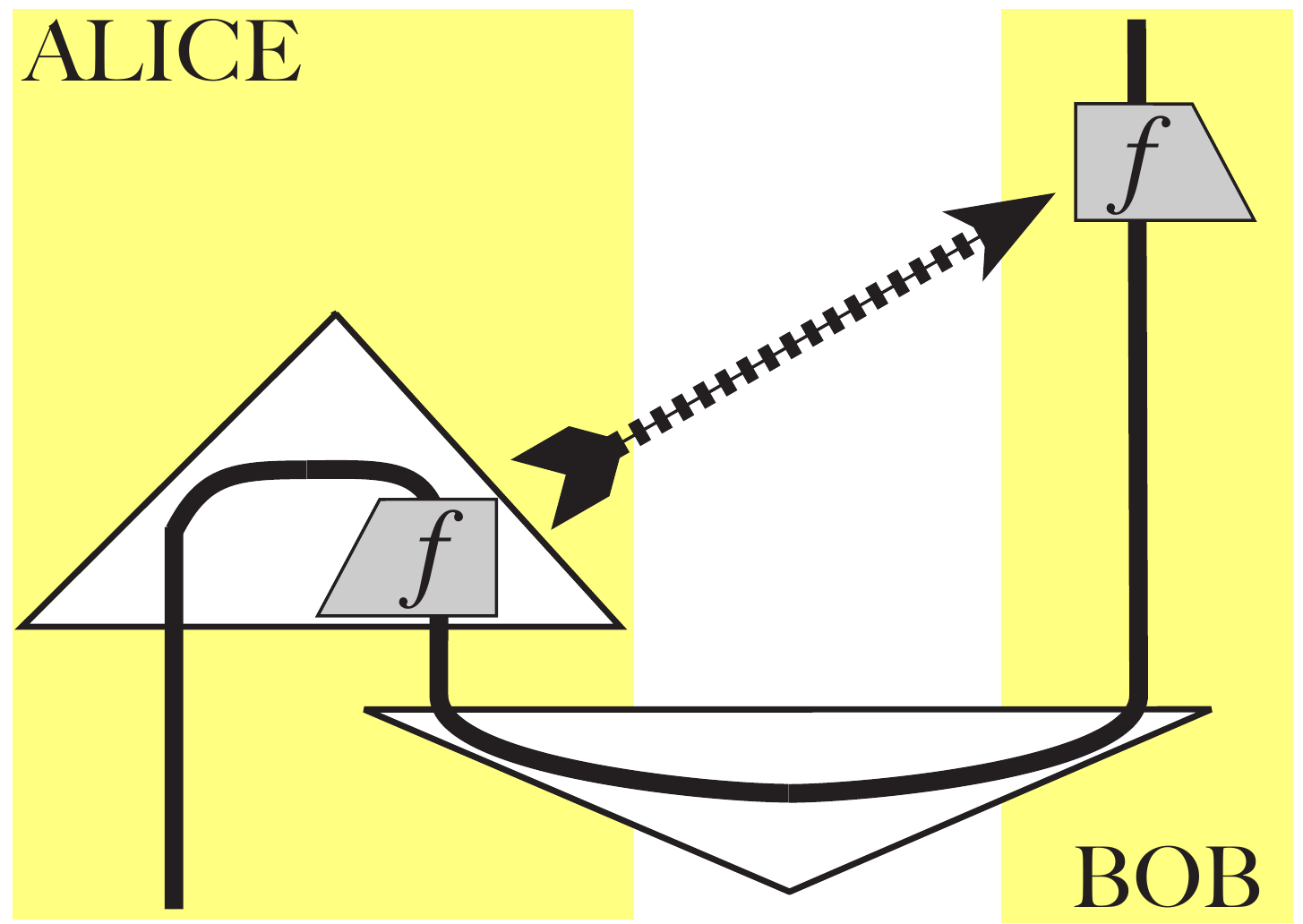,width=134pt}
  \end{center}
    \vspace{-10pt}
    \caption{We want to depict a classical channel (here indicated by a dotted arrow) also by a wire different from a quantum channel of course.}\label{pic:classwire}
    \vspace{-10pt}
\end{wrapfigure}\noindent
\bit
\item 
In our graphical description of teleportation in the previous section we mentioned 
that the fact that $f$ appears both at Alice's and Bob's site implied that they needed to communicate with each other.  A comprehensive diagrammatic presentation of this protocol should therefore have a second kind of wire which represents such a \em classical channel\em.
 \item The graphical description of teleportation included effects labelled by $f$, and we mentioned that $f$ may vary due to the non-deterministic nature of measurements.  But we didn't express which such effects together make up a measurement.  In other words, we have no diagrammatic descriptions of the projector spectra and eigenvectors of observables.
\eit
We only need one kind of additional graphical element to be able to articulate each of these graphically.  There are two complementary presentations of it, each pointing at distinct features. To one we refer  as \em spiders\em, and to the other  one as a \em copying-deleting-pair\em.  This in particularly involves a novel mathematical representation of orthonormal bases, so brace yourself for a fairly mathematical intermezzo, in a very different area of mathematics than you might be used to.  The results presented here appeared in joint papers with Pavlovic, Vicary and Paquette \cite{CPav,CPV,CPaq}.

\subsubsection{Spider presentation of non-degenerate observables}

A \em non-degenerate observable \em or \em basis \em for an object $A$ in a dagger symmetric monoidal category is a family of \em spiders \em with $n$ front and $m$ back legs, one for each $n,m\in\N$,  and depicted as  
\begin{center}
\raisebox{-0.25cm}{$\underbrace{\overbrace{\epsfig{figure=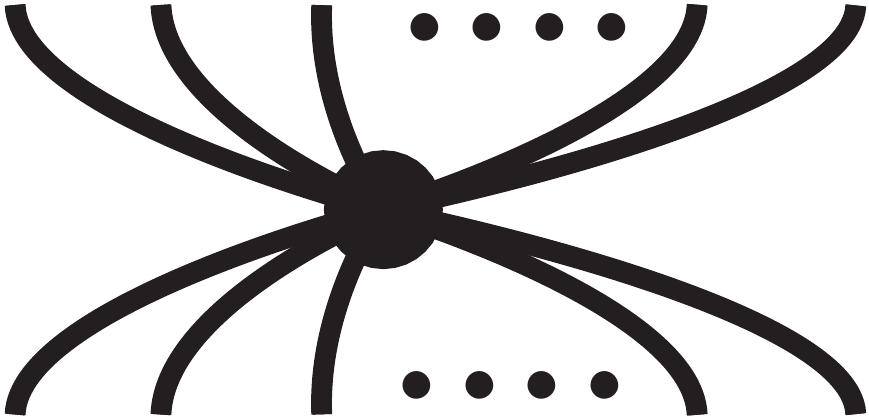,width=80pt}}^{\scriptstyle{m}}}_{\scriptstyle{n}}$} 
\end{center}
Symbolically, we  denote a spider as 
\[
A^{\otimes n}\rTo^{\delta_n^m} A^{\otimes m}\,.  
\]

\begin{wrapfigure}{r}{0.46\textwidth}
\vspace{-30pt}
  \begin{center}
  $
    \underbrace{\overbrace{\epsfig{figure=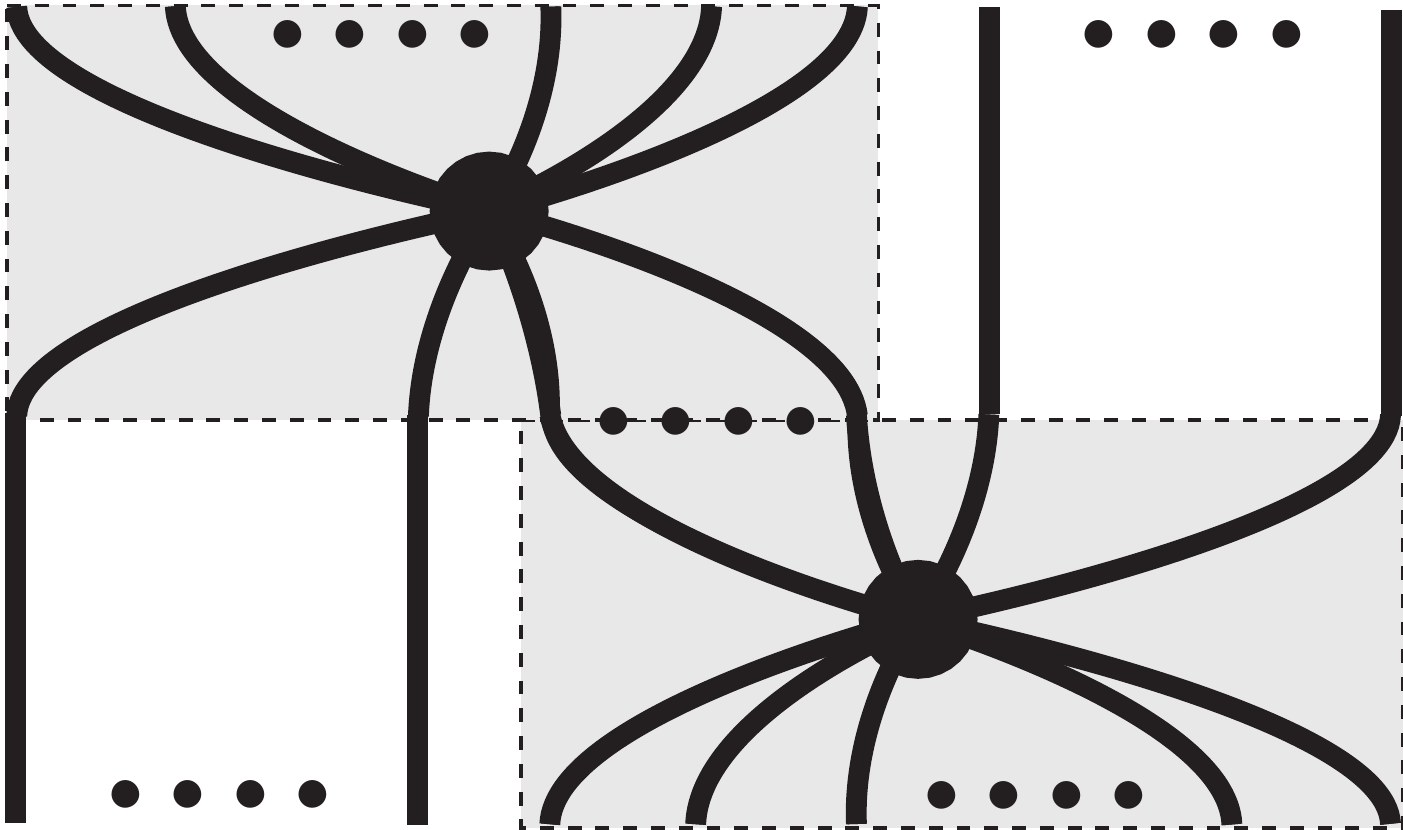,height=62pt}}^{\scriptstyle{m}}}_{\scriptstyle{n}}
   \ \  \epsfig{figure=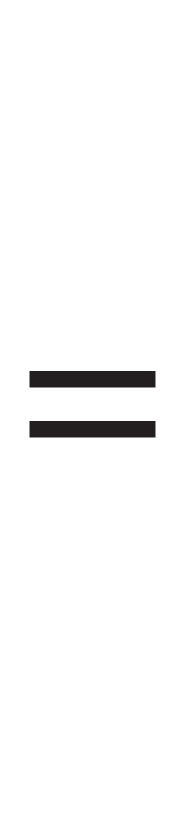,height=62pt}\ \ 
    \underbrace{\overbrace{\epsfig{figure=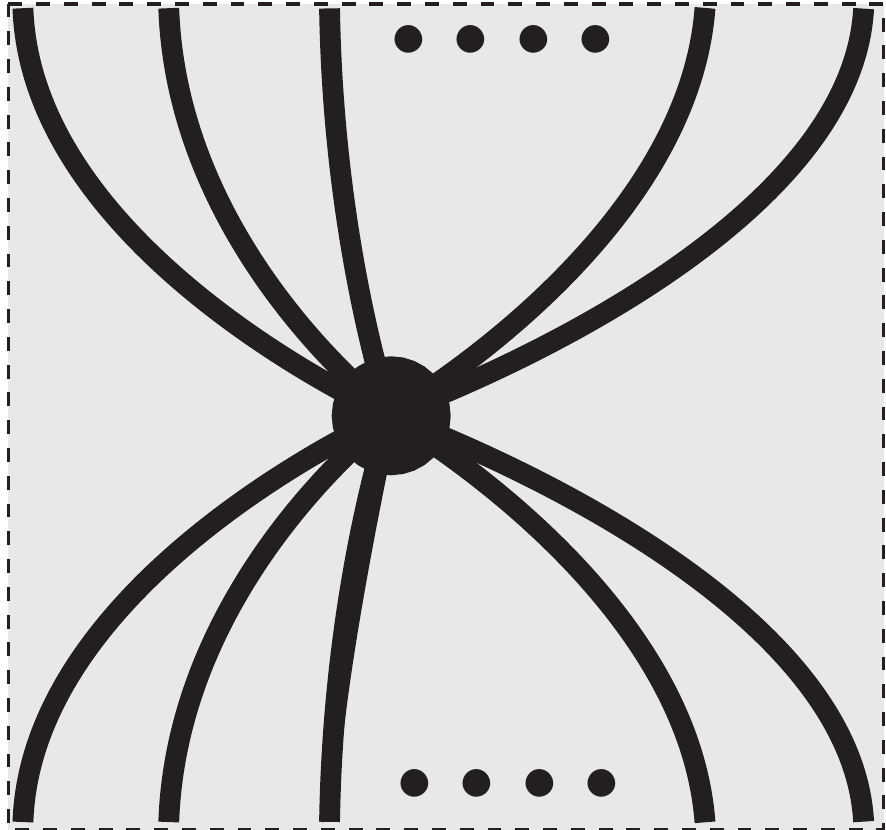,height=62pt}}^{\scriptstyle{m}}}_{\scriptstyle{n}}
    $
  \end{center}
    \vspace{-10pt}
    \caption{Rule for composing spiders.  It is essential that the spiders `shake hands/legs' i.e.~the two dots corresponding to the spiders' heads need to be connected via at least one wire.}\label{fig:spidercomp}
    \vspace{10pt}
      \begin{center}
  $
\epsfig{figure=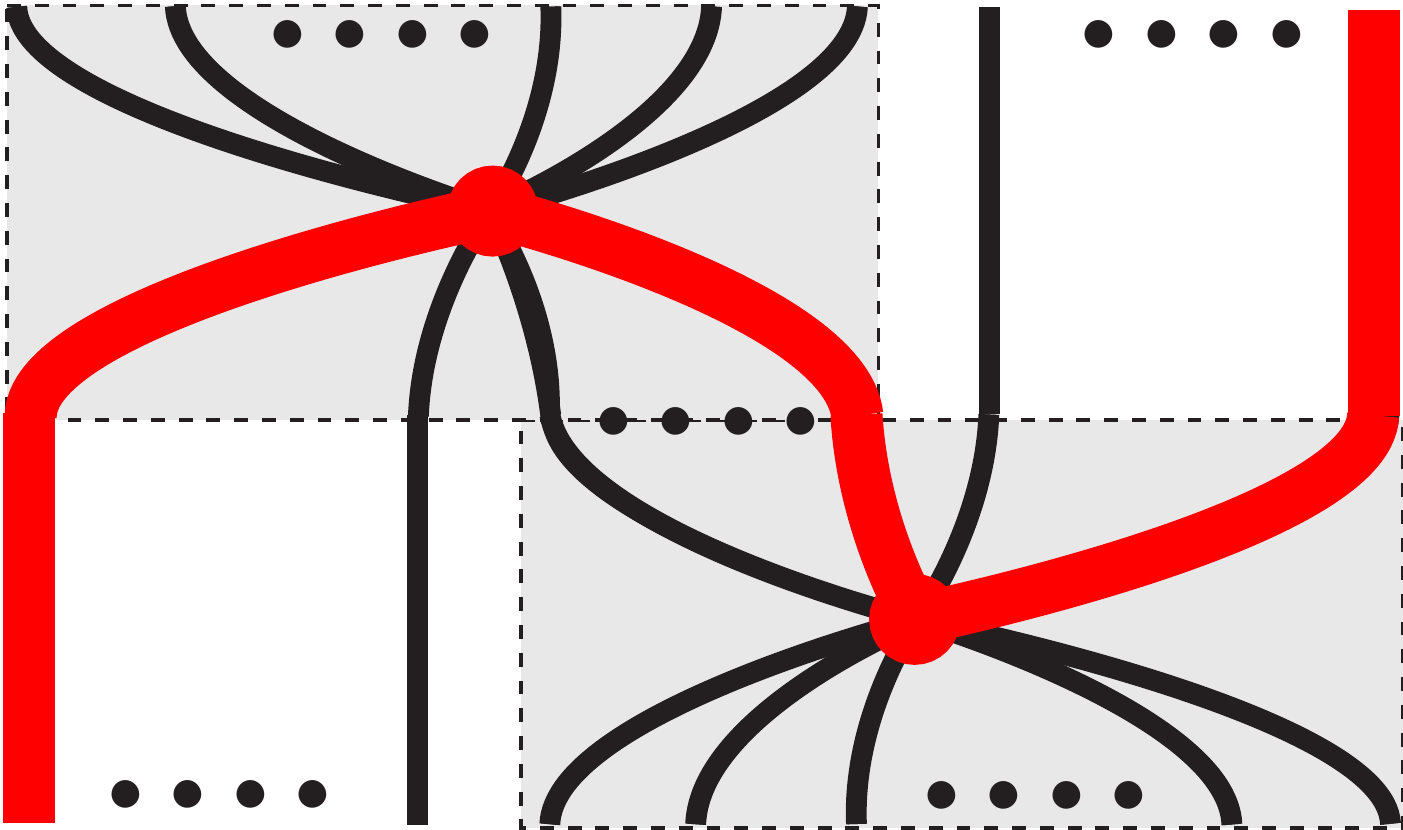,height=62pt}
   \ \  \epsfig{figure=SpiderComposition2,height=62pt}\ \ 
   \epsfig{figure=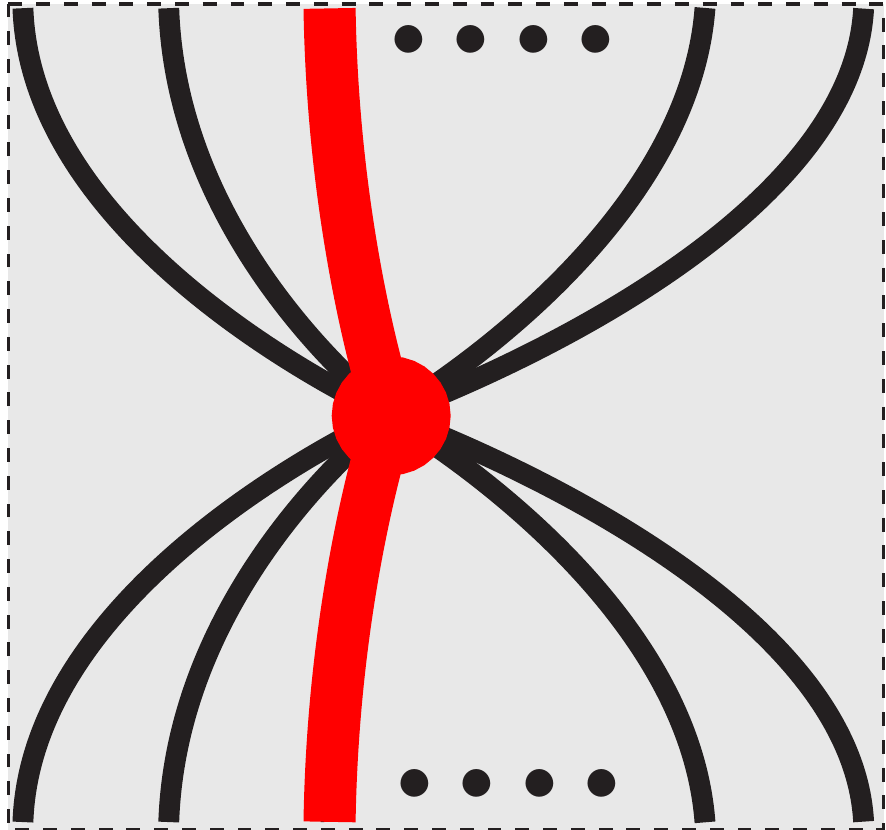,height=62pt}
    $
  \end{center}
    \vspace{-10pt}
    \caption{The rule for composing spiders subsumes the yanking rule. It generalizes it in a very powerful manner, as we shall see below.}\label{fig:spidercompred}
    \vspace{0pt}
\end{wrapfigure}\noindent
The composition axiom which governs these spiders is depicted on the right.  In words, whenever we have two spiders (1 and 2) such that at least one leg of spider 1 is connected to a leg of spider 2, then we can \em fuse \em them into a single spider. We also require the spider $\delta_1^1$ to be the identity, and that the set of spiders is invariant under upside-down flipping and leg-swapping.  Spiders and their composition rules generalise the $\cup'$'s,  $\cap$'s and their yanking rule of the previous section.  Indeed, when comparing Figure \ref{fig:analyanking} and Figure \ref{fig:spidercompred} one sees that one obtains cup's, cap's and their yanking rule by interpreting $\delta_0^2=\raisebox{-0.15cm}{\epsfig{figure=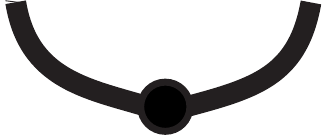,width=30pt}}$ as the cup and $\delta_2^0=\raisebox{-0.15cm}{\epsfig{figure=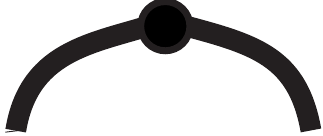,width=30pt}}$ as the cap.  So if on an object we have a non-degenerate observable then we automatically also have $\cup$'s and $\cap$'s.

You may rightfully ask yourself what the hell these spiders have to do with the observables of quantum theory.  The answer is given by the following not so trivial theorem.

\begin{theorem}\label{thm:CPV}{\rm\cite{CPV}}
In ${\bf FHilb}$ we have that non-degenerate observables $\{{\cal H}^{\otimes n}{\rTo^{\delta_n^m}} {\cal H}^{\otimes m}\}_{n,m}$ in the above sense exactly correspond with orthonormal bases on the underlying Hilbert space ${\cal H}$.
\end{theorem}

Referring to the discussion of quantum measurements and projectors at the beginning of Section \ref{sec:LA101}, non-degenerate observables $M$ on an $n$-dimensional Hilbert space can always be represented in in the form $r_1\cdot {\rm P}_1+\ldots +r_n\cdot {\rm P}_n$, with all $r_i$ non-equal, and ${\rm P}_i=|i\rangle\langle i|$ where $\{|1\rangle,\ldots, |n\rangle\}$ is some orthonormal basis.  So non-degenerate observables are in correspondence with orthonormal bases. Since Theorem \ref{thm:CPV} tells us that on a Hilbert space ${\cal H}$ in ${\bf FHilb}$  the non-degenerate observables that we defined in terms of spiders and ordinary orthonormal bases are one-and-the-same, we indeed showed that in ${\bf FHilb}$ our notion of non-degenerate observables in terms of spiders matches the usual notion of non-degenerate observables.

To establish which orthonormal  basis on a Hilbert space ${\cal H}$ corresponds to a given  non-degenerate observable $\{{\cal H}^{\otimes n}{\rTo^{\delta_n^m}} {\cal H}^{\otimes m}\}_{n,m}$, we will first pass to an alternative but equivalent presentation of non-degenerate observables in dagger symmetric monoidal categories.  From a pictorial point of view this alternative presentation is less attractive, but both from a physical and an algebraic point of view it makes a lot more sense.

\subsubsection{Copying-deleting-pair presentation of non-degenerate observables}

A \em non-degenerate observable \em or \em basis \em for an object $A$ in a dagger symmetric monoidal category consists of a \em copying \em operation $A\rTo^\delta A\otimes A$ and a \em deleting \em operation $A{\rTo^\varepsilon} {\rm I}$ which satisfy the following axioms:
\ben
\item $\varepsilon$ is a \em unit \em for (the \em comultiplication\em) $\delta$ i.e.~$(\varepsilon\otimes 1_A)\circ\delta=1_A$\,;
\item $\delta$ is \em coassociative \em i.e.~$(1_A\otimes\delta)\circ\delta=(\delta\otimes 1_A)\circ\delta$\,;
\item $\delta$ is \em cocommutative \em i.e.~$\sigma_{A,A}\circ\delta=\delta$\,;
\item $\delta$ is an \em isometry \em i.e.~$\delta^\dagger\circ\delta=1_A$\,;
\item $\delta$ satisfies the \em Frobenius law \em \cite{CarboniWalters} i.e.~$(\delta^\dagger\otimes 1_A)\circ(1_A\otimes\delta)= \delta\circ\delta^\dagger$. 
\een  

\begin{wrapfigure}{r}{0.29\textwidth}
\vspace{-0pt}
  \begin{center}
\epsfig{figure=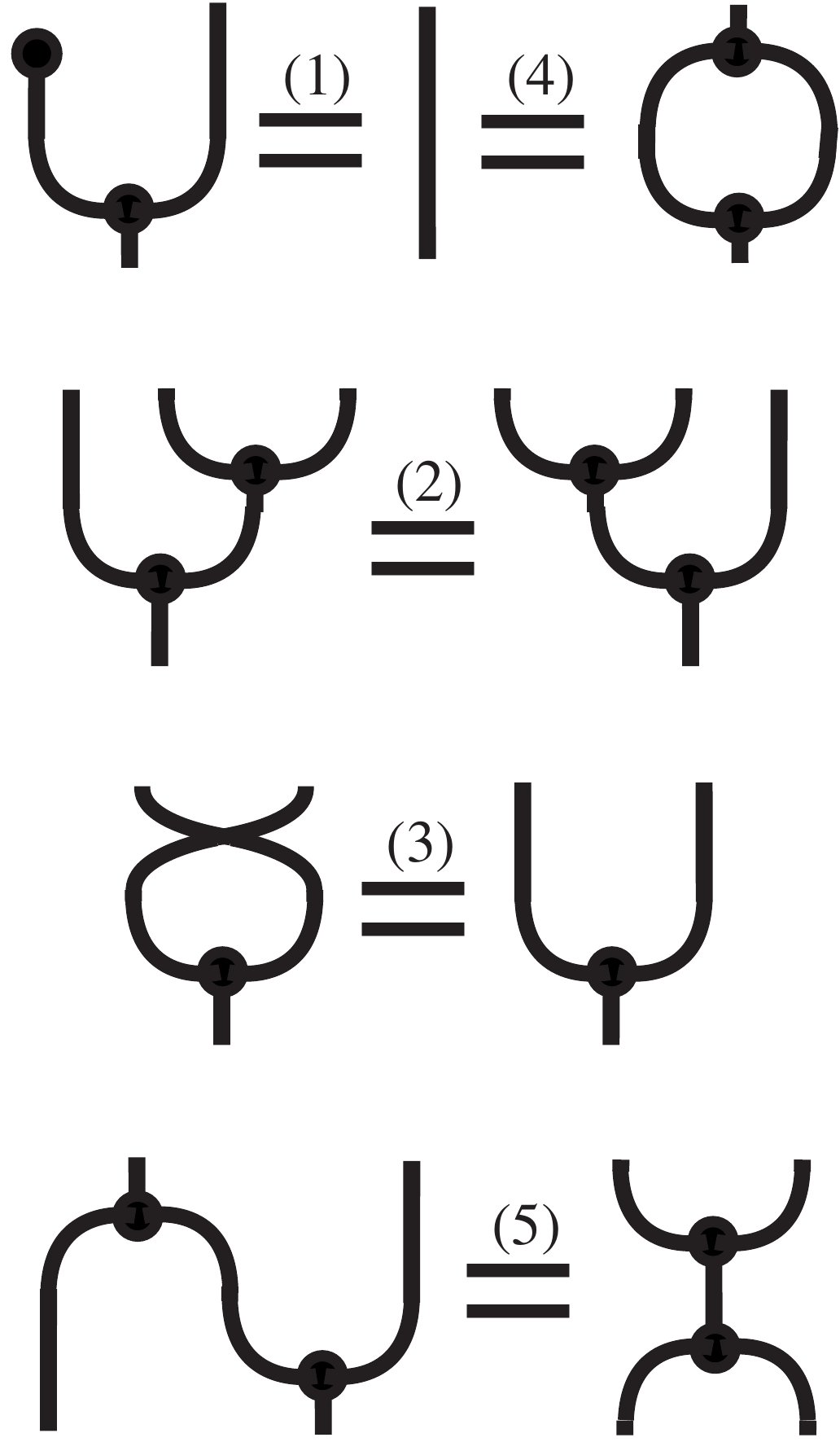,width=130pt}  \end{center}
    \vspace{-20pt}
\end{wrapfigure}\noindent
When introducing graphical elements 
\[
\delta=\raisebox{-0.15cm}{\epsfig{figure=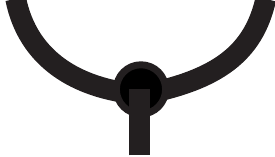,width=38pt}}\qquad\mbox{and}\qquad\varepsilon=\raisebox{-0.15cm}{\epsfig{figure=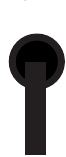,width=8.5pt}}
\]
we obtain the graphical rules depicted on the right.   In standard mathematical jargon all of these together mean that $(A,\delta,\varepsilon)$ is a so-called \em special dagger Frobenius commutative comonoid\em.  It is quite remarkable that this set of axioms exactly corresponds to the spiders discussed above. 
To pass from spiders to a copying-deleting-pair we set $\delta:=\delta_1^2$ and $\varepsilon:=\delta_1^0$.  Conversely, from the above axioms it follows that any composite of $\delta$'s, $\varepsilon$'s, their adjoints, identities, by using both composition and tensor, and provided its graphical representation is connected, only depends on the number of inputs $n$ and outputs $m$ \cite{Lack,CPaq}.  The spider $\delta_n^m$ then represents this unique morphism.

Turning our attention again to Theorem \ref{thm:CPV},  how does a copying-deleting-pair encode a basis? Given a basis $\{|i\rangle\}_i$  of a Hilbert space ${\cal H}$ we define the copying operation to be the linear map which `copies these basis vectors', and the deleting operation to be the linear map which `uniformly deletes these basis vectors' i.e.
\[
\delta:{\cal H}\to{\cal H}\otimes{\cal H}::|i\rangle\mapsto|ii\rangle\qquad\mbox{\rm and}\qquad\varepsilon:{\cal H}\to\C::|i\rangle\mapsto 1. 
\]
That these maps faithfully encode this basis, and no other basis, follows directly from the no-cloning theorem \cite{Dieks,WZ}; as the only vectors that can be copied by such an operation have to be orthogonal, they can only be the basis vectors we started from.  Explicitly put, with the above prescription of $\delta$, the only non-zero vectors $|\psi\rangle\in{\cal H}$ satisfying the equation $\delta(|\psi\rangle)=|\psi\rangle\otimes|\psi\rangle$ are the basis vectors $\{|i\rangle\}_i$.
Putting this in Dirac notation, while usually in quantum mechanics we represent a non-degenerate observable that corresponds to an orthonormal basis $\{|i\rangle\}_i$ by a linear operator $\sum_i r_i|i\rangle\langle i|$, we can represent this basis also by the linear map $\sum_i |ii\rangle\langle i|$.

More generally, in any dagger symmetric monoidal category one can define \em eigenvectors \em (or \em eigenstates\em) for an observable in the copying-deleting-pair sense,  as a state that is copied by $\delta$.  Graphically this means that these generalised eigenvectors  $\raisebox{-0.40cm}{\epsfig{figure=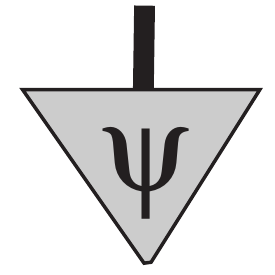,width=25pt}}$  satisfy the equation:\vspace{-3.5mm}
\beq\label{eq:eigenstate}
\raisebox{-0.60cm}{\epsfig{figure=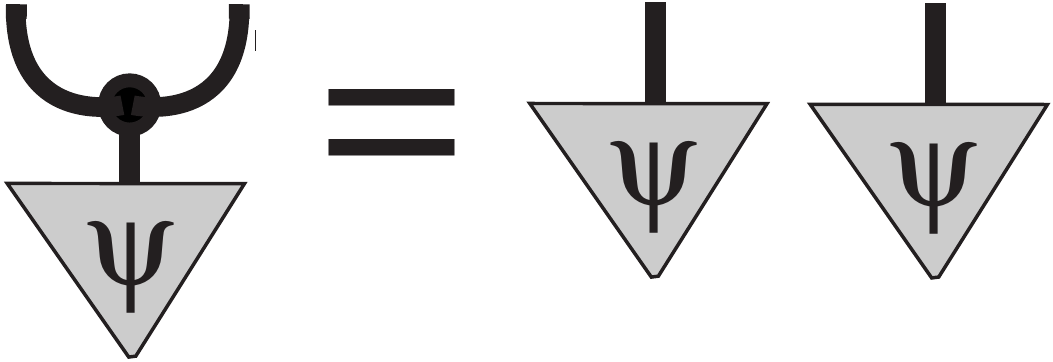,width=100pt}}  \qquad\qquad\mbox{i.e.}\qquad\qquad 
\delta\circ\psi=\psi\otimes\psi\,. \vspace{-1.5mm}
\eeq
This is a strong property since it means that the `connected' picture on the left can be replaced by the `disconnected' one on the right.  Obviously this has major implications in computations.  

The copying-deleting-pair presentation also points at  a physical interpretation of non-degenerate observables.  The copying and deleting maps witness those states that can be copied, and hence, again by the no-cloning theorem, those that happily live together within a classical realm.  This provides a perspective of the classical-quantum distinction which is somewhat opposite to the usual one: rather than constructing a quantum version of a classical theory via quantization, here we extract a classical version out of a quantum theory, via \em classicization\em.  

As a little break from the mathematical developments we philosophize a bit here.  The above argument suggests that there is some world out there, say the \em quantum universe\em, which we can probe by means of \em classical interfaces\em.  There are many different interfaces through which we can probe the quantum universe, and each of them can only reveal a particular aspect of that quantum universe.  Here one can start speculating.  For example, one could think that the change of state in a quantum measurement is caused by forcing part of the quantum universe to match the format of the classical interface by means of which we are probing it.  In other words, there is a very rich world out there, and we as human agents do not have the capability to sense it in its full glory.  We have no choice but to mould the part of that universe in which we are interested into a form that fits the much smaller world of our experiences.  This smaller world is what in physics we usually refer to as a space-time manifold.  But of course this is only my speculation.

\subsubsection{General observables}

So the observables defined in terms of spiders  are  all non-degenerate.  But one can define  degenerate counterparts to these which, in fact, more clearly elucidate their conceptual significance.  The main idea is that given spiders on $A$, or equivalently, a copying-deleting-pair on $A$, we will no longer think of it as the observable itself, but as the set of outcomes for some other observable which now can be degenerate.  We define these arbitrary  observables as certain morphisms $B\rTo^{m}A\otimes B$, subject to additional constraints.  So in this case $B$ stands for the quantum system, while $A$ stands for the classical data  for that observable, i.e.~the \em measured values \em or the \em spectrum\em.  Since $B$ appears both before and after the measurement we are considering non-demolition measurements here. The additional constraints that $B\rTo^{m}A\otimes B$ obeys are
\beq\label{eq:measurement}
\raisebox{-0.80cm}{\epsfig{figure=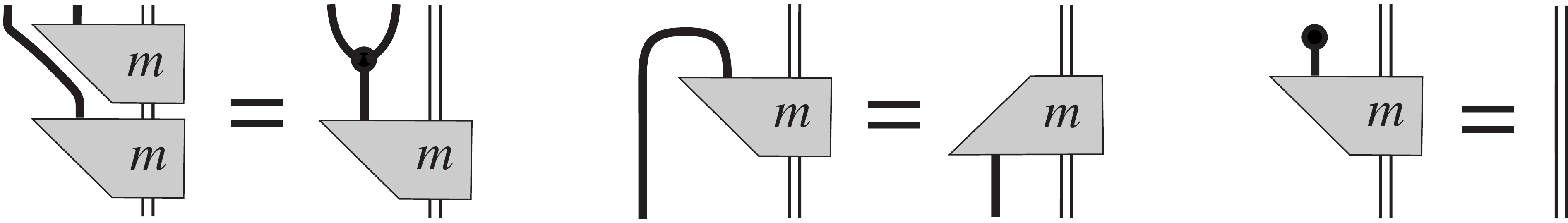,width=380pt}}
\eeq
where the single wire stands for the classical data $A$ while the double wire stands for the quantum system $B$ -- the structural reason for this single-double distinction is explained in \cite{CPaqPav}.  The first of these conditions states:  if after a measurement we perform the same measurement again, then this boils down to the same thing   as when we would have just copied the outcome obtained in the first measurement.  This of course is the same as: 
\bit
\item[-] we obtain the same outcome in the second measurement as in the first one;
\item[-] the second measurement does not alter the state of the system anymore.
\eit
The following theorem shows that we indeed recover the usual notion of a quantum measurement.

\begin{theorem}{\rm\cite{CPav}}
Let ${\cal H}_2$ be an $n$-dimensional Hilbert space together with a chosen basis, that is, by Theorem \ref{thm:CPV}, a copying-deleting-pair. Then linear maps $f: {\cal H}_1\to{\cal H}_2\otimes {\cal H}_1$ in  ${\bf FHilb}$ satisfying eqs.(\ref{eq:measurement}) exactly correspond to all projector spectra $\{{\rm P}_1, \ldots, {\rm P}_n\}$ of self-adjoint operators on ${\cal H}_1$.  Explicitly we have $f= \sum_i |i\rangle\otimes {\rm P}_i$ where each $|i\rangle$ represents an outcome.
\end{theorem}

One can verify that for the non-degenerate observables defined as triples $(A,\delta,\varepsilon)$, the morphism  $A\rTo^{\delta}A\otimes A$ provides an example of such a measurement, with $B:=A$. The fact that both the classical data and the quantum system are represented by the same symbol might look a bit weird at first, but poses no structural problem:  the classical values are represented by the triple $(A,\delta,\varepsilon)$ and not by $A$ alone. 
The analogy in Hilbert space quantum mechanics is that we think of the Hilbert space as the quantum system, while the pair consisting of a Hilbert space and an observable `thereon' represents the classical values for that observable.  Moreover, to avoid conceptual confusion, we could represent the quantum system by an isomorphic copy of $A$.  

So now we've got ourselves a graphical representation of arbitrary observables at hand.  As already mentioned at the beginning of this section, this will also allow us to reason about classical data flow diagrammatically -- cf.~the caption of Figure \ref{pic:classwire}.  We won't discuss this here but refer the interested reader to \cite{CPav, CPaqPav}, where also the role of \em decoherence \em in measurements is discussed.  All of this applies to all measurements in dagger symmetric monoidal categories.   

Rather than focussing how the classical and the quantum interact, we will now focus on how different quantum observables interact, all still within a diagrammatic realm of course.  These results were obtained in collaboration with Duncan in \cite{CD}. 

\subsection{A pair of complementary observables in pictures}

The most famous example of complementary observables are obviously the \em position \em and \em momentum \em observables. Here we will only consider finite dimensional Hilbert spaces.  Let $M$ and $M'$ be two non-degenerate observables acting on an $n$-dimensional Hilbert space, let $|\psi_1\rangle, \ldots,|\psi_n\rangle$ be 
\linebreak\vspace{-12pt}

\begin{wrapfigure}{r}{0.16\textwidth}
\vspace{4pt}
  \begin{center}
\begin{minipage}[b]{1\linewidth} 
\begin{picture}(30,90)
\centering\epsfig{figure=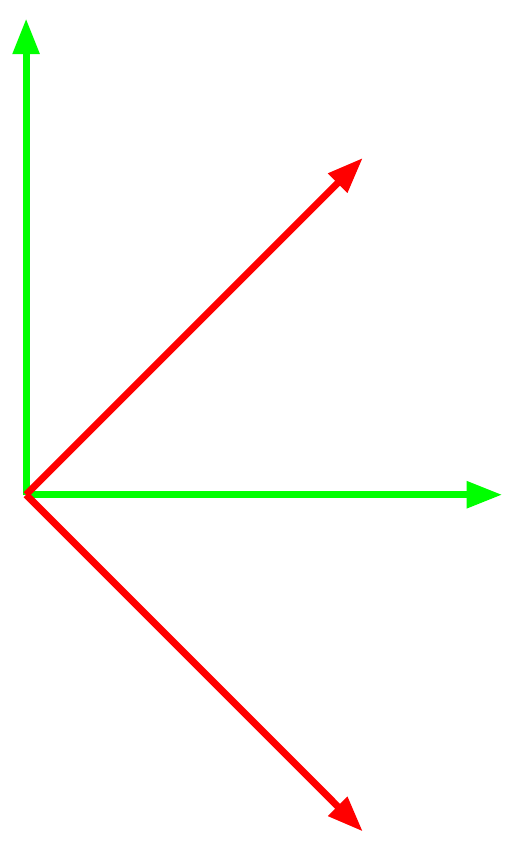,width=60pt}\hspace{-40pt}
\put(40,40){\bG$|0\rangle$\e}
\put(-22,100){\bG$|1\rangle$\e}
\put(22,85){\bR$|+\rangle$\e}
\put(22,-05){\bR$|-\rangle$\e}
\end{picture}
\end{minipage}
  \end{center}
  \vspace{-0pt}
  \caption{Eigenvectors for the complementary observables  $Z$ and $X$.}\label{fig:unibiasvect}
    \vspace{0pt}
\end{wrapfigure}\noindent
mutually orthogonal normalised eigenvectors of $M$, and let $|\psi_1'\rangle, \ldots,|\psi_n'\rangle$ be mutually orthogonal normalised eigenvectors of $M'$.  Then $M$ and $M'$ are \em complementary \em \cite{Kraus}  (or \em unbiased \em \cite{Schwinger})    
if for 
all $i$ and all $j$ we have that $|\langle\psi_i|\psi_j'\rangle|^2={1\over n}$.  The simplest example of complementary observables are the $Z$- and $X$-observables for a qubit.  For the $Z$-observable the eigenvectors are $|0\rangle$ and $|1\rangle$ while for the $X$-observable the eigenvectors are $|+\rangle:={1\over \sqrt{2}}\cdot(|0\rangle+|1\rangle)$ and $|-\rangle:={1\over \sqrt{2}}\cdot(|0\rangle-|1\rangle)$.  We can break down the definition of complementary or unbiased observables in terms of the notion of unbiased vectors.  We say that a normalized vector $|\psi'\rangle$ is \em unbiased \em  for an observable $M$ if for all $i$ we have that  $|\langle \psi_i|\psi'\rangle|^2={1\over N}$. This in particular means that when the system is in state $|\psi'\rangle$ and we measure observable $M$, all outcomes are equally probable, hence the term `unbiased'.    Hence two observables are  complementary or unbiased if the normalized eigenvectors for one are unbiased for the other.   
One could alternatively say that such  a pair of observables are `maximally non-classical', that is,  `maximally quantum', given that the eigenvectors of one fail to be an eigenvector of the other in an `extremal manner'.  Hence one would expect a substantial 
\linebreak\vspace{-12pt}

\begin{wrapfigure}{r}{0.62\textwidth}
\vspace{-16pt}
 \begin{center} 
\begin{tabular}{|c|c|c|} 
\bW\fbox{\bBl\bf$\!\!$observable$\!\!$\e}\e & \bW\fbox{\bBl\bf$\!\!$eigenvectors$\!\!$\e}\e & \bW\fbox{\bBl\bf unbiased states\e}\e \\
\hline
\bW\fbox{\bBl$Z$\e}\e & \bW\fbox{\bBl$|0\rangle, |1\rangle$\e}\e &    
\bW\fbox{\bBl$|0\rangle + e^{i\alpha} |1\rangle$  e.g.~$|+\rangle, |-\rangle$\e}\e \\
\hline
\bW\fbox{\bBl$X$\e}\e & \bW\fbox{\bBl$|+\rangle, |-\rangle$\e}\e &    
\bW\fbox{\bBl$|+\rangle + e^{i\alpha} |-\rangle$  e.g.~$|0\rangle, |1\rangle$\e}\e \\
\hline
\bW\fbox{\bBl$(A,\delta,\varepsilon)$\e}\e & 
\bW\fbox{\bBl $|\psi\rangle$ in eq.(\ref{eq:eigenstate})\e}\e & \bW\fbox{\bBl $|\psi\rangle$ in eq.(\ref{eq:unbiasedstate})\e}\e\\
\hline
\end{tabular}
\end{center} 
    \vspace{-18pt}
\end{wrapfigure}\noindent
chunk of quantum mechanical structure 
to be captured by complementary observables. A graphical account of these would substantially boost the power of the graphical calculus.  

It turns out that we can straightforwardly translate all the above to the more general graphical framework, and we will even obtain additional insights.   Firstly, unbiasedness of  a state for an observable $(A,\delta,\varepsilon)$ can be expressed in an arbitrary dagger symmetric monoidal category, and we depict these as follows:
\vspace{-1.5mm}
\beq\label{eq:unbiasedstate}
\raisebox{-0.60cm}{\epsfig{figure=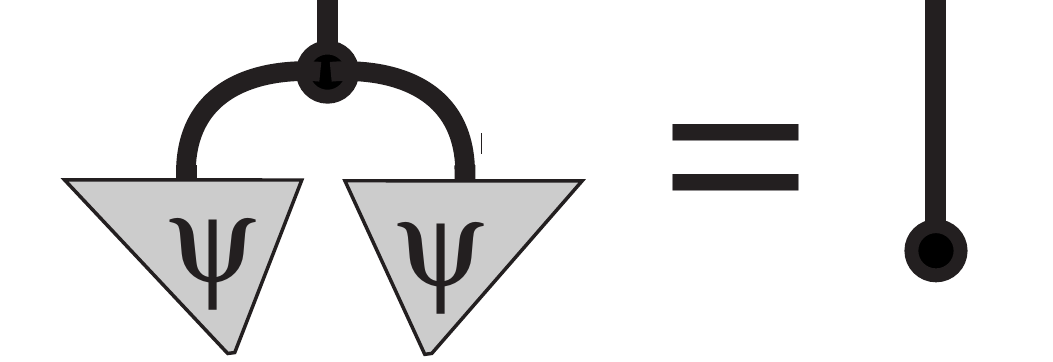,width=100pt}}
  \qquad\qquad\mbox{i.e.}\qquad\qquad\delta^\dagger\circ(\bar{\psi}\otimes\psi)=\varepsilon^\dagger\,. \vspace{-1.5mm}
\eeq
 In ${\bf FHilb}$,  taking $\delta$ to be the linear map which copies the vectors in $\{|i\rangle\}_{i=1}^{i=n}$, then 
\[
\delta^\dagger:{\cal H}\otimes{\cal H}\to{\cal H}:: 
\left\{\begin{array}{l}
|ii\rangle\mapsto|i\rangle\\
|ij\rangle\mapsto 0 \quad\mbox{when}\quad i\not=j\,.
\end{array}\right.
\]
So for $|\psi\rangle=(\psi_1,\ldots,\psi_n)=(\langle \psi|1\rangle, \ldots, \langle \psi|n\rangle)$ we obtain 
\[
\delta^\dagger(\overline{|\psi\rangle}\otimes|\psi\rangle)=(\bar{\psi}_1{\psi_1},\ldots,\bar{\psi}_n{\psi_n})=
\left(|\langle \psi|1\rangle|^2, \ldots, |\langle \psi|n\rangle|^2\right) \,.
\]
Furthermore $\varepsilon=\sum_i\langle i|$ so $\varepsilon^\dagger=\sum_i|i\rangle$, that is, written as a matrix, $\varepsilon^\dagger=(1,\ldots,1)$.  Hence 
\beq\label{eq:notnorm}
\left(|\langle \psi|1\rangle|^2, \ldots, |\langle \psi|n\rangle|^2\right)=(1,\ldots,1)\quad\mbox{i.e.}\quad
|\langle \psi|1\rangle|^2=1\ ,\, \ldots\ ,\, |\langle \psi|n\rangle|^2=1\,.
\eeq
The reason that the righthandside of these equations is $1$ rather than ${1\over n}$ is that the state $\psi$ is not normalized but that $\sum_i|\langle \psi|i\rangle|^2=n$, as a consequence of eq.(\ref{eq:unbiasedstate}).  To see this it suffices 
to add eqs.(\ref{eq:notnorm}), yielding $\langle \psi|\psi\rangle= \sum_i |\langle \psi|i\rangle|^2=\sum_i 1=n$.  So eqs.(\ref{eq:notnorm}) do imply that the normalized vector ${1\over\sqrt{n}}|\psi\rangle$ is indeed unbiased  relative to observable $({\cal H}, \delta,\varepsilon)$.  Also in general, the states obeying eq.(\ref{eq:unbiasedstate}) won't be normalised, but have the square-root of the dimension as length. 
What does this graphically mean, ``to have the square-root of the dimension as length"?  In ${\bf FHilb}$ it turns out that the dimension is exactly a $\cup$ post-composed with a $\cap$, hence a circle.  Given $(A,\delta,\varepsilon)$ with $\cup=\delta^2_0=\raisebox{-0.15cm}{\epsfig{figure=cupspider.pdf,width=30pt}}$ while $\cap=\delta^0_2=\raisebox{-0.15cm}{\epsfig{figure=capspider.pdf,width=30pt}}$\,, we define the \em dimension \em of $A$ to be $\delta^0_2\circ\delta^2_0$, which is equal to $\varepsilon\circ\varepsilon^\dagger$ since both correspond to $\delta_0^0$, the spider with no legs.  We have
\vspace{-1.5mm}
\[
\epsfig{figure=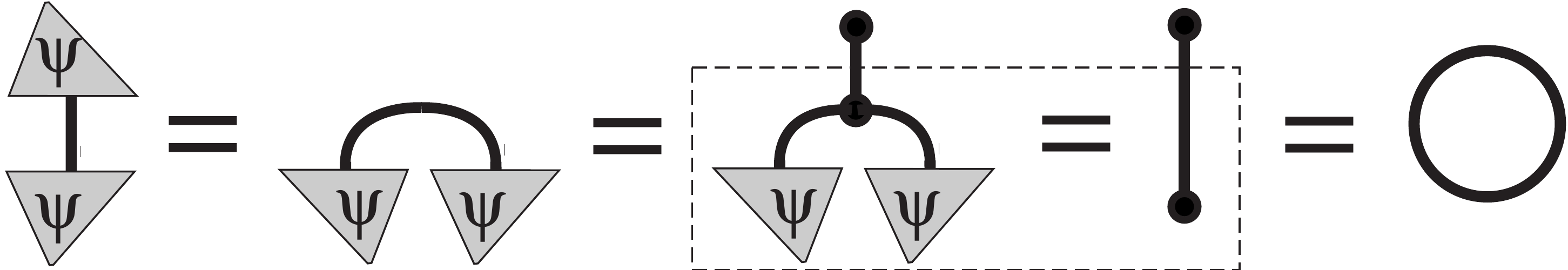,width=260pt}
 \vspace{-1.5mm}
\]
where the dotted area is eq.(\ref{eq:unbiasedstate}). Hence $\psi$ indeed has the square-root of the dimension as length.

Since we now both know what `eigenvector' --cf.~eq.(\ref{eq:eigenstate})-- and `unbiased' --cf.~eq.(\ref{eq:unbiasedstate})-- mean in arbitrary  dagger symmetric monoidal categories we can define complementarity for them:

\begin{definition}
Two observables $(A,\delta_Z,\varepsilon_Z)$ and $(A,\delta_X,\varepsilon_X)$ in a dagger symmetric monoidal category 
are \em complementary \em if the eigenvectors of one are unbiased for the other.
\end{definition}

Graphically, to distinguish between two observables, we will depict the `head of the spiders' of one in green and of the other one in red.  We obtain the following remarkable characterization of complementary, one of the most fascinating results our approach has thus far produced.

\begin{theorem}{\rm \cite{CD}}\label{thm:complementary}
If a dagger symmetric monoidal category has `enough states', then two observables $(A,\delta_Z,\varepsilon_Z)$ and $(A,\delta_X,\varepsilon_X)$ are complementary if and only if they satisfy:
\beq\label{eq:Hopf}
\raisebox{-0.70cm}{\epsfig{figure=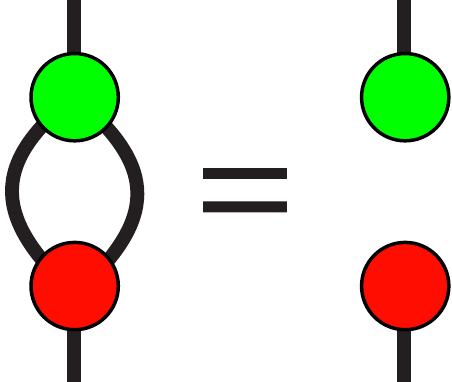,width=55pt}}
 \qquad\qquad\qquad\mbox{i.e.}\qquad\qquad\qquad
\delta^\dagger_Z\circ\delta_X=\varepsilon_Z\circ\varepsilon_X^\dagger\,.\vspace{-1mm}
\eeq
\end{theorem}
We won't spell out what it means to have `enough states', we just mention that it is a very weak requirement which holds in all example categories we are aware of.  Observe the radical topology change  from the lefthandside to the righthandside of the equation. The reader can easily verify that for the $Z$- and the $X$-observables, respectively defined as:
\[
\delta_Z::\left\{
\begin{array}{l}
|0\rangle\mapsto |00\rangle\\
|1\rangle\mapsto |11\rangle
\end{array}
\right.\quad
\varepsilon_Z::\left\{
\begin{array}{l}
|0\rangle\mapsto 1\\
|1\rangle\mapsto 1
\end{array}
\right.\quad
\delta_X::\left\{
\begin{array}{l}
|+\rangle\mapsto |++\rangle\\
|-\rangle\mapsto |--\rangle
\end{array}
\right.\quad
\varepsilon_X::\left\{
\begin{array}{l}
|+\rangle\mapsto 1\\
|-\rangle\mapsto 1
\end{array}
\right.
\]
this equation indeed holds in ${\bf FHilb}$, up to a scalar multiple. 

\subsection{Phases in pictures}

Without any further requirements, the general notion of observable $(A,\delta,\varepsilon)$ in dagger symmetric monoidal categories comes with a corresponding notion of  phase.  Let ${\cal S}(A,\delta,\varepsilon)$ be the set of all states ${\rm I}\rTo^\psi A$  that are unbiased for $(A,\delta,\varepsilon)$.  On the set ${\cal S}(A,\delta,\varepsilon)$ we define a multiplication as follows.  Given two states ${\rm I}\rTo^\psi A$ and ${\rm I}\rTo^\phi A$ their $\odot$-product is
\[
\psi\odot \phi:=\delta^\dagger\circ(\psi\otimes \phi)=\raisebox{-0.40cm}{\epsfig{figure=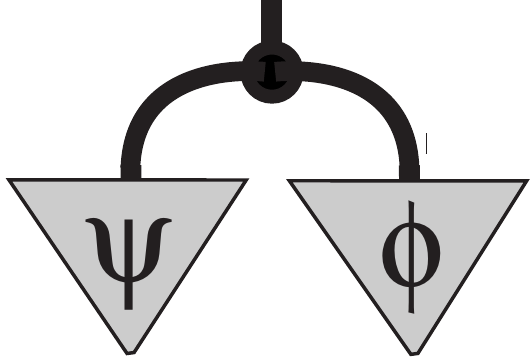,width=49pt}}\,.
\]  
Given any state ${\rm I}\rTo^\psi A$ we can also consider the morphism\vspace{-1mm}
\[
U_\psi:=\delta^\dagger\circ(\psi\otimes 1_A)=\raisebox{-0.46cm}{\epsfig{figure=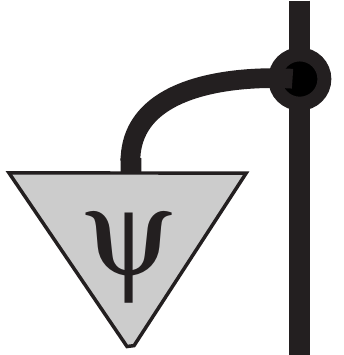,width=34pt}}
\]
and one can easily show that $U_\psi$ is unitary if and only if $\psi$ is unbiased for $(A,\delta,\varepsilon)$.  Let ${\cal U}(A,\delta,\varepsilon)$ be the set of  all unitary morphisms of the form $U_\psi=\delta^\dagger\circ(\psi\otimes 1_A)$.

\begin{theorem}{\rm\cite{CD}}
 For any observable $(A,\delta,\varepsilon)$ in a dagger symmetric monoidal category $({\cal S}(A,\delta,\varepsilon), \odot,\varepsilon^\dagger)$ and $({\cal U}(A,\delta,\varepsilon), \circ, 1_A)$ are isomorphic Abelian groups.  For ${\cal S}(A,\delta,\varepsilon)$ the inverse is provided by the conjugate and for $({\cal U}(A,\delta,\varepsilon), \circ, 1_A)$ the inverse is provided by the adjoint.
\end{theorem}

For the $Z$ observable $({\cal Q}, \delta_Z, \varepsilon_Z)$ on a qubit ${\cal Q}$ in ${\bf FHilb}$ we have \vspace{-1mm}
\[
{\cal S}(A,\delta,\varepsilon^\dagger)=\left\{|0\rangle + e^{i\alpha}|1\rangle\bigm| \alpha\in[0,1) \right\}
\qquad\mbox{and}\qquad
{\cal U}(A,\delta,\varepsilon)=
\left\{\left( 
  \begin{array}{cc}
1&0\\0&e^{i\alpha}
  \end{array}
\right)
\Biggm| \alpha\in[0,1) \right\}\vspace{-1mm}
\]

\begin{wrapfigure}{l}{0.19\textwidth}
\vspace{-16pt}
  \begin{center}
\epsfig{figure=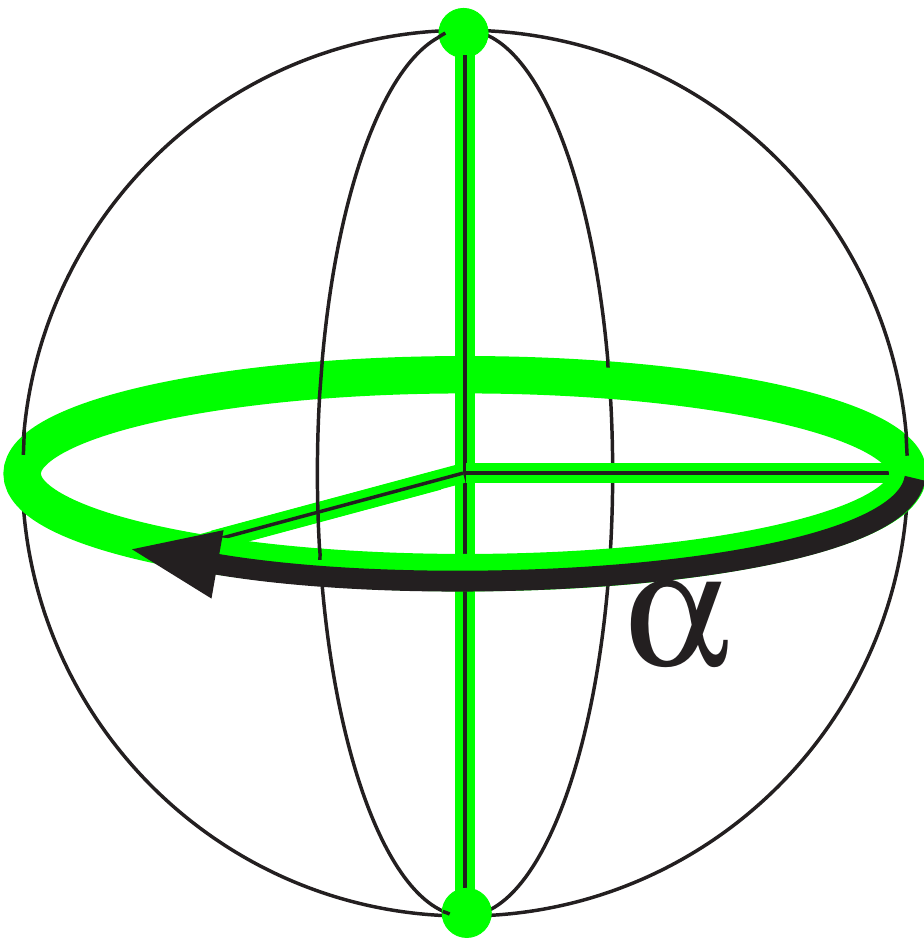,width=70pt}  
\end{center}
    \vspace{0pt}
  \vspace{-6pt}
  \caption{The phases for the $Z$-observable on a qubit in ${\bf FHilb}$.}\label{fig:unibiasvect}
    \vspace{-0pt}
\end{wrapfigure}\noindent
So we obtain \em phases \em and hence call this group the \em phase group\em.  Since \vspace{-0.5mm}
\[
(|0\rangle + e^{i\alpha}|1\rangle)\odot(|0\rangle + e^{i\alpha'}|1\rangle)= |0\rangle + e^{i(\alpha+\alpha')}|1\rangle \vspace{-0.5mm}
\]
the multiplication in the group corresponds to adding angles, and since  \vspace{-0.5mm}
\[
(|0\rangle + e^{i\alpha}|1\rangle)\odot(|0\rangle + e^{-i\alpha}|1\rangle)= |0\rangle + |1\rangle=\varepsilon^\dagger \vspace{-0.5mm}
\]
the inverse in the group  corresponds to reversing angles.  To hint that in the case of qubits in ${\bf FHilb}$, unbiased states  correspond to phase angles, we  will now denote unbiased states of non-degenerate
observables as $\alpha$, and correspondingly,   
\linebreak\vspace{-12pt}

\begin{wrapfigure}{r}{0.46\textwidth}
\vspace{-6pt}
\begin{center}
   \mbox{\epsfig{figure=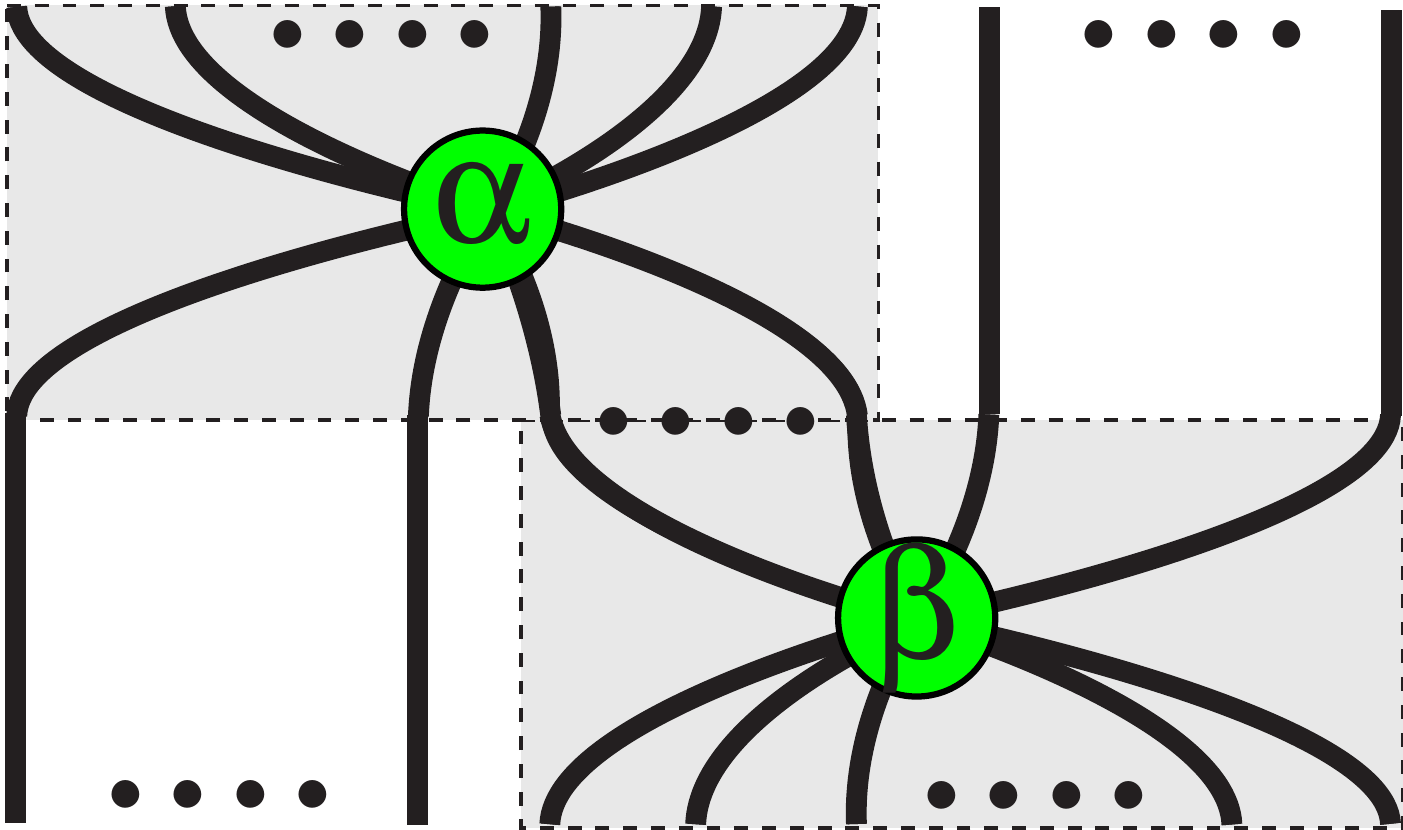,height=62pt}
   \ \  \epsfig{figure=SpiderComposition2,height=62pt}\ \ 
 \epsfig{figure=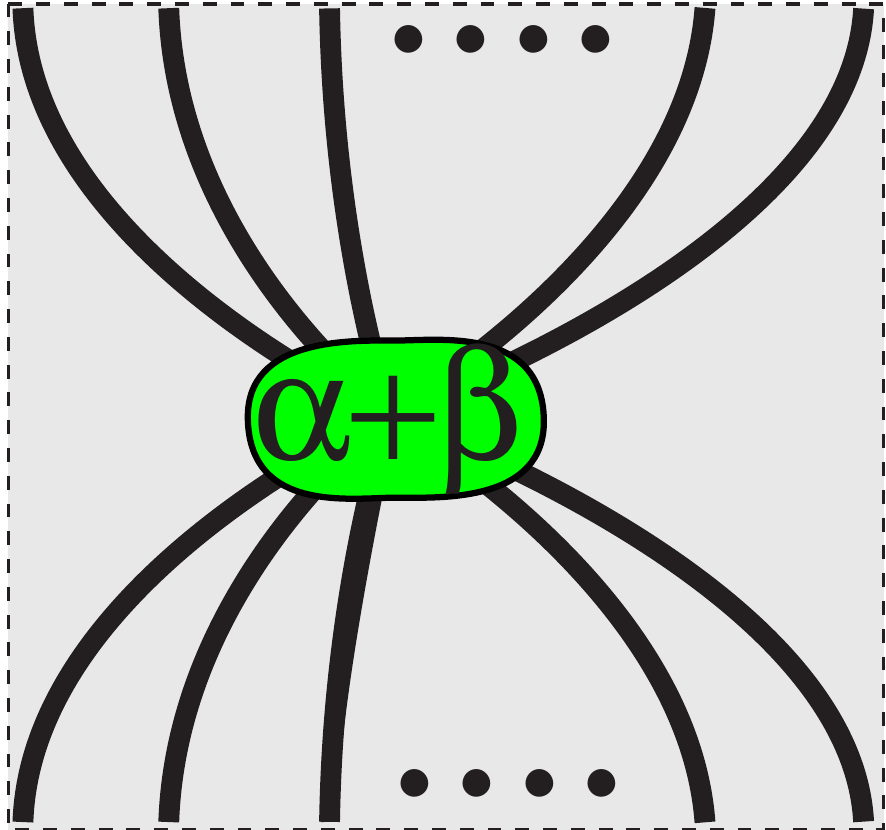,height=62pt}}
   \end{center}
    \vspace{-10pt}
    \caption{Spiders decorated with phases can still be fuzed together provided we add the phases.}\label{fig:spidercompbis}
    \vspace{-0pt}
\end{wrapfigure}\noindent
denote the group's multiplication as $+$. 

Due to the fact that these \em generalised phases \em are derivable from a non-degenerate observable in a dagger symmetric monoidal category, that is, a family of spiders, they interact particularly well with these spiders.  In fact, we obtain a much richer family of spiders, of which the heads are now \em decorated \em with these generalised phases.  Strictly speaking, the heads of these spiders shouldn't be symmetric since they are not invariant under conjugation, but given that we depict them in a particular way, i.e.~as circles enclosing a Greek letter, it should be clear to the reader that they change under conjugation.  Special examples of decorated spiders are the unbiased states $\alpha=\raisebox{-0.28cm}{\epsfig{figure=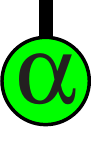,width=12pt}}$ and \em generalised phase gates \em $\delta^\dagger\circ(\alpha\otimes 1_A)=\raisebox{-0.28cm}{\epsfig{figure=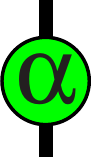,width=12pt}}$.  \vspace{-1mm}


\subsection{Example: information  flows in quantum computational models} 

We define the category ${\bf FHilb}_2$ to be the same as ${\bf FHilb}$, except for the fact that we restrict
\linebreak\vspace{-12pt}

\begin{wrapfigure}{r}{0.40\textwidth}
\vspace{-20pt}
\begin{center}
\epsfig{figure=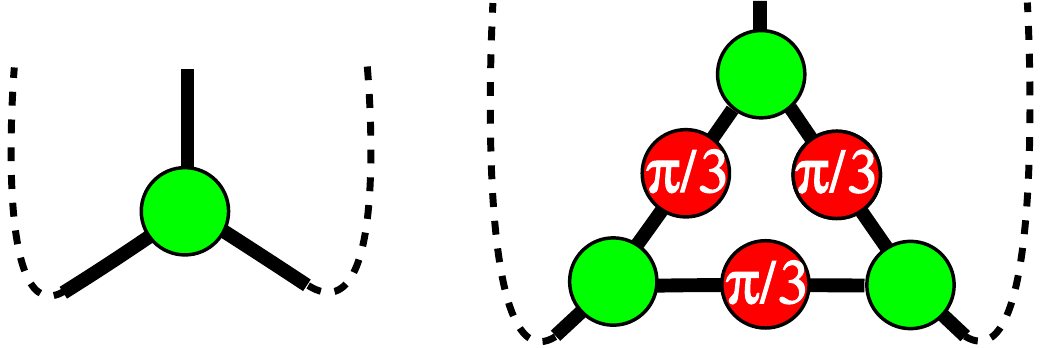,width=140pt}
   \end{center}
    \vspace{-10pt}
    \caption{The GHZ-state and the W-state \cite{W_GHZ} expressed in terms of decorated spiders.}\label{GHZpic}
    \vspace{-10pt}
\end{wrapfigure}\noindent
the objects to those Hilbert spaces of ${\bf FHilb}$, except for the fact that we restrict the objects to those Hilbert spaces of which the dimension is of the form $2^n$,  where $n$ can be 
any natural number, including $0$.  Put in physical terms, we only consider systems consisting of $n$ qubits where $n$ takes values within $\N$. Since in quantum computing we mainly work with qubits, ${\bf FHilb}_2$ constitutes the most relevant part of ${\bf FHilb}$ for quantum informatic purposes.  The following theorem tells us that the abstract language developed in the previous section, enables us to describe any morphism in ${\bf FHilb}_2$, and hence any state and physical process involving $n$ qubits.  For example, Figure \ref{GHZpic} shows how the so-called W-state looks in terms of decorated spiders.

\begin{theorem}
 Every linear map in ${\bf FHilb}_2$ can be expressed in the  language of a pair of complementary observables and corresponding phases in a dagger symmetric monoidal category.  Hence, every linear map in ${\bf FHilb}_2$ can be depicted using only red and green decorated spiders.
\end{theorem}

Here is a \em proof \em of this fact.  It is a standard result in quantum computing that any unitary operation from $n$ qubits to $n$ qubits can be expressed in terms of one-qubit unitaries and a two-qubit unitary, which typically is taken to be the $CX$-gate \cite{NielsenChuang}.  This $CX$-gate is also called the $CNOT$-gate (read: `controlled not'). 
For $x$ either $0$ or $1$ it is explicitly given by
\[
CNOT:{\cal H}\otimes{\cal H}\to{\cal H}\otimes{\cal H}::\left\{\begin{array}{l}
|0x\rangle\mapsto |0x\rangle\\ |1x\rangle\mapsto |1NOT(x)\rangle
\end{array}\right.\quad\mbox{ with}\quad\  NOT:{\cal H}\to{\cal H}::\left\{\begin{array}{l}
|0\rangle\mapsto |1\rangle\\ |1\rangle\mapsto |0\rangle
\end{array}\right..
\]
It turns out that the $CX$-gate naturally arises from a pair of complementary observables.   For the $Z$- and $X$-observable one can verify that  $\raisebox{-0.3cm}{\epsfig{figure=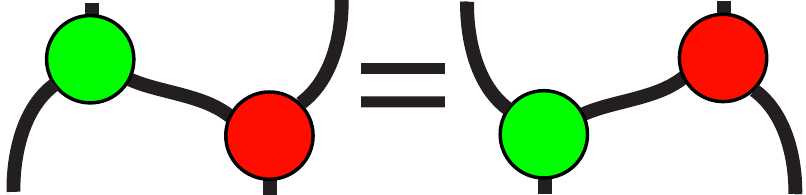,width=92pt}}$\,, which we therefore can depict as $\raisebox{-0.18cm}{\epsfig{figure=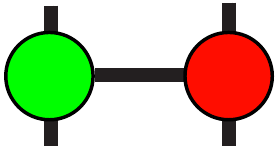,width=33pt}}$.  This is exactly the $CX$-gate, which can be verified by direct computation.  A more enlightening graphical proof is the following.  The states $|0\rangle$ and $|1\rangle$ are unbiased states for the $X$-observable,  so we can write them as decorated red spiders, namely  $|0\rangle=\raisebox{-0.22cm}{\epsfig{figure=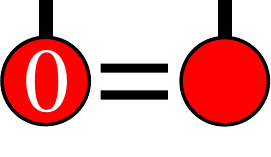,width=33pt}}$ and $|1\rangle=\raisebox{-0.22cm}{\epsfig{figure=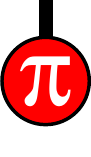,width=11pt}}$\,.  When we `plug' them in the green input of the $CX$-gate, then they are copied (as they are eigenvectors of the green observable), so we obtain
$\raisebox{-0.55cm}{\epsfig{figure=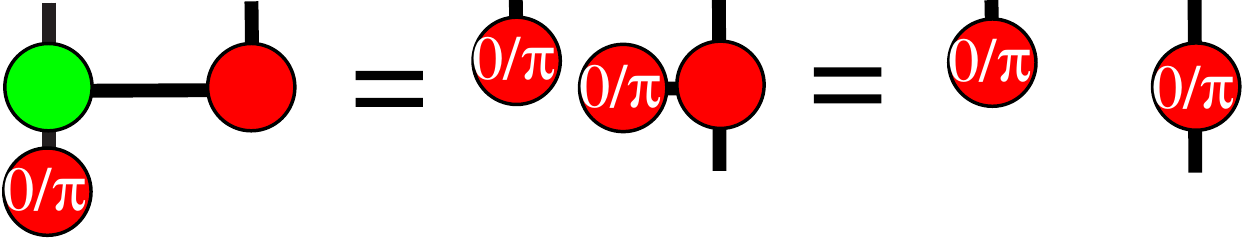,width=156pt}}$\,.  In the case we plugged in $|0\rangle=\raisebox{-0.22cm}{\epsfig{figure=zerostate.pdf,width=33pt}}$\,, the remaining input will act as the identity since $\raisebox{-0.22cm}{\epsfig{figure=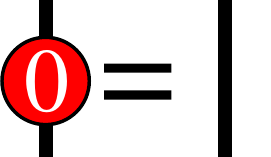,width=33pt}}=1_{\cal Q}$\,, while in the case we plugged in $|1\rangle=\raisebox{-0.22cm}{\epsfig{figure=pistate.pdf,width=11pt}}$\,, the remaining input will act as $\raisebox{-0.22cm}{\epsfig{figure=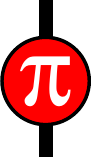,width=11pt}}=X=NOT$, matching perfectly the $CX$-gate.   So we also need arbitrary one-qubit unitaries.  These arise as follows.  Up to overall phase factors, the group $SU(2)$ of all one-qubit unitaries and the group $SO(3)$ of all orthogonal rotations are isomorphic.  Hence we can represent one-qubit unitaries as orthogonal rotations of the Bloch sphere, which we depicted in Figure \ref{fig:unibiasvect}.  All orthogonal rotations can be expressed in terms of a rotation of some angle $\alpha$ along one axis, then a rotation of some angle $\beta$ along a second axis which is orthogonal to the first one, and then a rotation of some angle $\gamma$ again along the first axis.  The $\alpha, \beta, \gamma$ are called the Euler angles.  Hence we can realize an arbitrary one-qubit unitary as $\raisebox{-0.62cm}{\epsfig{figure=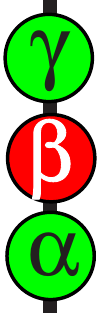,width=13pt}}$\,.  So we can indeed express any unitary operation on $n$ qubits using only red and green decorated spiders. By applying an appropriate unitary to some  $n$-qubit state, e.g.~$|+\ldots+\rangle=\raisebox{-0.14cm}{\epsfig{figure=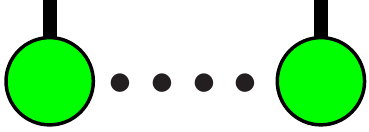,width=46pt}}$\,,  we can obtain any arbitrary $n$-qubit state. Finally,  we rely on map-state duality.  Using eq.(\ref{eq:mapstate}), we can obtain any linear map from $m$ qubits to $k$ qubits, from some $m+k$-qubit state. \hfill$\Box$

\bigskip
In joint work with Duncan  \cite{CD} we explored   how  to reason about algorithms and a variety of quantum computational models in this diagrammatic language.  There is also some related recent work in this area by Duncan and Perdrix \cite{DuncanPerdrix}.  We present two easy examples here.   

\subsubsection{Unitarity of the $CNOT$-gate}

The beauty of this example is that it uses the diagrammatic incarnation of complementarity of Theorem \ref{thm:complementary}.  We want to compute the composition of two $CNOT$-gates.  We have
\[
\left(\begin{array}{cccc}
1 & 0 & 0 & 0\\
0 & 1 & 0 & 0\\
0 & 0 & 0 & 1\\
0 & 0 & 1 & 0
\end{array}\right)
\circ
\left(\begin{array}{cccc}
1 & 0 & 0 & 0\\
0 & 1 & 0 & 0\\
0 & 0 & 0 & 1\\
0 & 0 & 1 & 0
\end{array}\right)=\ \raisebox{-0.48cm}{\epsfig{figure=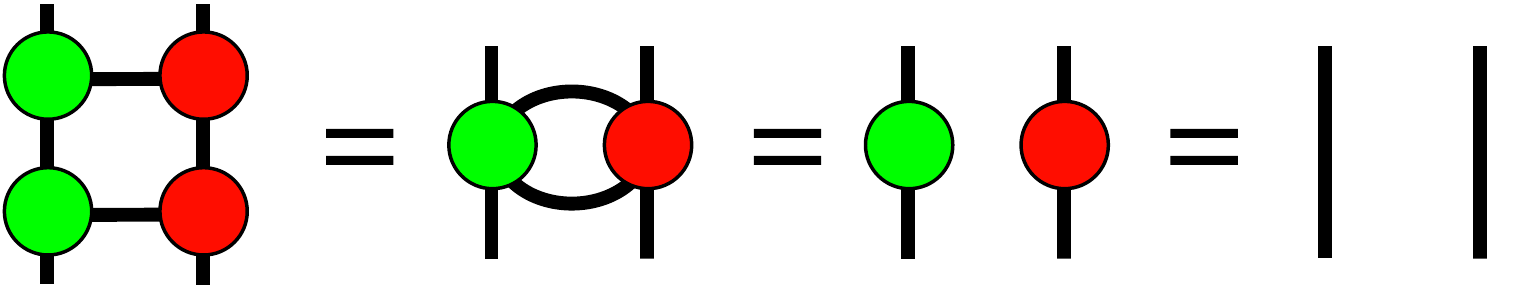,width=180pt}}
\]
where the second graphical step uses Theorem \ref{thm:complementary}.  So the $CNOT$-gate admits an inverse, namely itself. If we flip the $CNOT$-gate upside-down, we again obtain the $CNOT$-gate, so it is its own adjoint too.  Hence the adjoint to the $CNOT$-gate and the inverse to the $CNOT$-gate coincide, in both cases being the $CNOT$-gate itself, from which it follows that the  $CNOT$-gate is unitary.

\subsubsection{Universality of measurement-based quantum computing}

Measurement-based quantum computing is a new paradigm for quantum computing, which is based on the fact that a quantum measurement changes the state of the system, and hence can be used to process quantum states.  
There are several variants. Here we will focus on the one introduced by Briegel and Raussendorf  \cite{RBB}.
One starts with a number of qubits, all in a large entangled state, the so-called \em cluster state\em.  One then performs measurements on individual qubits.  Just like in quantum teleportation, it is required that one performs certain unitary corrections depending on the measurement outcomes.  For reasons of simplicity we assume that we obtained the desired measurement outcomes, so that we do not have to do those corrections.  

\begin{wrapfigure}{l}{0.28\textwidth}
\vspace{-12pt}
\begin{center}
\epsfig{figure=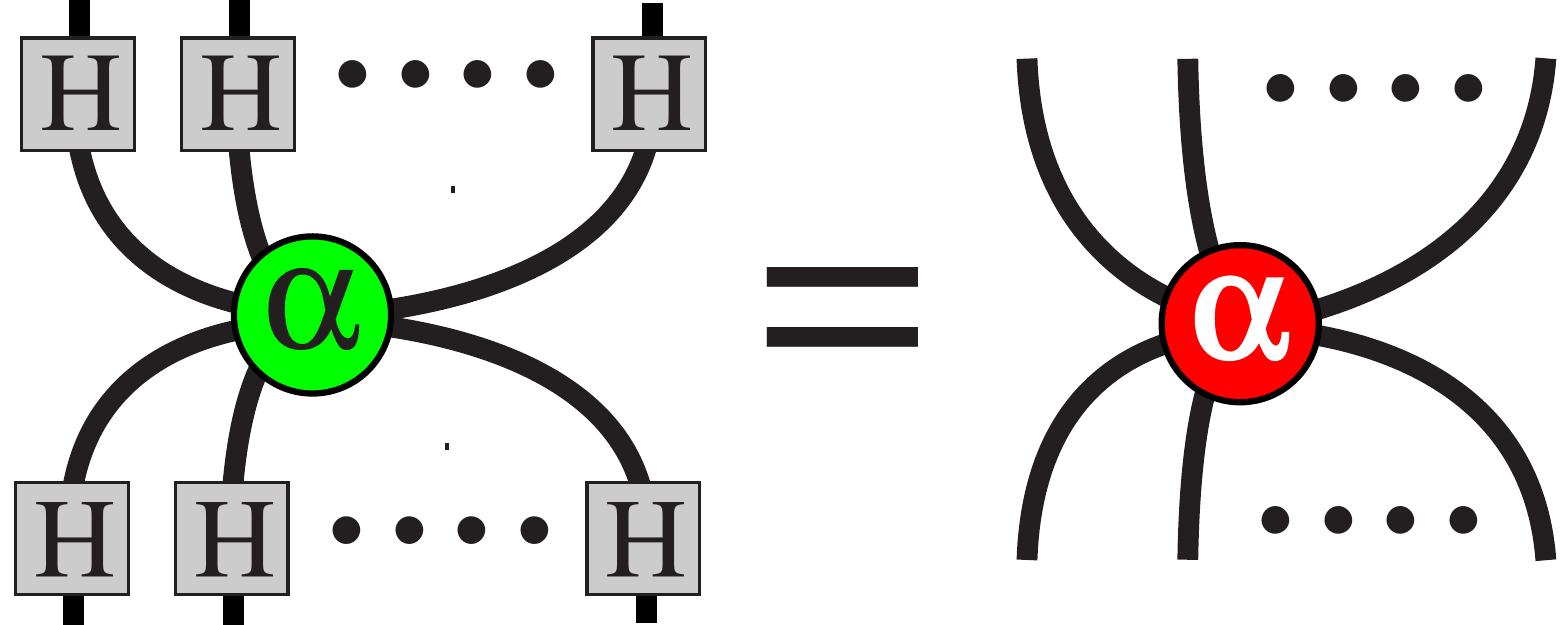,width=125pt}
   \end{center}
    \vspace{-20pt}
\end{wrapfigure}\noindent
What we will show is that by performing single qubit measurements on a cluster state, one can implement arbitrary one-qubit unitaries.   For this purpose it turns out to be useful to assume the existence of an operation that `changes the colors of spiders',  which is depicted on the left.
For qubits in ${\bf FHilb}$, the \em Hadamard gate \em plays this role, the matrix of which is 
${1\over\sqrt{2}}
\left(\begin{array}{cr}
1&1\\ 1&-1
\end{array}\right).$ 
To produce cluster states, one starts with all qubits in the $|+\rangle=\raisebox{-0.28cm}{\epsfig{figure=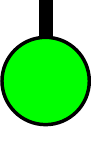,width=12pt}}$-state, and then applies the $CZ$-gate pairwise.
This $CZ$-gate, also called the  `controlled phase' gate, is explicitly given by
 \[
CZ:{\cal H}\otimes{\cal H}\to{\cal H}\otimes{\cal H}::\left\{\begin{array}{l}
|0x\rangle\mapsto |0x\rangle\\ |1x\rangle\mapsto |1Z(x)\rangle
\end{array}\right.\quad\mbox{ with}\quad Z:{\cal H}\to{\cal H}::\left\{\begin{array}{l}
|0\rangle\mapsto |0\rangle\\ |1\rangle\mapsto -|1\rangle
\end{array}\right..
 \]
It is easy to see that $CZ=(1\otimes H)\circ CX\circ(1\otimes H)$, so by relying on the `color change'-property of $H$ we obtain  $CZ=\raisebox{-0.46cm}{\epsfig{figure=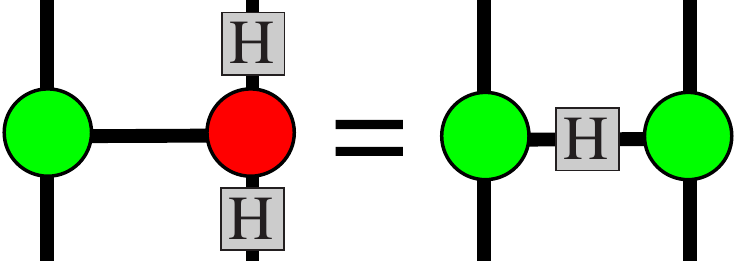,width=95pt}}$.  We can now easily derive how we can implement an arbitrary one-qubit unitary gate when the available resources are cluster states, which we can prepare with qubits in the $|+\rangle$-state and $CZ$-gates,  and bra's $\langle 0|+e^{-i\alpha}\langle 1|=\raisebox{-0.28cm}{\epsfig{figure=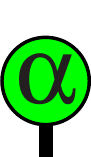,width=12pt}}$.  We have
\begin{center}
\epsfig{figure=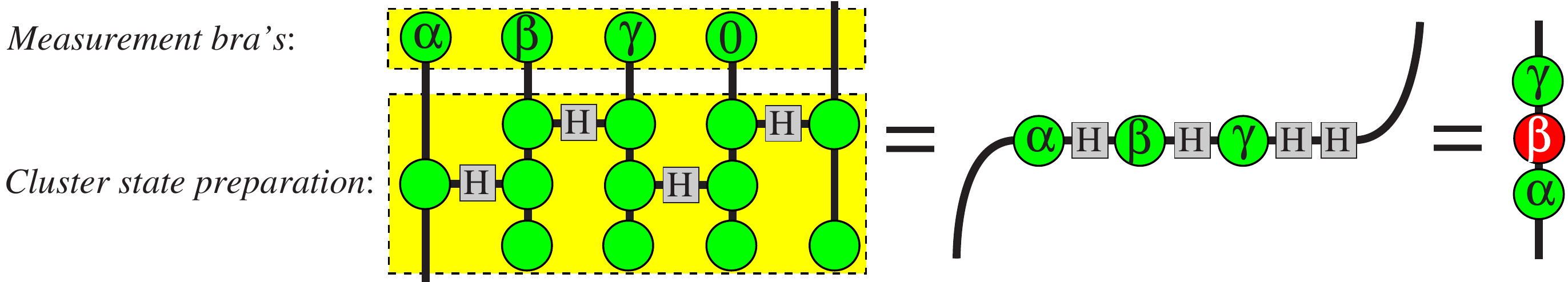,width=400pt}
   \end{center}
The first step fuses decorated spiders, and the second step is just the action of the colour changer. Firstly, it turns the $\beta$-gate red, and secondly, we apply $H\circ H=1_{\cal Q}$, which is nothing but the application of the color changer to the $\delta_1^1$-spider, that is, the identity.  The righthandside represents an arbitrary one-qubit unitary gate in terms of its Euler angle decomposition, which we discussed above.    
Since this computation involves 5 qubits, which are described in a $2^5=32$ dimensional Hilbert space, the matrix  description of this configuration requires $32\times 32$ matrices.

While this is of course a very simple example, it indicates the potential for simplifying far more complicated configurations. Implicit in the diagrammatic manipulations is the  transformation of a measurement based setup into a circuit.   Hence it also indicates the potential for translating implementations among quantum computational models.

\subsection{Example: the group-theoretic origin of quantum non-locality} 

In recent work Edwards and Spekkens \cite{CES}, we used the framework of physical theories casted as dagger symmetric monoidal categories to trace back non-local behaviors to the phase group and nothing but the phase group, and did this for a wide range of theories.  

We make precise what exactly we mean by non-local behaviors.  As mentioned in Section \ref{sec:introduction}, non-locality means that  measurements on far apart subsystems of a compound quantum systems exhibit correlations between the measured outcomes which cannot be explained as having been established 
in the past when the two systems were in close proximity. This phenomenon is typically known as the \em EPR-paradox \em \cite{EPR} or violation of \em Bell-inequalities \em \cite{Bell}. It was experimentally observed in 1982 \cite{Bellexp}. Establishing non-locality required analyzing the measurement statistics.  There is however a newer version of this story which does not involve probabilities at all.  When measuring subsystems of the compound system it only requires to look at which outcomes can occur together, and which can't.  The state which exhibits these non-local correlations is the tripartite GHZ-state $|000\rangle+|111\rangle$ \cite{GHZ}.  In our pictorial framework these GHZ-states are spiders with three front and no back legs i.e.~$\delta_0^3=\raisebox{-0.15cm}{\epsfig{figure=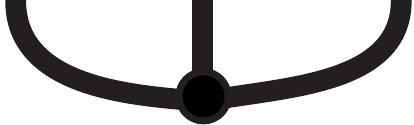,width=42pt}}$. The corresponding correlations, and more specifically,  the state the third system is in (after measuring the first two systems), is 
\[
(\psi^T\otimes \phi^T\otimes 1_{\cal Q})\circ\delta_0^3\ = \  \raisebox{-0.44cm}{\epsfig{figure=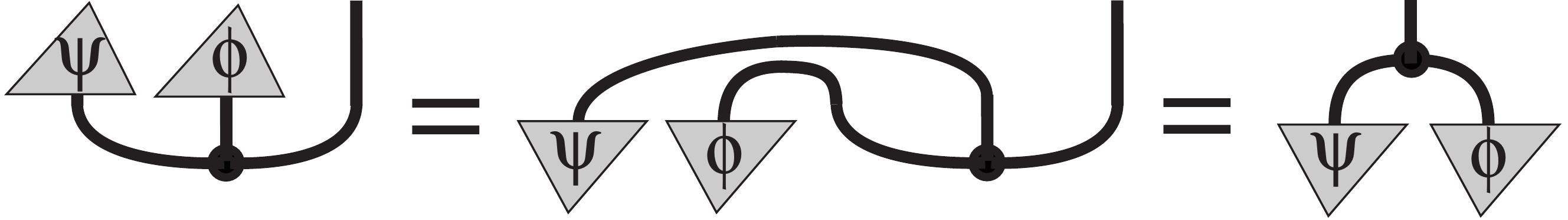,width=250pt}}\ =\ \psi\odot\phi
\]
where we now use the notation $\odot$ for arbitrary $\psi$ and $\phi$, and not just the unbiased states.  Our key theorem states that for a certain class of theories,  \em mutually unbiased theories\em, or in short MUTs, 
the correlations obtained in measurements are completely determined by the phase group.  By a MUT we mean a theory which is such that for each state $\psi$ of an elementary system $A$, and each observable $(A, \delta, \varepsilon)$,  $\psi$ is either an eigenvector or unbiased for $(A, \delta, \varepsilon)$.  While ${\bf FHilb}_2$ is not an MUT, an important fragment of ${\bf FHilb}_2$, namely \em qubit  stabilizer theory\em, is such an MUT.  
To demonstrate non-locality of quantum mechanics, this fragment suffices.

\begin{theorem}{\rm\cite{CES}}\label{thm:spek}
In any MUT the correlations obtained in measurements on the GHZ-state are completely determined by the phase group.  Hence non-local behaviors of (finitary) mutually unbiased theories are classified by the (finite) Abelian groups. 
\end{theorem}

Classifying the finite simple groups was one of the great achievements of mathematics at the end of the previous century.  Theorem \ref{thm:spek} tells us that this classification, restricted to Abelian groups,  carries over to the non-local behaviors which MUTs can exhibit.  Here we will consider MUTs with four-element phase groups.   There are exactly two irreducible four element Abelian groups, namely the cyclic four element group $Z_4$ and the Klein four group $Z_2\times Z_2$.   What are the theories they correspond to?  It turns out that $Z_4$ is the phase group of qubit  stabilizer theory, which, roughly put, is obtained by restricting the states of the qubit in quantum theory to the eigenvectors of the $Z$-, $X$- and $Y$-observables.  One can present this theory elegantly as a dagger symmetric monoidal category, which we called ${\bf Stab}$.  As already mentioned above, it is a non-local theory.  In fact, it is easily shown that having  $Z_4$ as a subgroup of the phase group is enough for a theory to be non-local, so non-locality of quantum theory is caused simply by this small four-element phase group.  Such a non-locality proof cannot  be derived from the group $Z_2\times Z_2$.
\linebreak\vspace{-12pt}

\begin{wrapfigure}{r}{0.38\textwidth}
\vspace{-10pt}
 \begin{center} 
\begin{tabular}{c|c|c|} 
 & phase group & theory\\ \hline
local: & $Z_2\times Z_2$ & ${\bf Spek}$\\ \hline
non-local: & $Z_4$ & ${\bf Stab}$\\ \hline
\end{tabular}
\end{center} 
   \vspace{-00pt}  \caption{The two possible non-local behaviors for mutually unbiased theories with four-element phase groups.}
    \vspace{-00pt}
\end{wrapfigure}\noindent 
A theory which has $Z_2\times Z_2$ as its phase group is the \em toy theory \em proposed by Spekkens \cite{Spekkens}.  This toy theory looks remarkably similar to quantum theory, and just like qubit  stabilizer theory, is also an MUT, however, it is  local.  We called  Spekkens' toy theory casted as a dagger symmetric monoidal category ${\bf Spek}$ \cite{Spek}.  A careful analysis shows that the groups $Z_4$ and $Z_2\times Z_2$ constitute the only difference between ${\bf Stab}$ and ${\bf Spek}$.

\section{Experimental verification: kindergarten quantum mechanics}

In physics and science in general, traditionally, claims have to be substantiated by experiments.  Is there any way we can substantiate our claims concerning the low-levelness of the quantum mechanical formalism via actual experiments?  Here is a sketch for such an experiment.
\par\bigskip\noindent
\textbf{\emph{Experiment.}}  Consider ten  children of ages between six and ten and consider ten high-school teachers of physics and mathematics.  The high-school teachers of physics and mathematics will have all the time they require to refresh their quantum mechanics background, and also to update it with regard to recent developments in quantum information.  The children on the other hand will have quantum theory explained in terms of the graphical formalism.  Both teams will be given a certain set of questions, for the children formulated in diagrammatic language, and for the teachers in the usual quantum mechanical formalism.  Whoever solves the most problems and solves them in the fastest time wins.  If the diagrammatic language is much more intuitive, it should in principle be possible for the children to win. 

\section{Contributors and key applications currently under development}\label{sec:ContKeyDevelopments}

The categorical axiomatisation of quantum theory, which provides the passage to the diagrammatic formalism, was initiated by Samson Abramsky and myself in \cite{AC1},  drawing inspiration from 
a theorem on diagrammatic reasoning for teleportation-like protocols in \cite{LE, Svetlichny}.  Other key contributions were made  by Peter Selinger in \cite{Selinger}, and in collaborations with Ross Duncan, Bill Edwards, Eric Paquette, Dusko Pavlovic, Simon Perdrix and Jamie Vicary \cite{Kindergarten,CPav,CD,CPV,Coecke-Paquette-Perdrix,CPaqPav}.  

Diagrammatic reasoning techniques for monoidal categories trace back to Penrose's work in the early 70's \cite{Penrose}. He used diagrams in a somewhat more informal way.  Our approach substantially relied on existing work mainly done by the `Australian School of category theory', namely by 
\linebreak\vspace{-12pt}

\begin{wrapfigure}{r}{0.44\textwidth}
\vspace{-10pt}
\begin{center}
\epsfig{figure=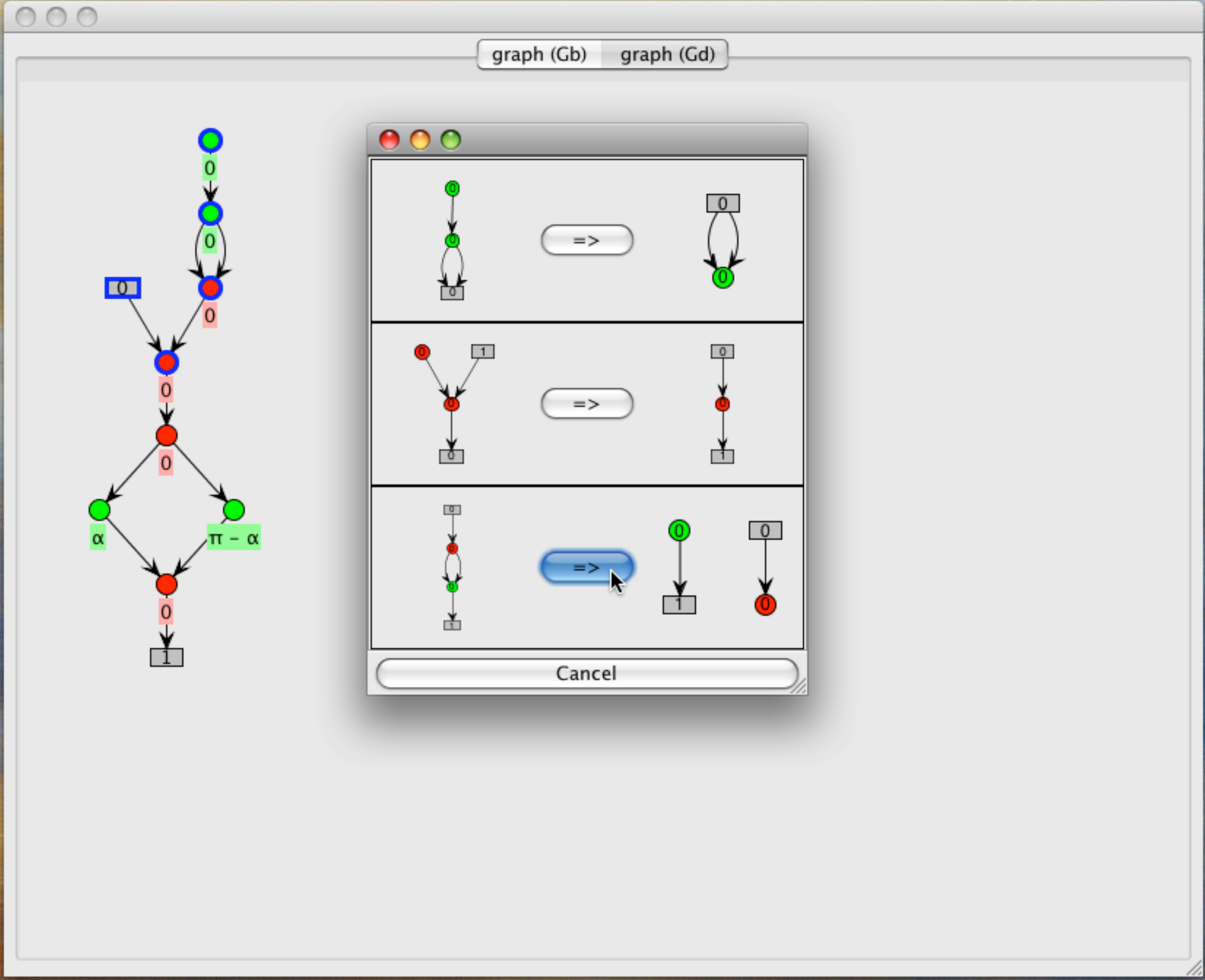,width=200pt}
   \end{center}
    \vspace{-10pt}
    \caption{{\tt quantomatic} software.}\label{pic:quanto}
    \vspace{-16pt}
\end{wrapfigure}\noindent
Kelly,  Carboni, Walters, Joyal, Street and Lack in \cite{KellyLaplaza,CarboniWalters,JoyalStreet,Lack}. Among other things, they provided a 
rigourous mathematical foundation for diagrammatic reasoning.  Related graphical  
methods have been around for a bit more than a decade now in mathematical physics and pure mathematics, for example in \cite{LNPBS, BaezDolan, Kuperberg, FuchsRunkelSchweigert, Morrison} and references therein. A proponent of these methods, John Baez, has several available postings \cite{ThisWeek}.

An important current development is \em automated reasoning\em, that is, to make a computer perform  the graphical 
reasoning rather than ourselves. Key to this is that these categorical structures are discrete, as opposed to the continuum of complex numbers.  Can we make a computer prove new theorems about quantum theory?  We think so. A team of researchers in Oxford and Edinburgh, Dixon, Duncan and Kissinger,  is currently in the process of producing such a piece of on pictures based automated reasoning software \cite{Dixon}, named {\tt quantomatic}.   
Results such as Theorem \ref{ThmCompletenessSelinger} are very important for these attempts to automate quantum reasoning.  They tell us the space of theorems which a `theorem prover' based on 
diagrammatic logic is able to prove.  

At the same time these pictures provide a new axiomatic foundation for quantum theory, with many degrees of structural freedom.  Hence it provides a canvas to study theories more general than quantum theory.  This enables us to understand what makes quantum theory so special.  
Since this axiomatic foundation is very flexible, it also has the potential for unification of quantum theory with other theories, hence for crafting new theories of physics.

\section*{Acknowledgements}

We are grateful for constructive critical feedback by Andreas Doering, Lucien Hardy, Basil Hiley, Chris Isham, Terry Rudolph and Rob Spekkens which affected our presentation here.  We thank  Jacob Biamonte, Peter Morgan, Rob Spekkens, James Whitfield and the anonymous reviewers for carefully reading this paper and providing appropriate feedback.  The author is supported by an EPSRC Advanced Research Fellowship entitled `The Structure of Quantum Information and its Ramifications for IT', by an FQXi Large Grant entitled `The Road to a New Quantum Formalism: Categories as a Canvas for Quantum Foundations', and by an EC-FP6-STREP entitled `Foundational Structures in Quantum Information and Computation'.


\begin{thebibliography}{10}
\markboth{Bob Coecke}{Contemporary Physics}

\bibitem{vN} 
J.~von Neumann, {\it Mathematische Grundlagen der Quantenmechanik}. Springer-Verlag,  1932.   
\bibitem{Birk} 
G.~Birkhoff, \em von Neumann and lattice theory\em.  Bull.  Am. Math. Soc. {\bf 64} (1958), pp. 50--56.

\bibitem{Redei}
M.~R\'edei,  \em Why John von Neumann did not like the Hilbert space formalism of quantum mechanics (and what he liked instead)\em. Stud. Hist. Phil. Mod. Phys. {\bf 27} (1997), pp. 493--510.

\bibitem{BvN} 
G.~Birkhoff and J.~von Neumann,  \em The logic of quantum mechanics\em.  Ann. Math.  {\bf 37} (1936), pp.  823--843. 

\bibitem{CMW}
B.~Coecke, D.~J.~Moore and A.~Wilce, \em Operational quantum logic: an overview\em,   in \em  Current Research in Operational Quantum Logic: Algebras, Categories and Languages\em,  
Springer,  2000, pp. 1--36.  arXiv:quant-ph/0008019

\bibitem{Bub}
J.~Bub, \em Interpreting the quantum world\em. Cambridge University Press,  1997.

\bibitem{Peres}
A.~Peres, \em Interpreting the quantum world\em. Stud. Hist. Phil. Mod. Phys. {\bf 29} (1998) pp. 611.  arXiv:quant-ph/9711003

\bibitem{BBC} 
C.~H.~Bennett, G.~Brassard, C.~Cr\'epeau, R.~Jozsa, A.~Peres and W.~K.~Wooters,  \em Teleporting an unknown quantum state via dual classical and Einstein-Podolsky-Rosen channels\em.  Phys. Rev. Lett. {\bf 70}  (1993), pp. 1895--1899.

\bibitem{Ekert91}
A.~Ekert, \em Quantum cryptography based on Bell's theorem\em. Phys. Rev. Lett. {\bf 67} (1991), pp. 661--663.

\bibitem{QKD}
http://en.wikipedia.org/wiki/Quantum$\underline{\ \, }$cryptography

\bibitem{Schrodinger}
E.~Schr\"odinger, \em Discussion of probability relations between separated systems\em. Proc. Cam. Phil. Soc. {\bf 31} (1935),  pp. 555--563.

\bibitem{SequentCalculus}
http://en.wikipedia.org/wiki/Sequent$\underline{\ \, }$calculus

\bibitem{Girard}
J.-Y.~Girard, \em Linear logic\em. Theor. Comp. Sc. {\bf 50} (1987), pp. 1--102.  See also
http://en.wikipedia.org/wiki/Linear$\underline{\ \, }$logic

\bibitem{Dieks}
D.~G.~B.~J.~Dieks, \em Communication by EPR devices\em. Phys. Lett. A {\bf 92} (1982), pp. 271--272.

\bibitem{WZ}
W.~Wootters and W.~Zurek, \em A single quantum cannot be cloned\em. Nature {\bf 299} (1982), pp. 802--803.

\bibitem{NoCloning}
http://en.wikipedia.org/wiki/No$\underline{\ \, }$cloning$\underline{\ \, }$theorem

\bibitem{Pati}
A.~K.~Pati and S.~L.~Braunstein, \em Impossibility of deleting an unknown quantum state\em. Nature {\bf 404} (2000), pp. 164--165. 

\bibitem{DanosReignier}
V.~Danos and L.~Reignier, \em The structure of multiplicatives\em. Arch. Math. Logic {\bf 28} (1989), pp. 181--203.

\bibitem{RossThesis}
R.~W.~Duncan, \em Types for Quantum Computing\em. D.Phil.~thesis. University of Oxford, 2006.

\bibitem{EilenbergMacLane} 
S.~Eilenberg and S.~MacLane (1945) \em General theory of natural equivalences\em. Transactions of the American Mathematical Society, {\bf 58}, pp. 231--294.

\bibitem{Cats}
B.~Coecke, \em Introducing categories to the practicing physicist\em, in: What is Category Theory?  Advanced Studies in Mathematics and Logic {\bf 30}, Polimetrica Publishing, 2006, pp. 45--74.   arXiv:0808.1032

\bibitem{Baez}
J.~C.~Baez, \em Quantum quandaries: a category-theoretic perspective\em, in: The Structural Foundations of Quantum Gravity, D.~Rickles, S.~French and J.~T.~Saatsi (Eds), Oxford University Press, 2006, pp. 240--266. arXiv:quant-ph/0404040

\bibitem{CPaqII}
B.~Coecke and  E.~O.~Paquette, \em Categories for the practising physicist\em, in: New Structures for Physics, B.~Coecke (ed.),  Springer Lecture Notes in Physics, 2009. arXiv:0905.3010 

\bibitem{LNPBS}
J.~Baez and M.~Stay, \em Physics, topology, logic and
computation: a Rosetta Stone\em, in: New Structures for Physics,
B.~Coecke (ed.),  Springer Lecture Notes in Physics, 2009. 

\bibitem{LNPAT}
S.~Abramsky and N.~Tzevelekos, \em Introduction to categories
and categorical logic\em, in: New Structures for Physics,
B.~Coecke (ed.),  Springer Lecture Notes in Physics, 2009. 

\bibitem{Selinger2}
P.~Selinger,  \em A survey of graphical languages for monoidal categories\em, in: New Structures for Physics,
B.~Coecke (ed.),  Springer Lecture Notes in Physics, 2009.

\bibitem{LawvereSchanuel}
F.~W.~Lawvere and S.~H.~Schanuel,
\em Conceptual Mathematics: A First Introduction to Categories\em. Cambridge UP.

\bibitem{MacLane}
S.~MacLane (1998) \em Categories for the Working Mathematician. 2nd edition\em. Springer-Verlag.

\bibitem{LE}
B.~Coecke,  \em The logic of entanglement\em. Research Report PRG-RR-03-12, 2003.  arXiv:quant-ph/0402014 (8 page short version)  http://web.comlab.ox.ac.uk/oucl/publications/tr/rr-03-12.html (full 160 page version) 

\bibitem{Laforest}
M. Laforest, J. Baugh, R. Laflamme, \em Time-reversal formalism applied to maximal bipartite entanglement: Theoretical and experimental exploration\em. Phys. Rev. A {\bf 73}  (2006), 032323.
arXiv:quant-ph/0510048

\bibitem{Gottesman}
D.~Gottesman and I.~L.~Chuang, \em Quantum teleportation is a universal computational primitive\em.  Nature {\bf 402}  (1999), pp. 390--393.   arXiv:quant-ph/9908010

\bibitem{Swap}
M.~\.{Z}ukowski, A.~Zeilinger, M.~A.~Horne and A.~K.~Ekert \em `Event-ready-detectors' Bell experiment via entanglement swapping\em. Phys. Rev. Lett. {\bf 71} (1993), pp. 4287--4290.

\bibitem{Benabou} 
J.~B\'enabou, \em Categories avec multiplication\em. Compt. Rend. S\'eanc. Acad\ Sci. Paris  {\bf  256} (1963), pp. 1887--1890.

\bibitem{Haag}
R.~Haag, \em Local Quantum Physics: Fields, Particles, Algebras\em. Springer-Verlag, 1992.

\bibitem{JoyalStreet}
A.~Joyal and R.~Street, \em The Geometry of
tensor calculus \em I.  Adv. Math. {\bf 88} (1991), pp. 55--112.

\bibitem{deLL}
B.~Coecke, \em De-linearizing linearity: projective quantum axiomatics from strong compact closure\em.  ENTCS {\bf 170} (2007), pp. 47--72.
arXiv:quant-ph/0506134

\bibitem{AC1}
S.~Abramsky and B.~Coecke, \em  A categorical semantics of quantum protocols\em, in: Proceedings of 19th IEEE conference on Logic in Computer Science, IEEE Press, 2004, pp. 415--425.  arXiv:quant-ph/0402130.

\bibitem{Coecke-Paquette-Perdrix}
B.~Coecke, E.~O.~Paquette and S.~Perdrix, \em Bases in
diagrammatic quantum protocols\em. ENTCS {\bf 218}  (2008), pp. 131--152.
arXiv:0808.1037

\bibitem{Selinger}
P.~Selinger,  \em Dagger compact closed categories and completely positive maps\em.  ENTCS {\bf 170} (2007),  pp. 139--163. 

\bibitem{CPMbob}
B.~Coecke \em Complete positivity without positivity and without  compactness\em. Research Report PRG-RR-07-05, 2007.
http://web.comlab.ox.ac.uk/oucl/publications/tr/rr-07-05.html

\bibitem{JSV}  
A.~Joyal, R.~Street  and D.~Verity, \em Traced monoidal categories\em. Proc. Camb. Phil. Soc. {\bf 119} (1996), pp. 447--468.

\bibitem{KellyLaplaza}
G.~M.~Kelly and M.~L.~Laplaza, \em Coherence for compact closed categories\em.  J. Pure Appl. Alg. {\bf 19}  (1980), pp. 193--213.

\bibitem{Atiyah}
M.~Atiyah, \em Topological quantum field theories\em. Inst.~Hautes \'Etudes Sci.~Publ.~Math. {\bf 68} (1989), pp. 175--186.

\bibitem{Kock}
J.~Kock, \em Frobenius Algebras and 2D Topological Quantum Field
Theories\em. Cambridge University Press, 2003. 


\bibitem{Plotkin}
M.~Hasegawa, M.~Hofmann and G.~Plotkin, \em Finite dimensional vector spaces are complete for traced symmetric monoidal categories\em, 
LNCS {\bf 4800}, 2008, pp. 367--385. 

\bibitem{Selinger3}
P.~Selinger,  \em Finite dimensional Hilbert spaces are complete for dagger compact closed categories\em.  In: Proceedings of the 5th International Workshop on Quantum Physics and Logic, B.~Coecke and P.~Panangaden, (eds), 2009. 

\bibitem{AbrClone}
S.~Abramsky (2008) \em No-Cloning in categorical quantum mechanics\em, in: Semantic Techniques for Quantum Computation, I.~Mackie and S.~Gay (eds), Cambridge University Press, to appear.

\bibitem{CPav}
B.~Coecke and D.~Pavlovic, \em Quantum measurements without sums\em, in: Mathematics of Quantum Computing and Technology, G.~Chen, L.~Kauffman, S.~Lamonaco (eds), Taylor and Francis, 2007, pp. 567--604.  arXiv:quant-ph/0608035.

\bibitem{CPV}
B.~Coecke, D.~Pavlovic, and J.~Vicary (2008) \em A new description of
orthogonal bases\em. arXiv:0810.0812 

\bibitem{CPaq}
B.~Coecke and E.~O.~Paquette, \em POVMs and {N}aimark's theorem without sums\em. ENTCS {\bf 210}  (2008), pp. 15--31. arXiv:quant-ph/0608072

\bibitem{CarboniWalters}
A.~Carboni and R.~F.~C.~Walters,  \em Cartesian bicategories {I}\em.  J. Pure Appl. Alg. {\bf 49} (1987), pp. 11--32.

\bibitem{Lack}
S.~Lack, \em Composing PROPs\em. Theor. Appl.
Cat. {\bf 13} (2004), pp. 147--163. 

\bibitem{CPaqPav}
B.~Coecke, E.~O.~Paquette and D.~Pavlovic, \em Classical and quantum structuralism\em, in: Semantic Techniques for Quantum Computation, I.~Mackie and S.~Gay (eds),  Cambridge University Press. arXiv:0904.1997 

\bibitem{CD}
B. Coecke and R. W. Duncan, \em Interacting quantum observables\em, in: 
Proceedings of the 35th International Colloquium on Automata, Languages and Programming, LNCS {\bf 5126}, Springer, 2008, pp. 298--310. Extended version: arXiv:0906.4725 

\bibitem{Kraus}
K.~Kraus, \emph{Complementary observables and uncertainty
  relations}.  Phys. Rev. D \textbf{35} (1987), pp. 3070--3075.

\bibitem{Schwinger}
J.~Schwinger, \emph{Unitary operator bases}. Proceedings of the
National Academy of Sciences of the U.S.A.~\textbf{46}, 1960. 

\bibitem{W_GHZ}
W.~D\"ur, G.~Vidal, J.~I.~Cirac, \em Three qubits can be entangled in two inequivalent ways\em. Phys. Rev. A 62  (2000), 062314.

\bibitem{NielsenChuang}
M.~A.~Nielsen and L.~Chuang, \em Quantum Computation and Quantum Information\em.  Cambridge University Press, 2000.

\bibitem{DuncanPerdrix}
R.~Duncan and S.~Perdrix, \em Graphs States and the necessity of Euler Decomposition\em. arXiv:0902.0500

\bibitem{RBB} 
R.~Raussendorf, D.~E.~Browne  and H.-J.~Briegel, \em Measurement-based quantum computation on cluster states\em. Phys. Rev.  AÊ {\bf 68} (2003), 022312.  arXiv:quant-ph/0301052.

\bibitem{CES}
B.~Coecke, B.~Edwards and R.~Spekkens, \em The group theoretic origin of non-locality for qubits\em.   web.comlab.ox.ac.uk/ publications/publication3026-abstract.html

\bibitem{EPR}
A.~Einstein, B.~Podolsky and N.~Rosen, \em Can quantum-mechanical description of physical reality be considered complete? \em  Phys. Rev. {\bf 47} (1935), pp. 777-Ð780.

\bibitem{Bell}
J.~Bell,  \em On the Einstein Podolsky Rosen paradox\em. Physics {\bf 1} (1964), pp. 195.

\bibitem{Bellexp}
A.~Aspect, J.~Dalibard and G.~Roger, \em Experimental test of Bell's inequalities using time-varying analyzers\em. Phys. Rev. Lett. {\bf 49} (1982), pp. 1804--1807.

\bibitem{GHZ} 
D.~M.~Greenberger, M.~A.~Horne, A.~Shimony and A.~Zeilinger (1990) \em Bell's theorem without inequalities\em.  Am. J. Phys. {\bf 58}, pp. 1131--1143.

\bibitem{Spekkens}
R.~Spekkens, \em Evidence for the epistemic view of quantum
states: A toy theory\em. Phys. Rev. A {\bf 75} (2007), 032110.

\bibitem{Spek}
B.~Coecke and B.~Edwards, \em Toy quantum categories\em.    arXiv:0808.1037

\bibitem{Svetlichny}
G.~Svetlichny, \em Tensor universality, quantum information flow, Coecke's theorem, and generalizations\em. Unpublised, 2006. arXiv:quant-ph/0601093

\bibitem{Kindergarten}
B.~Coecke,  \em Kindergarten quantum mechanics\em, in: Quantum Theory: Reconsiderations of the Foundations {\rm III}, AIP Press, 2005, pp.  81--98. arXiv:quant-ph/0510032

\bibitem{Penrose} 
R.~Penrose, \em Applications of negative dimensional tensor calculus\em,
in: Combinatorial Mathematics and its Applications, Academic Press, 1971, pp.
221--244. 


\bibitem{BaezDolan}
J.~C.~Baez and J.~Dolan, \em Higher-dimensional algebra and topological quantum field theory\em. J.  Math. Phys. {\bf 36} (1995),  pp. 6073--6105. arXiv:q-alg/9503002

\bibitem{Kuperberg}
G.~Kuperberg, \em Spiders for rank 2 Lie algebras\em. Comm. Math. Phys. {\bf 180} (1996), pp. 109--15. arXiv:q-alg/9712003

\bibitem{FuchsRunkelSchweigert}
J.~Fuchs, I.~Runkel and C.~Schweigert,
\em TFT construction of RCFT correlators I: Partition Functions\em, Nuc. Phys. B. {\bf 646} (2002), pp. 353--497. arXiv:hep-th/0204148

\bibitem{Morrison}
S.~E.~Morrison, \em A diagrammatic category for the representation theory of $U_q(sl_n)$\em.
PhD~thesis University of California at Berkeley, 2007.

\bibitem{ThisWeek}
J.~C.~Baez, \em This weeks finds in mathematical physics \em {\bf 1--256},  (1993--2009) .  http://www.math.ucr.edu/home/baez/ TWF.html --- see also the \em n-category caf\'e\em: http://golem.ph.utexas.edu/category/

\bibitem{Dixon}
L.~Dixon, R.~W.~Duncan and A.~Kissinger, http://dream.inf.ed.ac.uk/projects/quantomatic/.

\end{thebibliography}
\end{document}